%% file: PAHs_in_HII_rev1_clean.tex
\documentclass[longauth]{aa} 
\makeatletter
\let\linenumbers\relax

\makeatother

\usepackage{graphicx}
\usepackage{txfonts}
\usepackage[dvipsnames]{xcolor}
\usepackage[breaklinks, colorlinks, citecolor=Cerulean]{hyperref}
\usepackage{color}
\usepackage{lastpage}
\usepackage{tabularx}
\usepackage{tablefootnote} 

\newcommand{\SII}{[S~{\sc ii}]}
\newcommand{\SIII}{[S~{\sc iii}]}
\newcommand{\OIII}{[O~{\sc iii}]}
\newcommand{\NII}{[N~{\sc ii}]}
\newcommand{\HII}{H~{\sc ii}}

\newcommand{\HI}{H~{\sc i}}
\newcommand{\Ha}{H$\alpha$}
\newcommand{\Hb}{H$\beta$}

\newcommand{\ergs}{\,\mbox{erg}\,\mbox{s}^{-1}}

\newcommand{\JWST}{\textit{JWST}}
\newcommand{\SIIHa}{[S~{\sc ii}]/H$\alpha$}
\newcommand{\SIIISII}{[S~{\sc iii}]/[S~{\sc ii}]}

\newcommand{\OIIIHb}{[O~{\sc iii}]/H$\beta$}
\newcommand{\RPAH}{${\rm R_{PAH}}$}
\newcommand{\RPAHst}{${\rm R_{PAH}^*}$}

\newcommand{\be}{\begin{equation}}
\newcommand{\ee}{\end{equation}}

\def\revone{}
\begin{document} 

\newcommand{\numHII}{17151} % total number of HII regions
\newcommand{\numSNRs}{\revone{396}} %394 % total number of SNRs

   % \title{\note{Preliminary, better suggestions are welcome:}\newline Investigating of pathways of polycyclic aromatic hydrocarbons destruction and formation in relation to stellar feedback} % in star-forming regions as seen by JWST and MUSE observations.}
   \title{Polycyclic aromatic hydrocarbons destruction in star-forming regions across 42 nearby galaxies}
    \titlerunning{PHANGS-JWST and MUSE: PAHs in star-forming regions}
%   \subtitle{}
    \include{authors}

   \date{Received ????? ??, 2025; accepted ????? ??, ????}

% \abstract{}{}{}{}{} 
% 5 {} token are mandatory
 
  \abstract
  {
  %Modified Janice's version:
  Polycyclic aromatic hydrocarbons (PAHs) are widespread in the interstellar medium (ISM) of $\sim$Solar metallicity galaxies, where they play a critical role in ISM heating, cooling, and reprocessing stellar radiation. The PAH fraction, the abundance of PAHs relative to total dust mass, is a key parameter in ISM physics. Using JWST and MUSE observations of 42 galaxies from the PHANGS survey, we analyze the PAH fraction in over 17 000 \HII\ regions spanning a gas-phase oxygen abundance of $\rm 12+\log(O/H) = 8.0$–8.8 ($Z \sim 0.2$–1.3~$Z_\odot$), and $\sim400$ isolated supernova remnants (SNRs). We find a significantly lower PAH fraction \revone{toward} \HII\ regions compared to a reference sample of diffuse ISM areas at matched metallicity. At $\rm 12+\log(O/H) > 8.2$, the PAH fraction \revone{toward} \HII\ regions is strongly anti-correlated with the local ionization parameter, suggesting that PAH destruction is correlated with ionized gas and/or hydrogen-ionizing UV radiation. %Projection effects from surrounding diffuse gas may contribute to observed variations. 
  At lower metallicities, the PAH fraction declines steeply in both \HII\ regions and the diffuse ISM, likely reflecting less efficient PAH formation in metal-poor environments. Carefully isolating dust emission from the vicinity of optically-identified supernova remnants, we see evidence for selective PAH destruction from measurements of lower PAH fractions, which is, however, indistinguishable at $\sim 50$ pc scales. % We find no significant difference in PAH fraction between SNRs and the diffuse ISM at $\sim 50$ pc resolution, suggesting that shocks from supernovae do not strongly affect PAHs in unresolved studies, although we do see signatures of selective PAH destruction at finer scales after background subtraction. 
  Overall, our results point to ionizing radiation as the dominant agent of PAH destruction within \HII\ regions, with metallicity playing a key role in their global abundance in galaxies.

}

   \maketitle
%
%-------------------------------------------------------------------

\section{Introduction}
\label{sec:intro}

Polycyclic aromatic hydrocarbons (PAHs) are carbon-based nanoparticles that are ubiquitous in the interstellar medium (ISM). Together with stochastically heated very small ($20-30$~\AA) dust grains, PAHs are dominant contributors to the mid-infrared (mid-IR) spectra of galaxies \citep[e.g.][]{Draine2007}. PAHs absorb ultraviolet (UV) and optical photons and re-emit energy in the IR range in the form of continuum emission and several strong emission features at 3.3, 6.2, 7.7, 8.6, 11.3, 12.7, and 17~$\mu$m \citep{Tielens2008, Li2020}. As a result, PAHs are responsible for reprocessing up to 20\% of all stellar UV/optical radiation \citep{Smith2007}. Observationally, PAHs are considered good tracers of both cold gas \citep{Gao2019, Gao2022, Leroy2023, Sandstrom2023, Whitcomb2023, Chown2024phangs} and heating processes due to their association with star formation activity \citep{Peeters2004, Calzetti2007, Calapa2014, Cluver2017, Pathak2024, Gregg2024}.

PAHs are important for the evolution of the ISM and star formation in galaxies: they provide shielding from UV radiation and dominate photoelectric heating in neutral gas \citep{Wolfire1995, Wolfire2003, Li2020, Draine2021}, and thus influence how much gas is available to form molecular clouds.  Meanwhile, the mechanisms of their formation and destruction are not fully understood. Possible scenarios include PAH formation in the atmospheres of evolved stars \cite[e.g.][]{Latter1991, Cherchneff1992}, or directly in molecular clouds \citep[e.g.][]{Greenberg2000, Sandstrom2010, Chastenet2019, Burkhardt2021, McGuire2021} via shattering of larger dust grains \cite[e.g.][]{Jones1996, Hirashita2009, Seok2014, Wiebe2014}. Ionization and photodestruction by hard UV radiation \citep[e.g.][]{Allain1996, Montillaud2013, Pavlyuchenkov2013, Wenzel2020, Egorov2023}, sputtering and fragmentation in the ionized gas \citep{Micelotta2010ig, Bocchio2012}, and shocks \citep{OHalloran2006, Jackson2006, Micelotta2010, Zhang2022} are the main mechanisms considered for PAH destruction. %These processes should reduce the abundance of PAHs in star-forming regions. 

Observations of nearby galaxies suggest that the fraction of the total dust mass in PAHs, $q_{\rm PAH}$, remains relatively constant across their diffuse ISM, with a typical value of $q_{\rm PAH} \sim 5$\% for the Milky Way and near Solar metallicity conditions \citep[e.g.][]{Draine2007b, Chastenet2019,Aniano2020, Sutter2024}. However, a prominent decrease in tracers of $q_{\rm PAH}$ is typically observed \revone{toward} \HII\ regions \citep{Helou2004, Relano2009, Anderson2014, Chastenet2019, Chastenet2023, Egorov2023, Pedrini2024, Sutter2024}. 

Resolved and unresolved studies of low-metallicity galaxies show a deficit of PAHs in their ISM. 
% Unresolved studies of kpc-sized star-forming complexes or entire galaxies also demonstrate that the PAH fraction decreases in low-metallicity environments. 
Such a decrease of the PAH fraction in a low-metallicity environment is typically explained through the effects of harder ionizing radiation or more ubiquitous supernova shocks efficiently destroying PAHs at lower metallicities, or that the deficiency of cold dense gas and gas-phase carbon prevents \revone{PAH formation} under such conditions \citep[e.g.][]{Engelbracht2005, Madden2006, Engelbracht2008, Draine2007b, Galliano2008, Khramtsova2013, Remy-Ruyer2015, Whitcomb2024}. %Meanwhile, particular physical processes regulating such metallicity dependence remain unclear.  
% These observational studies indicate that the properties of star-forming regions (e.g. age, metallicity, gas density) define the local balance between the processes of PAH formation and destruction.

Analyzing PAHs and their connection to local ISM conditions and ionizing sources in a large and representative sample of resolved star-forming regions is a key to understanding the mechanisms of their formation and destruction.
\revone{Multiwavelength} observations of star-forming regions with high spatial resolution ($<100$~pc scales) sufficient to isolate individual \HII\ regions and associated photo-dissociation regions (PDR) from the surrounding diffuse ISM and other \HII\ regions are crucial to disentangle between the local and global effects on the PAH evolution. Before launching JWST, such resolved studies of PAHs were only possible in our Galaxy \cite[e.g.][]{Povich2007, Anderson2014, Binder2018}, and a handful of very nearby galaxies \cite[e.g.][]{Bolatto2007, Sandstrom2010, Wiebe2011, Lebouteiller2011, Chastenet2017, Maragkoudakis2018, Chastenet2019, Mallory2022}. %, which are not representative of the diversity of physical conditions in the ISM. 
JWST observations now enable investigations of PAHs \revone{toward} individual \HII\ regions, supernova remnants (SNRs), star clusters and their immediate surroundings in relatively distant galaxies \citep[e.g.][]{Dale2023, Egorov2023, Sutter2024, Chastenet2023, Sandstrom2023b, Pedrini2024, Gregg2024, Baron2024a, Ujjwal2024, Baron2025}, or in unresolved star-forming regions in galaxies at redshifts up to $z\sim2$ \citep{Shivaei2024}. %Combining JWST observations and optical integral field spectroscopy (IFS) for the large and representative samples of resolved star-forming regions, we can better understand how different physical conditions of the ISM and properties of stellar sources regulate the evolution of PAHs in different environments.

Using the JWST images and MUSE integral-field spectroscopic (IFS) observations for four galaxies from the Physics at High Angular Resolution in Nearby Galaxies Survey (PHANGS; \citealt{Leroy2021}), \citet{Egorov2023} investigated the mechanisms of PAH destruction in $\sim 1500$ high-metallicity \HII\ regions. In that study, they found that the PAH fraction \revone{toward} the \HII\ regions is tightly correlated with the ionization parameter. They suggested that this might indicate that the PAH destruction in the \HII\ regions is regulated by hydrogen-ionizing radiation. In this paper, we extend the analysis from \citet{Egorov2023} to all 42 PHANGS-JWST galaxies with available MUSE data (either from PHANGS or archival observations) providing us with a sample of $\sim 17200$ \HII\ regions with oxygen abundance (a proxy for gas-phase metallicities) spanning a range of $\rm 12 + \log(O/H) = 8.0-8.8$ ($Z = 0.2-1.3Z_\odot$), with a few regions outside this range. This contains a large sample of $\sim 5600$ marginally resolved (most luminous) \HII\ regions for which we can minimize the contamination by diffuse ISM. We also extend the analysis to a sample of more than $400$ isolated SNRs from the \citet{Li2024} catalog. The result is by far the largest systematic measurement of band ratios tracing the PAH fraction in spectroscopically characterized ionized nebulae. With these samples, we investigate the role of ionizing radiation and shocks in selective destruction of PAHs and how the effects of the relevant physical processes change with metallicity. 

The paper is organized as follows. In Section~\ref{sec:obs} we describe the underlying observational data. Section~\ref{sec:analysis} provides details on the methods and criteria implemented for selection of the \HII\ regions and deriving the properties of gas, dust and young stars there. Section~\ref{sec:results} describes results from the comparison of our tracer for the PAH fraction and properties of the \HII\ regions and SNRs. In Section~\ref{sec:discussion}, we discuss these results and consider the potential observational biases in this study. Section~\ref{sec:summary} summarizes our results. 

% \note{some random notes and useful references to include:}

% \cite{Zhang2022} -- see PAHs destruction by shocks in AGNs

% \cite{Whitcomb2024} -- PAHs vs metallicity for 4 galaxies from the Spitzer-IRS spectra. They find that different PAH species (i.e. band-to-total and band-to-band) do not change in HII regions compared to non-HII regions, but their total fraction reduces (in agreement with our results). They also see strong decline in PAH ratio starting from 12+log(O/H)$\sim8.5$, which is much higher than in our results. Evolution of grain sizes with metallicity.

% \cite{Rigopoulou2024} -- study PAHs features in several galaxies hosting AGNs using the JWST MIRI-MRS and NIRSpec-IFU data. Do not see differences in grain sizes among the spatially resolved regions of the galaxies studied there. There is no evidence for preferential destruction of the smallest grains, contrary to earlier findings. Neither extinction nor dehydrogenation play a crucial role in setting the observed PAH bands, but PAH charge plays a significant role in PAH interband variations. 

% \cite{Jackson2006} -- We propose that the lack of diffuse 8 $\mu$m emission in low-metallicity systems may be due to the destruction of dust grains by supernova shocks, assuming a long timescale to regrow dust.

\section{Observations}
\label{sec:obs}
\subsection{JWST}\label{sec:obs_jwst}

\begin{table*}[!htbp]
\centering
\caption{Properties of the galaxies}
\label{tab:sample}
{\small
\begin{tabular}{ccccccccccccc}
\hline
Galaxy & $D^{1}$ & $R_{25}^{2}$ & $\log M_*^{3}$  & SFR$^{3}$   & 12+log(O/H) range & $\mathrm{N_{HII}}$ & $\mathrm{N_{SNRs}}^{4}$ & JWST & MUSE\\
       & (Mpc) & (kpc) & ($M_\odot$) &  ($M_\odot\ \mathrm{yr}^{-1}$) &  &  & & source & source \\
\hline
IC1954 & 12.8 & 5.57 & 9.67 & 0.36 & $8.28 - 8.56$ &  593 & $-$ &  {\scriptsize Cycle~2$^6$ } & {\scriptsize 0108.B-0249 PI: Kreckel$^5$ } \\ 
IC5273 & 14.2 & 6.31 & 9.72 & 0.54 & $8.31 - 8.51$ &  204 & $-$ &  {\scriptsize Cycle~2$^6$ } & {\scriptsize 099.B-0242 PI: Carollo } \\ 
IC5332 & 9.0 & 7.95 & 9.67 & 0.41 & $8.23 - 8.56$ &  203 & 6 &  {\scriptsize Cycle~1$^7$ } & {\scriptsize PHANGS-MUSE LP$^8$ } \\ 
NGC628 & 9.8 & 14.15 & 10.34 & 1.75 & $8.4 - 8.65$ &  550 & 16 &  {\scriptsize Cycle~1$^7$ } & {\scriptsize PHANGS-MUSE LP$^8$ } \\ 
NGC1068 & 14.0 & 12.41 & 10.91 & 43.33 & $8.58 - 9.15$ &  366 & $-$ &  {\scriptsize Cycle~2$^6$ } & {\scriptsize 094.B-0298 PI: Walcher; 094.B-0321 PI: Marconi } \\ 
NGC1087 & 15.8 & 6.85 & 9.93 & 1.31 & $8.27 - 8.54$ &  741 & 14 &  {\scriptsize Cycle~1$^7$ } & {\scriptsize PHANGS-MUSE LP$^8$ } \\ 
NGC1097 & 13.6 & 20.87 & 10.76 & 4.74 & $8.57 - 9.11$ &  37 & $-$ &  {\scriptsize Cycle~2$^6$ } & {\scriptsize 60.A-9487, 097.B-0640 PI: Gadotti } \\ 
NGC1300 & 19.0 & 16.41 & 10.62 & 1.17 & $8.35 - 8.66$ &  487 & 4 &  {\scriptsize Cycle~1$^7$ } & {\scriptsize PHANGS-MUSE LP$^8$ } \\ 
NGC1317 & 19.1 & 8.53 & 10.62 & 0.48 & $8.54 - 9.1$ &  56 & $-$ &  {\scriptsize Cycle~2$^6$ } & {\scriptsize 109.2332.001 PI: Belfiore } \\ 
NGC1365 & 19.6 & 34.22 & 10.99 & 16.90 & $8.36 - 8.68$ &  453 & 8 &  {\scriptsize Cycle~1$^7$ } & {\scriptsize PHANGS-MUSE LP$^8$ } \\ 
NGC1385 & 17.2 & 8.53 & 9.98 & 2.09 & $8.31 - 8.57$ &  732 & 15 &  {\scriptsize Cycle~1$^7$ } & {\scriptsize PHANGS-MUSE LP$^8$ } \\ 
NGC1433 & 18.6 & 16.78 & 10.87 & 1.13 & $8.49 - 8.67$ &  236 & 1 &  {\scriptsize Cycle~1$^7$ } & {\scriptsize PHANGS-MUSE LP$^8$ } \\ 
NGC1512 & 18.8 & 23.10 & 10.71 & 1.28 & $8.47 - 8.67$ &  217 & 1 &  {\scriptsize Cycle~1$^7$ } & {\scriptsize PHANGS-MUSE LP$^8$ } \\ 
NGC1566 & 17.7 & 18.60 & 10.78 & 4.54 & $8.42 - 8.66$ &  1213 & 53 &  {\scriptsize Cycle~1$^7$ } & {\scriptsize PHANGS-MUSE LP$^8$ } \\ 
NGC1672 & 19.4 & 17.36 & 10.73 & 7.60 & $8.33 - 8.64$ &  704 & 5 &  {\scriptsize Cycle~1$^7$ } & {\scriptsize PHANGS-MUSE LP$^8$ } \\ 
NGC1808 & 12.8 & 161.11 & 9.36 & 1.28 & $8.55 - 8.9$ &  47 & $-$ &  {\scriptsize Cycle~2$^6$ } & {\scriptsize 0102.B-0617 PI: Fluetsch } \\ 
NGC2775 & 23.2 & 14.30 & 11.07 & 0.87 & $8.53 - 8.95$ &  94 & $-$ &  {\scriptsize Cycle~2$^6$ } & {\scriptsize 0104.B-0404 PI: Erwin } \\ 
NGC2835 & 12.2 & 11.40 & 10.00 & 1.24 & $8.24 - 8.6$ &  481 & \revone{19} &  {\scriptsize Cycle~1$^7$ } & {\scriptsize PHANGS-MUSE LP$^8$ } \\ 
NGC2903 & 10.0 & 17.37 & 10.63 & 3.08 & $8.53 - 8.84$ &  70 & $-$ &  {\scriptsize Cycle~2$^6$ } & {\scriptsize 111.24KE.001 PI: Barnes } \\ 
NGC3239 & 10.9 & 5.71 & 9.17 & 0.39 & $7.91 - 8.24$ &  330 & $-$ &  {\scriptsize Cycle~2$^6$ } & {\scriptsize 0108.B-0249 PI: Kreckel$^5$ } \\ 
NGC3351 & 10.0 & 10.47 & 10.36 & 1.32 & $8.52 - 8.73$ &  175 & 6 &  {\scriptsize Cycle~1$^7$ } & {\scriptsize PHANGS-MUSE LP$^8$ } \\ 
NGC3368 & 11.2 & 13.47 & 10.74 & 0.72 & $8.56 - 8.65$ &  3 & $-$ &  {\scriptsize Cycle~2$^6$ } & {\scriptsize 0104.B-0404 PI: Erwin } \\ 
NGC3521 & 13.2 & 16.02 & 11.02 & 3.72 & $8.52 - 8.69$ &  205 & $-$ &  {\scriptsize Cycle~2$^6$ } & {\scriptsize 099.B-0242 PI: Carollo } \\ 
NGC3596 & 11.3 & 5.98 & 9.66 & 0.30 & $8.38 - 8.68$ &  666 & $-$ &  {\scriptsize Cycle~2$^6$ } & {\scriptsize 0108.B-0249 PI: Kreckel$^5$ } \\ 
NGC3626 & 20.0 & 8.59 & 10.46 & 0.21 & $8.6 - 9.16$ &  29 & $-$ &  {\scriptsize Cycle~2$^6$ } & {\scriptsize 0104.B-0404 PI: Erwin } \\ 
NGC3627 & 11.3 & 16.93 & 10.83 & 3.84 & $8.47 - 8.64$ &  624 & 18 &  {\scriptsize Cycle~1$^7$ } & {\scriptsize PHANGS-MUSE LP$^8$ } \\ 
NGC4254 & 13.1 & 9.59 & 10.42 & 3.07 & $8.42 - 8.69$ &  1686 & 70 &  {\scriptsize Cycle~1$^7$ } & {\scriptsize PHANGS-MUSE LP$^8$ } \\ 
NGC4303 & 17.0 & 17.02 & 10.52 & 5.33 & $8.45 - 8.69$ &  1519 & 65 &  {\scriptsize Cycle~1$^7$ } & {\scriptsize PHANGS-MUSE LP$^8$ } \\ 
NGC4321 & 15.2 & 13.48 & 10.75 & 3.56 & $8.42 - 8.71$ &  851 & 36 &  {\scriptsize Cycle~1$^7$ } & {\scriptsize PHANGS-MUSE LP$^8$ } \\ 
NGC4424 & 16.2 & 7.16 & 9.91 & 0.30 & $8.49 - 8.55$ &  5 & $-$ &  {\scriptsize Cycle~2$^6$ } & {\scriptsize 097.D-0408 PI: Anderson } \\ 
NGC4457 & 15.1 & 6.13 & 10.42 & 0.31 & $8.56 - 8.9$ &  58 & $-$ &  {\scriptsize Cycle~2$^6$ } & {\scriptsize 109.2332.001 PI: Belfiore } \\ 
NGC4496A & 14.9 & 7.29 & 9.53 & 0.61 & $8.18 - 8.61$ &  528 & $-$ &  {\scriptsize Cycle~2$^6$ } & {\scriptsize 0108.B-0249 PI: Kreckel; 0100.B-0116 PI: Carollo } \\ 
NGC4535 & 15.8 & 18.69 & 10.53 & 2.16 & $8.47 - 8.69$ &  457 & 17 &  {\scriptsize Cycle~1$^7$ } & {\scriptsize PHANGS-MUSE LP$^8$ } \\ 
NGC4548 & 16.2 & 13.08 & 10.69 & 0.52 & $8.58 - 8.64$ &  2 & $-$ &  {\scriptsize Cycle~2$^6$ } & {\scriptsize 109.22VU.001 PI: Erwin } \\ 
NGC4694 & 15.8 & 4.57 & 9.86 & 0.16 & $8.26 - 8.5$ &  4 & $-$ &  {\scriptsize Cycle~2$^6$ } & {\scriptsize 109.2332.001 PI: Belfiore } \\ 
NGC4731 & 13.3 & 12.22 & 9.48 & 0.60 & $8.04 - 8.41$ &  316 & $-$ &  {\scriptsize Cycle~2$^6$ } & {\scriptsize 0111.C-2109 PI: Egorov$^5$ } \\ 
NGC4781 & 11.3 & 6.10 & 9.64 & 0.48 & $8.22 - 8.55$ &  759 & $-$ &  {\scriptsize Cycle~2$^6$ } & {\scriptsize 0111.C-2109 PI: Egorov$^5$ } \\ 
NGC4941 & 15.0 & 7.32 & 10.17 & 0.44 & $8.51 - 9.02$ &  49 & $-$ &  {\scriptsize Cycle~2$^6$ } & {\scriptsize 096.B-0309 PI: Carollo } \\ 
NGC5068 & 5.2 & 5.66 & 9.40 & 0.28 & $8.16 - 8.51$ &  667 & \revone{39} &  {\scriptsize Cycle~1$^7$ } & {\scriptsize PHANGS-MUSE LP$^8$ } \\ 
NGC5248 & 14.9 & 8.81 & 10.41 & 2.29 & $8.52 - 8.69$ &  90 & $-$ &  {\scriptsize Cycle~2$^6$ } & {\scriptsize 097.B-0640 PI: Gadotti } \\ 
NGC7456 & 15.7 & 9.39 & 9.64 & 0.37 & $8.11 - 8.49$ &  344 & $-$ &  {\scriptsize Cycle~2$^6$ } & {\scriptsize 0111.C-2109 PI: Egorov$^5$ } \\ 
NGC7496 & 18.7 & 9.12 & 10.00 & 2.26 & $8.3 - 8.64$ &  300 & 3 &  {\scriptsize Cycle~1$^7$ } & {\scriptsize PHANGS-MUSE LP$^8$ } \\ 

\hline
\end{tabular}
}
\begin{footnotesize}
References: 
$^{1}$\cite{Anand2021a,Anand2021b,Shaya2017,Kourkchi2020}; $^{2}$\cite{Paturel2003}; $^{3}$\cite{Leroy2021}; $^{4}$\cite{Li2024}; \\ $^{5}$Egorov et al. in prep.; 
$^{6}$ ID 03707 PI: Leroy \citep{Chown2024}; $^{7}$ ID 02107 PI: Lee (\citealt{Lee2023, Williams2024}); $^8$ \cite{Emsellem2022} 
\end{footnotesize}
\end{table*}

We use the \JWST\ data obtained in Cycle 1 and Cycle 2 as part of the PHANGS-JWST survey (programs GO 02107; PI: J.~C.~Lee and GO 03707; PI: A.~Leroy). In Cycle~1, 19 nearby star-forming galaxies were imaged with the NIRCam (with F200W, F300M, F335M and F360M filters) and the MIRI (F770W, F1000W, F1130W and F2100W) instruments. The first public data release and the description of the data reduction process are presented in \cite{Williams2024} (see also \citealt{Lee2023}). In Cycle~2, 55 additional galaxies from the PHANGS survey were observed with a modified filter set: F150W, F187N (replaced by F200W for a few galaxies with higher systemic velocities), F300M, F355M for NIRCam, and F770W, F2100W for MIRI. \citet{Chown2024phangs} describe the basic observational setup and processing of this Cycle 2 survey. Data reduction for all objects from both Cycle~1 and Cycle~2 is done in a uniform way with the package \textsc{pjpipe}\footnote{\url{https://github.com/PhangsTeam/pjpipe}}, which is described in \citet{Williams2024} and is based on the official JWST pipeline \citep{Bushouse2023}. In this paper, we use the F770W, F1130W, F2100W, and F300M images \revone{(including the associated uncertainties maps)} for all 19 galaxies from Cycle~1, and F770W, F2100W, and F300M images for 23 galaxies from Cycle~2 (see Table~\ref{tab:sample}). These 23 galaxies were selected from the entire sample of Cycle~2 galaxies because they have available MUSE data and derived products of quality similar to that of the PHANGS-MUSE data \citep{Emsellem2022}, which overlap with the $19$ Cycle 1 PHANGS-JWST targets (see Sec.~\ref{sec:obs_muse}).

We convolved all the images to the PSF of the F2100W images (FWHM~$\simeq0.67''$, corresponding to 28--64 pc at the distance of our targets), as produced by the \textsc{stpsf} \citep{WebbPSF} modeling tool\footnote{\url{https://stpsf.readthedocs.io/en/latest/}}. The convolution kernels were created following the procedure from \cite{Aniano2011}. We note that our analysis relies on the integrated measurements within footprints of the \HII\ regions (see Sec.~\ref{sec:analysis}), and therefore the exact angular resolution of the data (both the JWST and MUSE datasets) is not critical for the analysis.
%The images in all four MIRI bands were convolved to a common angular resolution of $\sim 0.67''$ (that for the F2100W band) corresponding to 28--64 pc spatial resolution at the distance of our targets. 
% Standard calibrations are applied, with minor modifications. A detailed description of the complete data reduction is presented in \citet{Williams2024} and \citet{Lee2023}.

\subsection{MUSE}\label{sec:obs_muse}

All 42 galaxies in our study were previously observed with the optical integral field spectrograph MUSE (VLT). The PHANGS-MUSE large program (1100.B-0651, PI: Schinnerer) provided data for 19 galaxies from PHANGS-JWST Cycle~1 with a coverage up to 2 effective radii for all targets, which is in general consistent with the footprint of JWST observations. The other 23 galaxies (hereafter extended sample) were observed within several complementary programs, or recovered from the ESO archive (see Table~\ref{tab:sample}). Due to the heterogeneity of the extended sample, MUSE covers only a small (typically -- central) part of the JWST footprint for about half of these galaxies, while the coverage of the other half of the extended sample is comparable to that of PHANGS-MUSE galaxies. 

The typical angular resolution of the MUSE data in this study is $\sim 1''$ (45--95~pc), which is sufficient to isolate individual \HII\ regions from each other and minimize contribution from the surrounding diffuse ionized gas \citep{Congiu2023}. The MUSE data span the spectral range $4800-9200$ \AA\ and include several strong emission lines (\Ha, \Hb, \OIII\ 4959, 5007\AA, \SII\ 6717, 6731\AA, \NII\ 6548, 6584\AA, \SIII\ 9069\AA) that are used in the current work. 

The reduced data for the PHANGS-MUSE large program are publicly available\footnote{\url{https://archive.eso.org/scienceportal/home?data_collection=PHANGS}}, and details of the observations and data reduction are given in \cite{Emsellem2022}. The data for galaxies from the extended sample are reduced and analyzed in a similar way as for the main PHANGS-MUSE sample. That is, we used the python package \textsc{pymusepipe}\footnote{\url{https://github.com/emsellem/pymusepipe}} to produce data cubes from the raw data. \textsc{pymusepipe} is a wrapper around the MUSE data processing pipeline software \citep{Weilbacher2020} and we adopted it since it allows for efficient mosaicking of multiple data cubes. For data analysis (i.e., producing the maps in emission lines by fitting the simple stellar population templates and single-component Gaussians to the observed spectra), we used the PHANGS data analysis pipeline \textsc{DAP}\footnote{\url{https://gitlab.com/francbelf/ifu-pipeline}}, which relies on the \textsc{gist} code \citep{Bittner2019}. Both \textsc{pymusepipe} and \textsc{DAP} are described in detail in \citet{Emsellem2022}. A detailed description of the data reduction, as well as presentation of the data for the 23 galaxies from the extended sample, will be presented in future work. %given in other papers elsewhere. 

% \cite{Groves2023} presented a catalog of $\sim 30,000$ nebulae in 19 PHANGS-MUSE galaxies containing their spatial masks (defining the borders and sizes of each nebula), integrated strong emission line and the oxygen abundances among other things. The nebulae were selected using \textsc{HIIphot} \citep{Thilker2000}. The emission line fluxes are corrected for the effects of dust extinction using the Balmer decrement. Oxygen abundance is derived using `Scal' method 
% \HII\ regions are selected from the nebulae catalog based on the \cite{BPT} diagnostics considering \OIIIHb\ vs \NIIHa\ and \SIIHa\ line ratios, as described in Groves et al.\ (submitted, see also \citealt{Kewley2019} for a review). The present analysis is restricted to only those \HII\ regions  that meet stringent surface brightness and signal-to-noise (S/N) criteria (see Sec.~\ref{sec:hii_selection}). 

\section{Analysis}
\label{sec:analysis}

In this work, we investigate how the PAH fraction changes in the presence of ionized gas in and around star-forming regions. The analysis consists of four main steps: (1) identifying \HII\ regions and SNRs in the MUSE data; (2) Reprojecting the spatial masks defining the borders of these nebulae onto the JWST grid to isolate the regions affected by ionizing radiation or SNR shocks from the diffuse ISM; (3) Measuring observational and physical properties of the ionized nebulae based on the integrated MUSE spectra of the regions; (4) Deriving JWST band ratios sensitive to the PAH mass fraction from the fluxes integrated in the apertures corresponding to the ionized nebulae or diffuse ISM. \revone{Our measurements are reported in the online catalog described in Appendix~\ref{app:catalog}.}

\subsection{Identifying \HII\ regions and measuring the properties of the ionized gas}
\label{sec:analysis_hii_snrs}

% \begin{figure}
%     \centering
%     \includegraphics[width=\linewidth]{figs/placeholder.png}
%     \caption{This is just a placeholder. This cat will draw a figure that will show some region in some of the sample galaxies (preferentially -- from Cycle2?) in several filters + MUSE-Halpha with overlaid HII region masks.}
%     \label{fig:HIIregs}
% \end{figure}

Following the same approach as in \cite{Egorov2023}, we selected \HII\ regions in JWST images based on the masks derived from the MUSE data using the package \textsc{hiiphot} \citep{Thilker2000}. The procedure is described in detail by \cite{Groves2023} where these masks and the corresponding nebular catalog are presented for 19 PHANGS-MUSE galaxies. The catalog contains measurements of emission line fluxes for each nebula from their integrated spectra (within the borders defined by \textsc{hiiphot}). The line fluxes are corrected for extinction based on the observed Balmer decrement. No correction for the contribution of the surrounding diffuse emission was applied. We also use measurements of the gas phase oxygen abundance (as a proxy for the metallicity), $12+\log(\rm O/H)$, and circularized radii of the nebulae, $r_{\rm circ}$, taken from the \cite{Groves2023} catalog for each \HII\ region, or measured here in a similar way for the extended sample. The oxygen abundance was derived from the integrated emission spectra made using the strong emission line calibration \revone{``Scal''} from \cite{Pilyugin2016}, which relies on the measurements of \Hb, \Ha, \OIII$\lambda$5007, \NII$\lambda$6584 and \SII$\lambda$6717,6731 emission lines and is usually consistent with the measurements made using the \revone{``direct''} $T_e$ method \citep[e.g.][]{Brazzini2024}. This calibration adopts $\rm 12+\log(O/H) = 8.69$ \citep{Asplund2009} as a reference value for the Solar oxygen abundance.  

The \cite{Groves2023} catalog contains properties of the ionized nebulae for all 19 PHANGS-MUSE galaxies. Here we analyze also the new MUSE data for 23 other galaxies from the extended sample, which were observed with JWST in Cycle 2. For these galaxies, we analyzed the MUSE data, produced the ionized nebulae spatial masks and the nebular catalog with the measured emission lines, metallicity etc. in the same way as described in \cite{Groves2023} and summarized above. Similarly to \citet{Groves2023}, we classified a nebula as \HII\ region based on its position on the diagnostic BPT diagram \citep{BPT}. We use the \HII\ region spatial masks and derived properties of the nebulae in the remaining analysis. The new catalogs for the extended sample will be published separately in future papers. %in dedicated papers elsewhere. %  (Canal i Saguer/Egorov et al., in prep., for the 7 lowest-mass galaxies, and Huber/Kreckel et al., in prep., for other objects \note{???}). 

The spatial masks for each \HII\ region were reprojected onto the grid of the JWST/MIRI F2100W images. Given that our original MUSE data typically have coarser resolution than F2100W images, we can expect that some of the pixels in the JWST images falling into the nebular masks in fact correspond to the surrounding PDRs and diffuse ISM heated by the \HII\ regions or even by the interstellar radiation field (ISRF). Moreover, the MUSE resolution is not sufficient for reliable measurements of the sizes of \HII\ regions (which are typically smaller than the PSF of our MUSE data). The derived sizes are systematically overestimated by a factor of 3 compared to what is measured from the better resolution HST \Ha\ images (\citealt{Barnes2022}, Barnes et al., subm.). Therefore, we expect that reprojected spatial masks on the JWST / MIRI F2100W grid isolate \HII\ regions from each other well, but have a non-negligible contribution of diffuse ISM. This contribution is expected to be higher in the case of the faint (hence small) \HII\ regions, and closer to negligible for the luminous (and thus larger) \HII\ regions. This is further discussed in Sec.~\ref{sec:HIIregs}. To minimize contamination by diffuse ISM, we also consider separately in the following analysis the subsample of the marginally resolved (luminous) \HII\ regions, which have size exceeding 2 PSF of MUSE data (i.e. $r_{\rm circ}> \mathrm{FWHM_{PSF}}$). This mostly corresponds to the extinction-corrected \Ha\ luminosities $L({\rm H}\alpha) > 10^{37}\ergs$.

We excluded from further analysis all the \HII\ regions residing in the galactic centers, as star-forming regions in such locations are typically very crowded and the MUSE resolution does not allow us to reliably separate them from each other. Furthermore, the gas and dust excitation there can be strongly affected by processes different from what is observed in normal star-forming regions in galactic \revone{disks} (e.g. active galactic nuclei (AGNs), nuclear starbursts, large-scale shocks due in the circumnuclear rings etc.). Recent studies in fact reveal destruction of PAHs by shocks and anomalous PAH-to-CO and PAH band ratios in the central regions of galaxies \citep[e.g.][]{Zhang2022, Chown2024phangs, Pathak2024, Baron2025}. To exclude central regions, we used environmental masks from \cite{Querejeta2021}. These masks were created with the Spitzer~3.6 $\mu$m images and distinguish between different environments in each galaxy in our sample. Therefore, our results presented below are valid for normal star-forming regions outside the galactic centers. For three galaxies that do not have published environmental masks, we applied elliptical masks derived by eye based on the Spitzer~3.6 $\mu$m (for NGC~1808) or JWST NIRCAM/F330M (for NGC~1068, NGC~3368) images to exclude the central and bar regions from the analysis.  

Finally, we consider here only those \HII\ regions with signal-to-noise ratio exceeding 10 in all important emission lines and JWST bands (i.e. \SIII$\lambda$9069\AA, \SII$\lambda$6717,6731\AA, $F770W$, $F1130W$, $F2100W$) in their integrated spectra or photometry. In total, we analyze a sample of \numHII\ \HII\ regions. The number of the selected \HII\ regions for each galaxy is given in Table~\ref{tab:sample}.

In this study, we rely on the reddening-corrected \mbox{\SIII$\lambda$9069, 9532\AA/\SII$\lambda$6717,6731\AA} line ratio as a proxy for the ionization parameter, commonly defined as the ratio of the number densities of ionizing photons and hydrogen atoms \citep[e.g.][see also Appendix~\ref{sec:app:cloudy}]{Diaz1991, Kewley2019, Mingozzi2020}. The emission line \SIII$\lambda$9532\AA\ resides beyond the MUSE wavelength range. Therefore, we use the well-defined theoretical ratio between these lines: \SIII$\lambda$9532\AA{} = $2.5\times$\SIII$\lambda$9069\AA. Hereinafter, \SII\ and \SIII\ refers to the sum of the corresponding doublets, unless otherwise specified.

\subsection{Identifying SNRs and measuring their properties}
\label{sec:analysis_snrs}

We aim here to analyze how the PAH fraction in star-forming regions is affected not only by ionizing radiation, but also by shocks. We study this by measuring the PAH fraction \revone{toward} the sample of SNRs identified in 19 galaxies from our sample. \citet{Li2024} found 1166 SNRs in 19 PHANGS-MUSE galaxies (outside their central parts) based on the combination of several indicators (different shock-sensitive line ratios and gas velocity dispersion). Their sample includes 964 objects isolated from any \HII\ region, as compared with the same \citet{Groves2023} \HII\ region catalog used in our work. We therefore consider this sample of isolated SNRs as the sites where the supernova shocks may dominate over photoionization in its influence on the dust and PAHs in the ISM. We integrated the optical spectra and fluxes in JWST bands in circular apertures with 50 pc diameter\footnote{This size roughly corresponds to the resolution of the JWST MIRI F2100W images for the most distant galaxies in our sample; using diameters of 100 pc, as in \citealt{Li2024} paper, or 25 pc does not affect the results presented below} and analyzed them in the same way as for \HII\ regions above to recover emission line fluxes and JWST photometry. Note that the Scal empirical metallicity calibration is not valid for the regions dominated by shocks. Therefore, similarly to \cite{Li2024}, we took the values corresponding \revone{to the SNR locations} from the spatially-interpolated 2D metallicity maps derived by \cite{Williams2022} based on Scal measurements for the \HII\ regions.

After excluding SNRs outside the JWST footprint and applying the same signal-to-noise criteria as for the \HII\ regions (i.e. $S/N>10$ in all emission lines and JWST bands significant for our analysis), we proceed with a sample of 398 isolated SNRs in 19 PHANGS-MUSE galaxies spanning metallicity range of $\rm 12+\log(O/H) = 8.19 - 8.63$.

 % The \citet{Li2024} catalog does not contain SNRs with oxygen abundance $\rm 12+\log(O/H) < 8.3$ \note{TBC!} because of its focus on relatively massive galaxies. In order to facilitate the analysis of PAHs in SNRs at the lower metallicities, we consider here an additional subset of 22 SNRs identified in 7 other lowest-mass galaxies from our sample (with $\log(M_*) < 9.7$, see Table~\ref{tab:sample}). To identify them, we applied the same criteria to MUSE data as in \citep{Li2024},  and then visually inspected all candidates to select a small subset of the most confident isolated SNRs. Therefore, this additional sample is far from complete (identification of the SNRs in low-mass galaxies and their detailed analysis are beyond the scope of this work and will be a subject of future studies), but rather served as a small test sample to verify that our findings remain consistent within the entire metallicity range. Since 2D metallicity maps are not available for the seven galaxies containing these additional SNRs, we used the median Scal oxygen abundance of \HII\ regions located within 100~pc of each SNR as a proxy of its metallicity. The total number of SNRs per galaxy considered in the further analysis is given in Table~\ref{tab:sample}.

\subsection{Tracing PAH-to-dust mass fraction}
\label{sec:analysis_rpah}

Measurement of the mass fraction of PAHs in dust, $q_{\rm PAH}$, requires SED modeling from mid- to far-IR wavelengths. Previous studies have shown that the ratio of intensity in filters dominated by PAH emission to those dominated by the small dust grain continuum can be a good observational tracer of $q_{\rm PAH}$ \citep[e.g.][]{Engelbracht2008, Marble2010, Sandstrom2010, Sutter2024, Chastenet2025}. For example, Spitzer studies relied on the $F_{8\mu m}/F_{24\mu m}$ (i.e. IRAC4/MIPS24) band ratio as a tracer of $q_{\rm PAH}$. With JWST it is possible to use several bands sensitive to different sizes and ionization states of PAHs (e.g. 7.7~$\mu$m and 11.3~$\mu$m). Following previous works \citep{Egorov2023, Chastenet2023, Sutter2024}, we use ${\rm R_{PAH}} \equiv \rm  (F770W_{\rm ss} + F1130W)/F2100W$ as a tracer of $q_{\rm PAH}$. The measurements in corresponding JWST filters here are flux densities in MJy sr$^{-1}$.  Based on theoretical modeling and observations of 7 galaxies, \citet{Sutter2024} demonstrated that \RPAH\ correlates with $q_{\rm PAH}$ in cases where full far-IR coverage is available. For reference, ${\rm R_{PAH}} \sim 4$ corresponds to $q_{PAH}\sim 5$\%, according to their paper. %radiation field intensities typical for $10-100$ pc scales in galaxies.
%linearly correlates with $q_{\rm PAH}$ when $q_{\rm PAH} < 4.5$\%, which is a typical value for galaxies in the nearby Universe \citep[e.g.][]{Draine2007, Khramtsova2013, Chastenet2019, Aniano2020}. 

In this definition of \RPAH, $\rm F770W_{\rm ss}$ corresponds to flux density of the non-stellar emission in the $\rm F770W$ band, i.e. to the total flux density measured in this band corrected for the contribution from stellar continuum. Stellar populations produce non-zero continuum emission at the wavelength around 8~$\mu$m \citep[e.g.][]{Helou2004, Marble2010, Sutter2024}, and its removal is required for proper measurements of \RPAH\ in areas of high stellar surface density and low gas density. This correction is very minor in gas/dust-rich regions ($\sim 1$\%; \citealt{Baron2025}), such as the star-forming regions considered here. Therefore, no significant differences are expected in \RPAH\ compared to previous measurements where such a correction was not implemented \citep{Egorov2023, Chastenet2023}. In diffuse gas, especially \revone{toward} galaxy centers where the starlight is concentrated, the starlight continuum contributes more, resulting in a higher correction, typically $10-30$\% \citep{Baron2025}. The more extreme values (up to $75$\%) are observed only in star formation deserts, which are not considered in this work. 
We calculate the starlight contribution based on the NIRCam F300M images convolved at the MIRI $\rm F770W$ resolution as described in \citet{Sutter2024}: $\rm F770W_{\rm ss} = F770W - (0.22 \pm 0.08) \times F300M$. The resulting $\rm F770W_{\rm ss}$ are further convolved at the $\rm F2100W$, as was described earlier in Sec.~\ref{sec:obs_jwst}.  %An alternative approach (and slightly more preferable due to higher S/N of the continuum) is to use F200W images, which are however not available for all galaxies in our sample.  \citet{Sutter2024} found good agreement between these two methods, with the only environments showing clear differences being those dominated by an older stellar population, and therefore not an issue for this work. 

The fluxes measured in apertures corresponding to \HII\ regions and SNRs are contaminated by diffuse ISM not only due to the imperfect isolation of the fainter (smaller) regions (as mentioned in Sec.~\ref{sec:analysis_hii_snrs}), but also due to the contribution of background or foreground emission. Careful subtraction of this component is challenging due to its inhomogeneity, but necessary for analysis of the compact sources (such as SNRs) where its contribution can dominate the total flux measured in the aperture. To perform such a correction, we measured the median brightness in each photometric band in circular annuli  with inner and outer radii $1.5\times r_{\rm circ}$ and $3\times r_{\rm circ}$, respectively, and centered on the \HII\ region or SNR. For SNRs, we assumed constant $r_{\rm circ} = 50$~pc. Before measuring the median brightness, we masked all pixels that reside within \HII\ region masks. \revone{The uncertainties of the background measurements, calculated as the standard deviation within the apertures, were propagated to the estimates of \RPAH.} The effect of background correction on the measured fluxes and derived \RPAH\ is investigated in the Appendix~\ref{sec:app:fluxes} and Sec.~\ref{sec:shocks}.
The local background does not significantly affect our measurements for \HII\ regions, and therefore for further analysis we use mostly original, non-corrected values. However, we consider background-subtracted fluxes investigating the effect of SNRs and in some additional applications where such correction is essential.

The measurements of \RPAH\ described here require images in the $\rm F1130W$ band, which are not available for 23 galaxies from Cycle~2. Meanwhile, the ratios of individual PAH-sensitive bands to the dust continuum (i.e. $\rm F770W$/$\rm F2100W$ or $\rm F1130W$/$\rm F2100W$, which are similar to the previously widely used Spitzer IRAC4/MIPS24 and WISE W3/W4 ratios) were also shown to be good tracers of PAH fraction. Thus, \citet{Sutter2024} showed that ${\rm R_{PAH}^*} = 2.57 \times \rm F770W_{\rm ss}/F2100W$ is generally equal to ${\rm R_{PAH}}$ for diffuse ISM if there are no significant variations in PAH charge or hardness of the interstellar radiation field expected. 
Based on our measurements for 19 galaxies where both $\rm F770W$ and $\rm F1130W$, we consider in Appendix~\ref{sec:app:calibration} whether ${\rm R_{PAH}^*}$ measured this way agrees well with ${\rm R_{PAH}}$ for \HII\ regions. 
We find that $\rm F770W_{\rm ss}$/$\rm F1130W$ ratio grows with the \Ha\ luminosity of a region (Fig.~\ref{fig:F770_F1130_vs_Ha}; see also \citealt{Belfiore2023}), %\note{(maybe we should move this figure here from appendix? Opinions?)},
which is likely related to the differences in ratio between the ionized and neutral PAHs, or in the hardness of the radiation field \citep{Chastenet2023b, Baron2025}. In turn, this leads to a systematically higher ${\rm R_{PAH}^*}$  (by $\sim 7-17$\%) than \RPAH\ in bright star-forming regions (Fig.~\ref{fig:RPAH_vs_RPAH*}). We demonstrate that this offset can be removed with a small correction non-linearly depending on $\rm F770W_{\rm ss}$ brightness. Therefore, here we use the parameterization of \begin{equation}
\label{eq:rpah_star}
    {\rm R_{PAH}^*} = 2.57/C \times \rm F770W_{\rm ss}/F2100W ,
\end{equation}
where 
\begin{multline}
\label{eq:rpah_corr}
    C = -0.00832\log(\rm L_{\nu}(F770W_{\rm ss}))^2+\\ +0.4881\log(L_{\nu}(F770W_{\rm ss})) -6.01, 
    % C = -0.01697\log(\rm L_{\nu}(F770W_{\rm ss}))^2+\\ +0.9063\log(L_{\nu}(F770W_{\rm ss})) -10.98, 
    % C = -0.021\log(\rm L_{\nu}(F770W_{\rm ss}))^2 +0.208\log(L_{\nu}(F770W_{\rm ss})) +0.61, (old, for flux density)
\end{multline}
where $L_{\nu}(F770W_{\rm ss})$ is the monochromatic luminosity of the non-stellar emission in the F770W band. The latter is measured in erg s$^{-1}$ Hz$^{-1}$ calculated from flux density $\rm F770W_{ss}$ (converted from MJy to erg s$^{-1}$ cm$^{-2}$ Hz$^{-1}$), and distance to the galaxy $D$ (converted to cm) as $L_{\nu}(F770W_{\rm ss}) = 4\pi D^2 \times F770W_{\rm ss}$. The resulting parameterization shows a better agreement with ${\rm R_{PAH}}$ for relatively bright \HII\ regions than the one from \citet{Sutter2024} and is in good agreement with it for the low-luminosity end. The best fit for the background-subtracted measurements slightly differ leading to close to a constant factor correction (see Fig.~\ref{fig:RPAH_vs_RPAH*_bgrsub}):
\begin{multline}
\label{eq:rpah_corr_bgrsub}
    C_{\rm bgrsub} =\rm  -0.02009\log(\rm L_{\nu}(F770W_{\rm ss}))^2 + \\
    +1.0597\log(\rm L_{\nu}(F770W_{\rm ss})) -12.81, 
    % C_{\rm bgrsub} =\rm  -0.01954\log(\rm L_{\nu}(F770W_{\rm ss}))^2 + \\
    % +0.991\log(\rm L_{\nu}(F770W_{\rm ss})) -11.39, 
    % C_{\rm bgrsub} =\rm  -0.011\log(F770W_{\rm ss})^2 +0.09\log(F770W_{\rm ss}) +0.98, % (old, for flux density) 
\end{multline}
assuming that all fluxes here and in Eq.~\ref{eq:rpah_star} are background-subtracted.

For further analysis, we consider \RPAHst\ derived from Eqs.~(\ref{eq:rpah_star}, \ref{eq:rpah_corr}) as the reliable tracer of PAH fraction across all 42 galaxies in our sample. When considering only the 19 PHANGS-JWST galaxies from our Cycle~1 program, we rely on \RPAH\ instead to avoid unnecessary empirical correction. 

% For further analysis, we consider \RPAH\ as the most reliable tracer of PAH fraction across the 19 PHANGS-JWST galaxies from our Cycle~1, and extend the analysis by using $R_{\rm PAH}^*$ (from Eq.~\ref{eq:rpah_star}) for all 42 galaxies in our sample. 

% Therefore, we also consider another $q_{PAH}$ tracer: $R_{\rm PAH}^* = 2.57 \times F770W_{\rm ss}/F2100W$, which is similar to what was widely accepted for the analysis of the Spitzer and WISE data. \cite{Sutter2024} demonstrated that \RPAH\ and $R_{\rm PAH}^*$ are consistent with each other and derived the factor of 2.57 as optimal for 1-to-1 agreement between these two parameters considering the diffuse ISM \note{(Jessica, is this correct?)}. %Here we use the same factor, however note the systematic differences between \RPAH\ and $R_{\rm PAH}^*$ in the bright star-forming regions as shown in Fig.~\ref{fig:RPAH_vs_RPAH*}. 

\section{Results}
\label{sec:results}

\subsection{Metallicity dependence of the PAH fraction}
\label{sec:metallicity}

Previous studies of star-forming regions and galaxies at kpc-scales with Spitzer and WISE demonstrated that the PAH fraction rapidly decreases in the low-metallicity regime \citep[e.g.][]{Engelbracht2005, Madden2006, Draine2007b, Galliano2008, Khramtsova2013, Remy-Ruyer2015, Whitcomb2024}. This is further supported by a still scarce number of resolved studies of nearby low-metallicity dwarf galaxies measuring the PAH fraction to be much lower than in the more massive galaxies \citep[e.g.][]{Chastenet2019, Chown2025}. 
These studies find a threshold between the low- and high-metallicity regimes (typically at $\rm 12+\log(O/H) \sim 8.0-8.3$, although some studies report higher value; see Section~\ref{sec:discussion:oh}), with a particularly large scatter of the PAH fraction around these values. 
All of these studies argue that gas-phase metallicity (or something associated with it) is among the key factors regulating the PAH life cycle, although the exact transition threshold, the particular form of the metallicity dependence, and the role of different physical mechanisms responsible for these remain uncertain.

\begin{figure}%[!htbp]
    \centering
    \includegraphics[width=\linewidth]{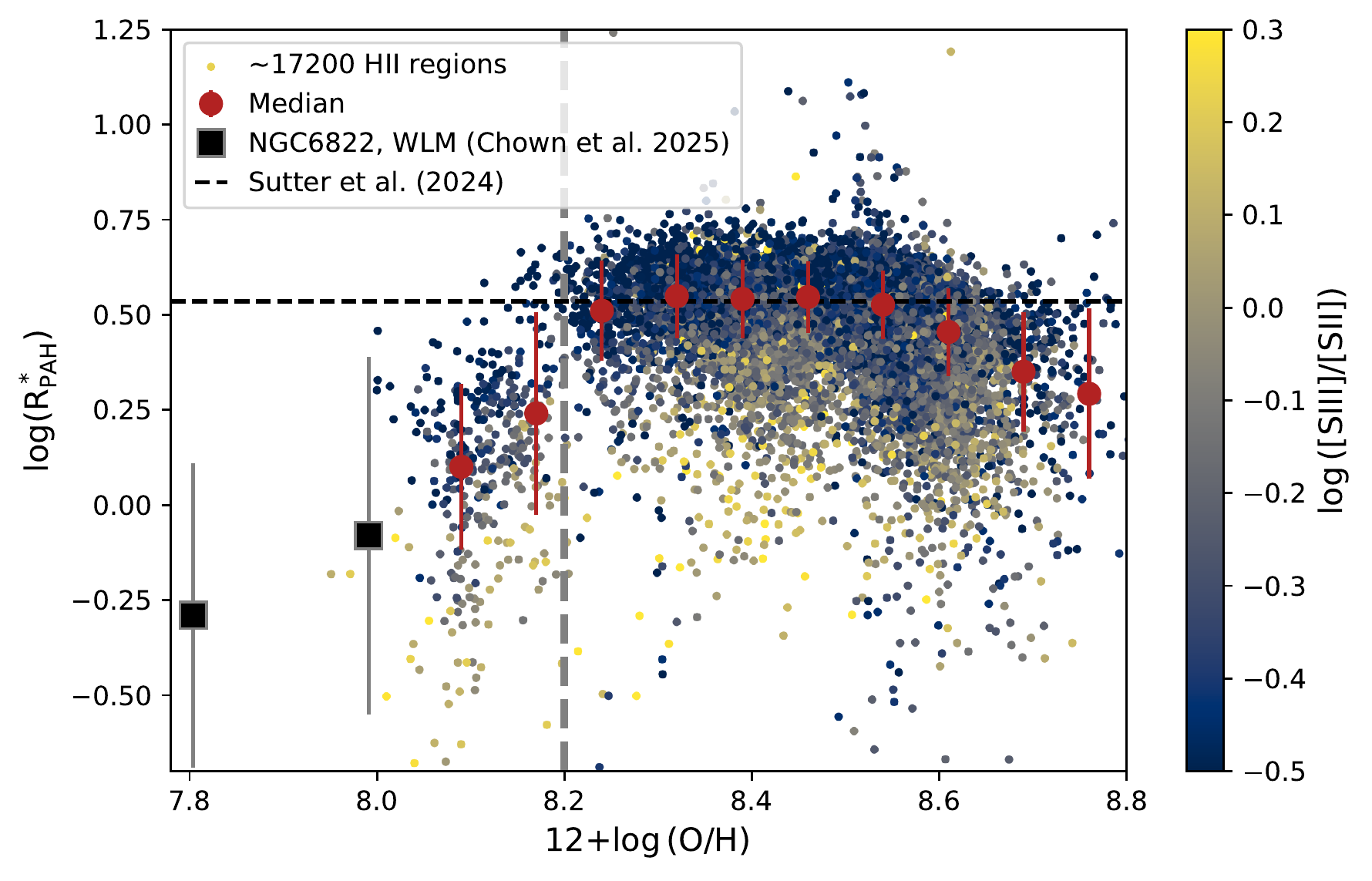}
    \caption{Distribution of $\rm \log({\rm R_{PAH}^*})$ vs oxygen abundance color-coded by $\rm \log([SIII]/[SII])$ (tracer of the ionization parameter) for all \HII\ regions in the sample. Red circles show average values in different metallicity bins (reported in Tab.~\ref{tab:Rpah_vs_z_bins}). Vertical gray dashed line marks $\rm 12 + \log(O/H) = 8.2$ as a threshold between the low- and high-metallicity regimes. The horizontal dashed line correspond to the average \RPAH\ for diffuse ISM measured by \citet{Sutter2024}. Black squares correspond to measurements for WLM and NGC6822 (from low- to high-metallicity) made by \citep{Chown2025} using JWST F770W and F2100W images. The PAH fraction drops significantly in the low-metallicity regime. }
    % \includegraphics[width=\linewidth]{figs/RPAH_vs_ip2_v16.pdf}
    % \includegraphics[width=\linewidth]{figs/RPAH_vs_met_DR12_v16.pdf}
    % \caption{PAH fraction steeply decrease stop following the changes in the ionization parameter in the low-metallicity regime. Panel (a) shows \RPAHst\ vs $\rm \log([SIII]/[SII])$ (ionization parameter tracer) for $\sim 5900$ resolved \HII\ regions in 26 galaxies. Color denotes oxygen abundance.  Contours show a probability density distribution of the data points (levels correspond to 50, 65, 80, 95 and 99 percentile intervals). Red dashed line correspond to Eq.~\ref{eq:RPAH_vs_ip}. Panel (b) shows $\rm \log(R_{PAH}^*)$ vs oxygen abundance color-coded by $\rm \log([OIII]/H\beta)$ (tracer of the UV hardness). Gray dashed line marks $\rm 12+\log(O/H)=8.2$ as a threshold between the low- and high-metallicity regimes.}
    \label{fig:RPAH_vs_met}
\end{figure}

Fig.~\ref{fig:RPAH_vs_met} shows how $\log$(\RPAHst) changes with metallicity for our sample of \numHII\ \HII\ regions spanning the metallicities \mbox{$\rm 12+\log(O/H) \sim 8.0 - 8.8$} \revone($Z \sim 0.2-1.3\ Z_\odot$). The average values per O/H bin clearly demonstrate a drop of \RPAHst\ at metallicities \mbox{$\rm 12+\log(O/H) < 8.2$} and a high-metallicity plateau. This trend is in very good agreement with the estimates for low-metallicity dwarf galaxies WLM and NGC~6822 (shown by black squares) derived as $2.57 \times \rm F770W/F2100W$ based on the average band ratios reported by \citet{Chown2025}, despite slightly different treatment of the stellar continuum subtraction and calculation of \RPAHst. % Note that they did not perform stellar continuum subtraction for $\rm F770W$ band, and we did not apply the \RPAHst\ correction with a $\rm F770W$-dependent coefficient (i.e. with Eq.~\ref{eq:rpah_corr}), and therefore the real \RPAHst\ values for these dwarf galaxies might appear to be even 0.1--0.2~dex lower. 

\begin{table}
\caption{\revone{Average values and standard deviations of \RPAHst\ measured in different metallicity bins in Fig.~\ref{fig:RPAH_vs_met}.}}
    \centering
    \small
    \begin{tabular}{cc}
    \hline
        $\rm 12+\log(O/H)$ & \RPAHst \\
        \hline
        8.09 & $1.26\pm0.63$ \\
        8.17 & $1.74\pm1.07$ \\
        8.24 & $3.24\pm0.98$\\
        8.32 & $3.54\pm0.90$\\
        8.39 & $3.48\pm0.83$\\
        8.46 & $3.52\pm0.77$\\
        8.54 & $3.36\pm0.70$\\
        8.61 & $2.85\pm0.76$ \\
        8.69 & $2.24\pm0.81$\\
        8.76 & $1.96\pm1.01$\\
        % 8.08 & $1.43\pm0.63$ \\
        % 8.17 & $1.86\pm1.04$ \\
        % 8.25 & $3.34\pm0.95$\\
        % 8.34 & $3.58\pm0.90$\\
        % 8.42 & $3.49\pm0.83$\\
        % 8.50 & $3.53\pm0.72$\\
        % 8.59 & $3.09\pm0.76$\\
        % 8.67 & $2.46\pm0.82$ \\
        % 8.76 & $2.16\pm0.95$\\
        \hline
    \end{tabular}
    \label{tab:Rpah_vs_z_bins}
\end{table}

\begin{figure}
    \centering
    \includegraphics[width=\linewidth]{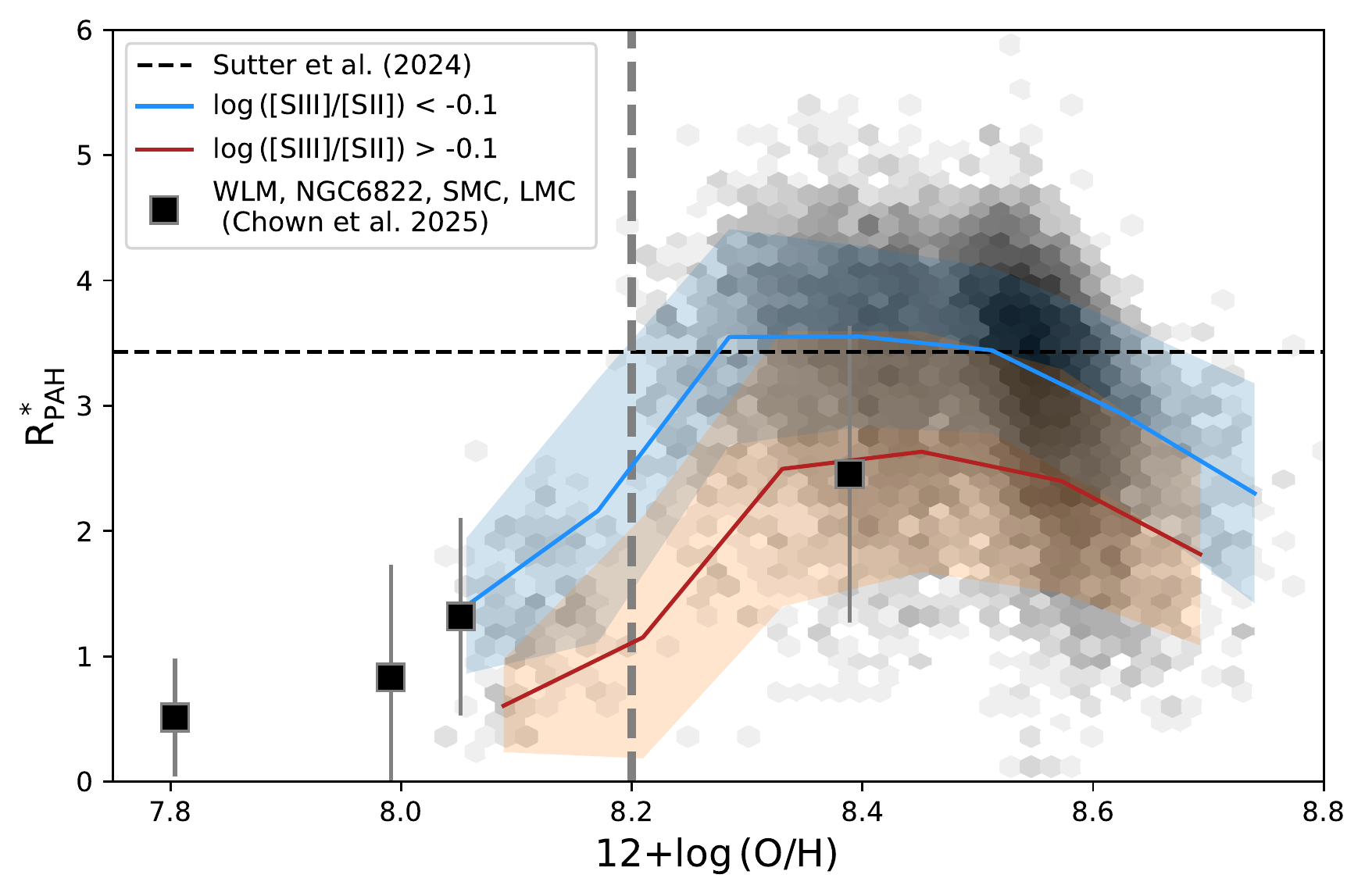}
    \caption{Distribution of \RPAHst\ vs oxygen abundance for the entire sample of \HII\ regions (gray histogram with intensity tracing the probability density), and average distribution for two bins in $\log$(\SIIISII). Dashed lines are the same as in Fig.~\ref{fig:RPAH_vs_met}. Black squares represent measurements for WLM, NGC6822 (same as in Fig.~\ref{fig:RPAH_vs_met}), SMC and LMC from \citet{Chown2025}, listed in the order of increasing metallicity. Spitzer IRAC4 and MIPS24 filters were used to estimate \RPAHst\ for the SMC and LMC  \citep{Chown2025}.}
    \label{fig:rpah_met_bins}
\end{figure}

\begin{figure}
    \centering
    \includegraphics[width=0.95\linewidth]{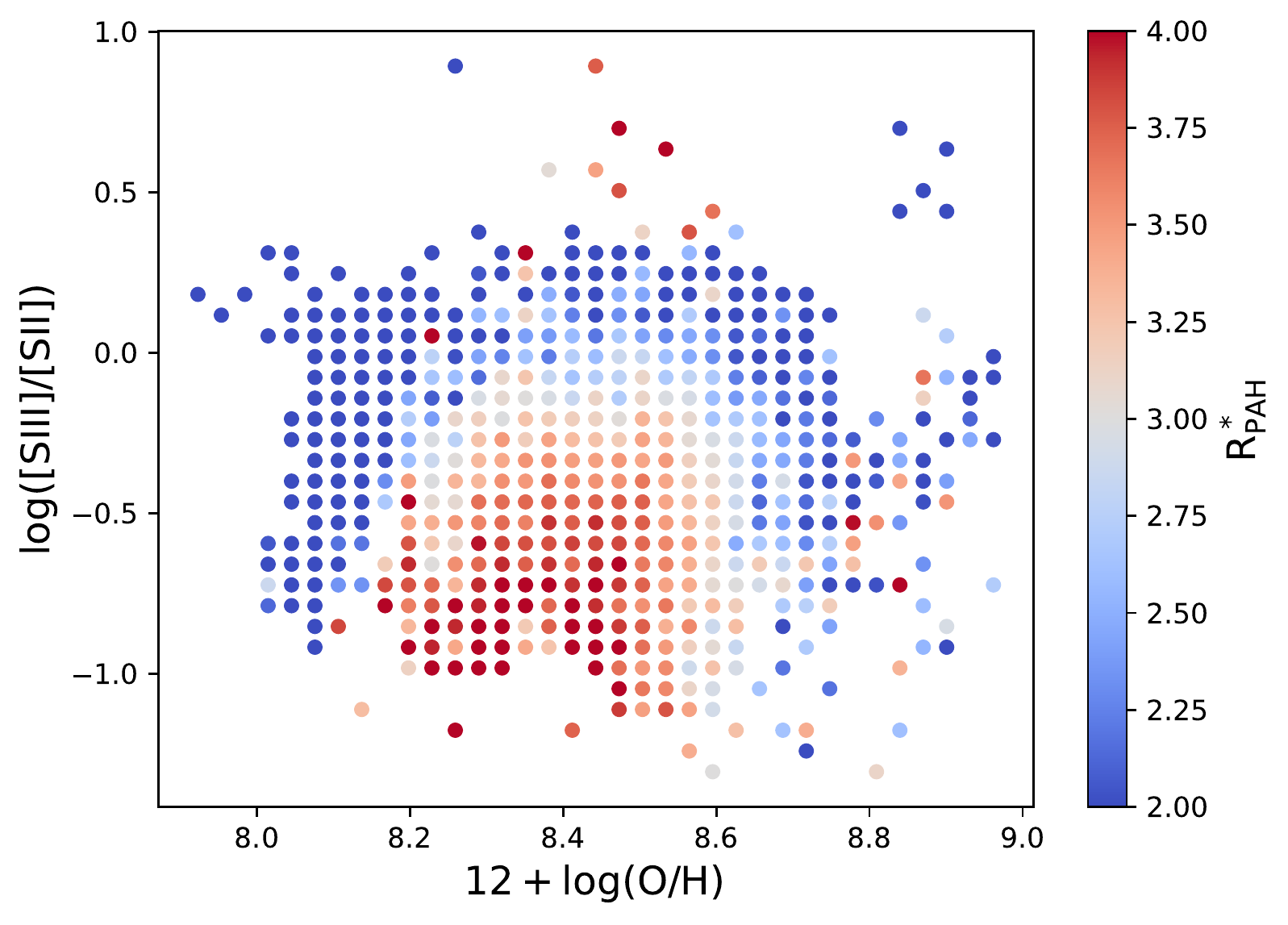}
    \caption{\RPAHst\ (shown by color) calculated on the regularly distributed bins with unique O/H and \SIIISII\ values. Both metallicity and ionization parameter are both important for setting up the PAH fraction.}
    \label{fig:RPAH_2D}
\end{figure}

The high-metallicity plateau in Fig.~\ref{fig:RPAH_vs_met} exhibits a large scatter of \RPAHst\ values, which can be associated with the differences in \SIII/\SII\ emission line ratio (color-coded in this Figure) -- a good proxy of the ionization parameter (see Section~\ref{sec:analysis_hii_snrs} and Appendix~\ref{sec:app:cloudy}). Indeed, splitting our data into two subsamples with different \SIIISII\ line ratios, we find that the average \RPAHst\ vs O/H  trends have very similar shapes with a clear offset (Fig.~\ref{fig:rpah_met_bins}). Furthermore, splitting our measurements into regular bins with different $\log$(\SIIISII) and $\rm 12+\log(O/H)$, we can see that the median \RPAHst\ values in these bins form a rather smooth 2D shape (Fig.~\ref{fig:RPAH_2D}). These results indicate that both environmental conditions (metallicity) and local ionization conditions are both important in setting the PAH fraction.

To check that the steep decrease of PAH fraction at low metallicities in our data is not just due to the different properties of the ionizing sources, we also measured \RPAHst\ in diffuse ISM for different metallicity bins. To do that, we compiled a comparable sample of mock diffuse regions following a procedure similar to what \cite{Dale2025} performed for their analysis of star clusters and associations. We masked out all \HII\ regions using the spatial masks (see Sec.~\ref{sec:analysis_hii_snrs}), and then randomly selected the positions of circular apertures in the residual footprint. The diameter of the aperture was fixed at $\rm 3\times PSF_{MUSE}$ for each galaxy to match the size of \revone{``resolved''} \HII\ regions (after rejecting the masked pixels). We discarded the regions where more than 50\% pixels in the JWST MIRI/F2100W grid overlap with \HII\ region masks. Repeating this procedure, we compiled a sample of the mock diffuse regions of comparable size and number per galaxy as for the \HII\ regions. Then, we derived \RPAH\ and \RPAHst\ in the same way as for \HII\ regions (Sec.~\ref{sec:analysis_rpah}). We assumed that these diffuse regions have a metallicity equal to the median for five nearest \HII\ regions. 

\begin{figure}%[!htbp]
    \centering
    \includegraphics[width=\linewidth]{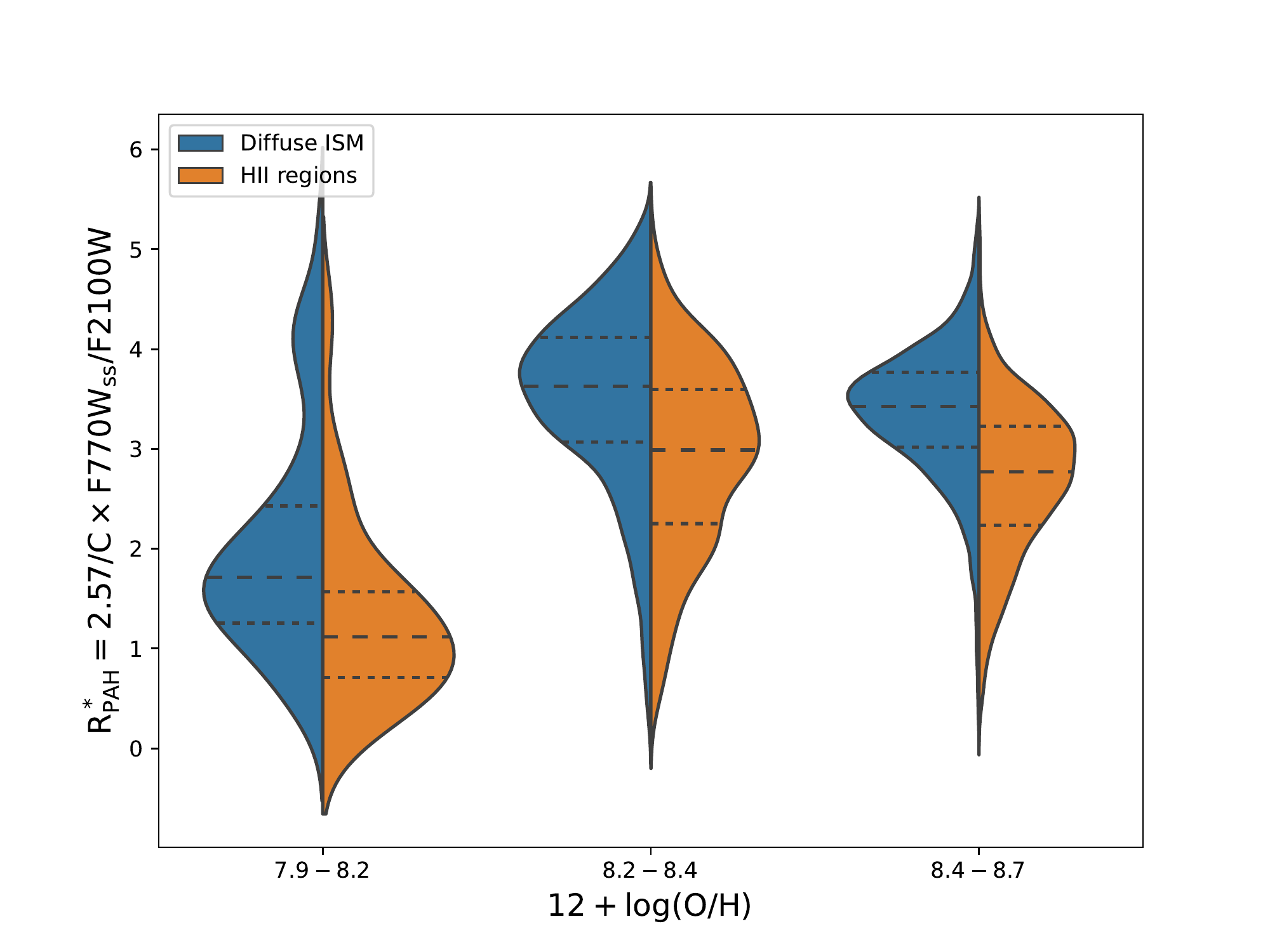}
    \caption{Distribution of \RPAHst\ for \revone{``resolved''} (luminous) \HII\ regions (orange) and the regions of the similar size from diffuse ISM (blue) for several gas-phase metallicity bins. Dashed and two dotted lines on each distribution show median and 25 and 75 percentile quartiles, respectively. Both \RPAH\ and \RPAHst\ are lower in \HII\ regions than in diffuse ISM. A strong decrease of \RPAHst\ in the lowest metallicity bin is seen for both \HII\ regions and diffuse ISM.}
    \label{fig:diff_vs_hii}
\end{figure}

We divide all measurements for the diffuse and \HII\ regions into several metallicity bins: $\rm 12+\log(O/H) = 7.9-8.2$, $8.2-8.4$, and $8.4-8.7$. Fig.~\ref{fig:diff_vs_hii} shows distributions of \RPAHst\, for each metallicity bin separately for the diffuse ISM and luminous (\revone{``resolved''}) \HII\ regions. Table~\ref{tab:rpah_met_bins} reports the median values of \RPAHst\ and standard deviations for both types of regions per metallicity bin. 

%A significant decrease of \RPAHst\ in \HII\ regions compared to diffuse ISM is consistent with the scenario of PAH destruction in \HII\ regions. Meanwhile, no difference is seen between the two highest metallicity bins, which is in agreement with the plateau in Fig.~\ref{fig:RPAH_vs_met}. 
%The lowest metallicity bin exhibits a significant decrease of \RPAHst\ in both diffuse ISM and \HII\ regions, in agreement with the observed steep drop of \RPAHst\ (Fig.~\ref{fig:RPAH_vs_met}). Therefore, the low-metallicity environment alters the PAH fraction in the diffuse ISM in the same way as for \HII\ regions.  

% Fig.~\ref{fig:diff_vs_hii} shows a singificant decrease of \RPAHst\ in \HII\ regions compared to the diffuse ISM, consistent with a scenario where PAH gets destroyed in these nebulae. 
% Despite this, \RPAHst shows the same type of behaviour with metallicity as we see in Fig.~\ref{fig:RPAH_vs_met}, with a plateau at high metallicity accompanied by a significant drop at low metallicities, in both the diffuse ISM and \HII\ regions.
%Therefore, the low-metallicity environment alters the PAH fraction in both environments in the same way.
Fig.~\ref{fig:diff_vs_hii} shows a significant decrease of \RPAHst\ in \HII\ regions compared to the diffuse ISM, consistent with a scenario where PAHs get destroyed in these nebulae. The two higher metallicity bins behave very similarly, reflecting the plateau in \RPAHst\ at high metallicity.  The lowest metallicity bin exhibits a significant decrease of \RPAHst\ in both \HII\ regions and the diffuse ISM. The differences between the lowest and moderate metallicity bins are in agreement with what was previously measured for the Magellanic Clouds \citep{Chastenet2019}. Therefore, the low-metallicity environment alters the PAH fraction in both environments in the same way. 

\begin{table}
  \centering
  \caption{\revone{Median \RPAHst\ and its 16th and 84th percentile intervals for different metallicity bins from Fig.~\ref{fig:diff_vs_hii}.}}
  \begin{tabular}{ccc}
    \hline
    $12+\log(\text{O/H})$ & Diffuse ISM & \HII\ Regions \\ 
    \hline
    % 7.9--8.2   & $1.75{\raisebox{0.5ex}\tiny\substack{+1.19 \\ -0.79}}$   & $1.13{\raisebox{0.5ex}\tiny\substack{+0.92 \\ -0.51}}$ \\ 
    % 8.2--8.4   & $3.61{\raisebox{0.5ex}\tiny\substack{+0.69 \\ -1.03}}$   & $2.95{\raisebox{0.5ex}\tiny\substack{+0.84 \\ -1.03}}$  \\ 
    % 8.4--8.7   & $3.41{\raisebox{0.5ex}\tiny\substack{+0.53 \\ -0.65}}$  & $2.76{\raisebox{0.5ex}\tiny\substack{+0.66 \\ -0.85}}$     \\ 
    7.9--8.2   & $1.72{\raisebox{0.5ex}\tiny\substack{+1.48 \\ -0.65}}$   & $1.12{\raisebox{0.5ex}\tiny\substack{+0.86 \\ -0.50}}$ \\ 
    8.2--8.4   & $3.63{\raisebox{0.5ex}\tiny\substack{+0.70 \\ -0.95}}$   & $2.99{\raisebox{0.5ex}\tiny\substack{+0.89 \\ -1.05}}$  \\ 
    8.4--8.7   & $3.43{\raisebox{0.5ex}\tiny\substack{+0.51 \\ -0.64}}$  & $2.77{\raisebox{0.5ex}\tiny\substack{+0.66 \\ -0.86}}$     \\ 
    \hline
  \end{tabular}
  \label{tab:rpah_met_bins}
\end{table}

\subsection{Correlation between PAH fraction and ionization parameter at 50--100~pc scales}
\label{sec:HIIregs}

Observations of nearby galaxies with JWST revealed a fairly uniform distribution of ${\rm R_{PAH}}$ in their diffuse ISM \citep{Sutter2024}. At the same time, the PAH fraction drops in \HII\ regions or central regions with AGNs \citep{Chastenet2023, Egorov2023, Pedrini2024, Baron2024a, Baron2025}. This drop in \HII\ regions at any metallicity (within the considered range) is clearly seen from the comparison of our sample of luminous \HII\ regions with the diffuse ISM areas shown in Fig.~\ref{fig:diff_vs_hii}. From the analysis of the vicinity of \HII\ regions in four galaxies from the PHANGS sample, \citet{Egorov2023} discovered a strong anti-correlation between ${\rm R_{PAH}}$ and $\log({\rm [SIII]/[SII]})$, which is a good tracer of the ionization parameter (see Appendix~\ref{sec:app:cloudy}). This dependence is also clearly seen in Fig.~\ref{fig:RPAH_2D}.

\begin{figure*}[!htbp]
    \centering
    \includegraphics[width=0.44\linewidth]{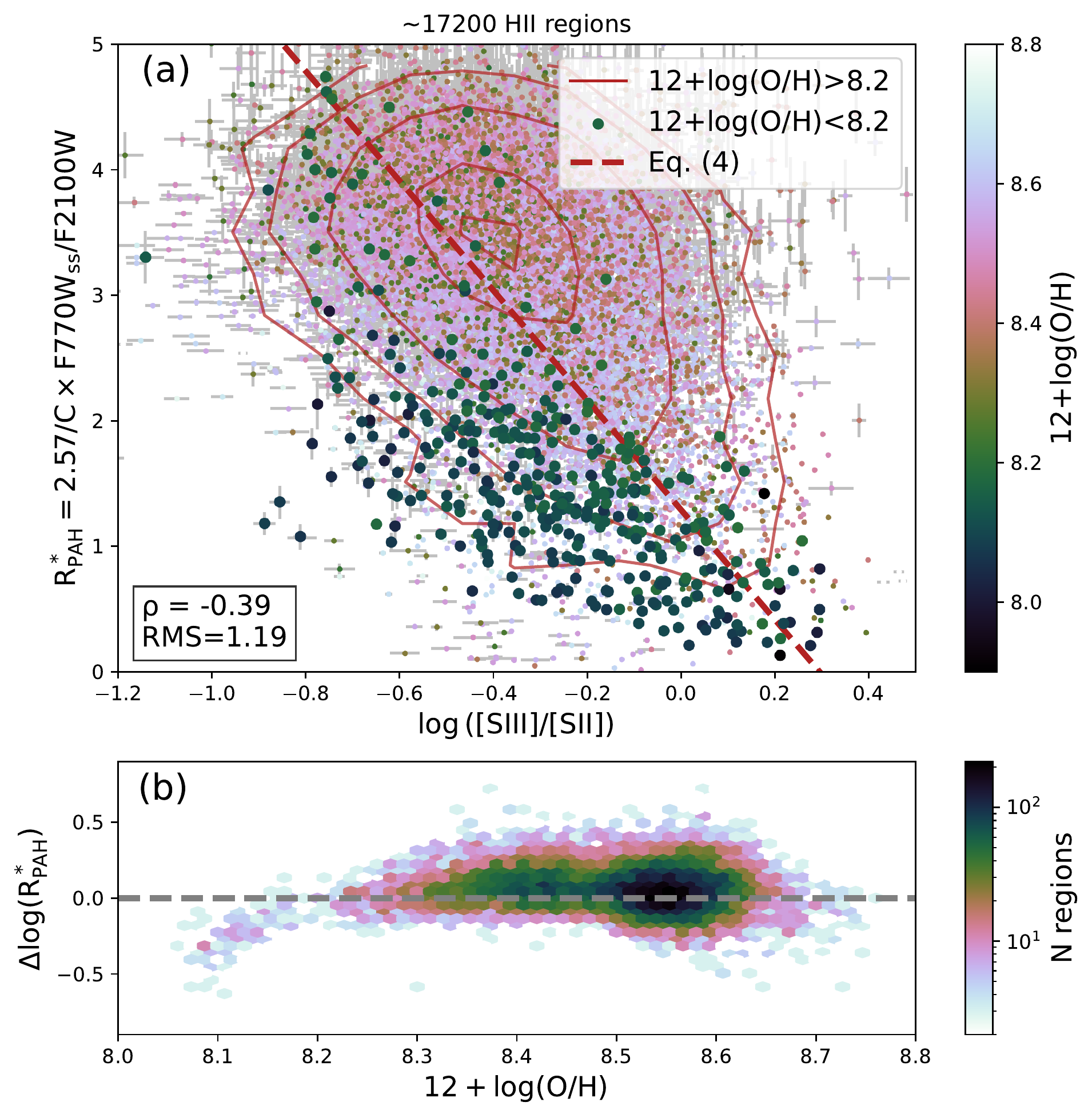}
\includegraphics[width=0.44\linewidth]{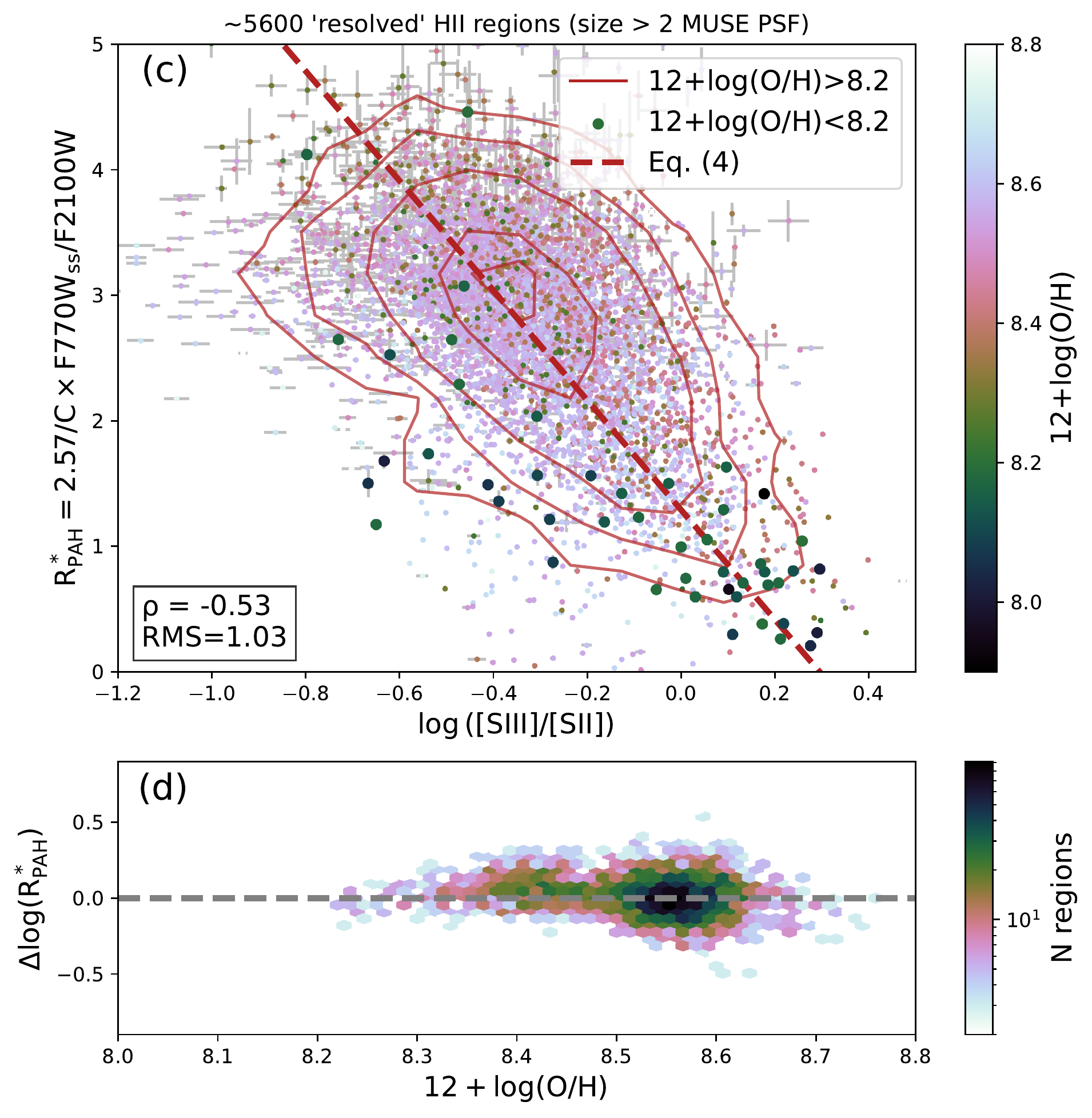}
    \caption{PAH fraction anti-correlates with $\log$(\SIII/\SII) (ionization parameter tracer) for thousands \HII\ regions in 42 PHANGS-JWST galaxies. Panel (a, b) shows all $\sim 17200$ analyzed \HII\ regions, while  $\sim 5600$ brightest \revone{``resolved''} \HII\ regions are shown on Panels (c, d). The dashed red line shows the best-fit linear regression to the \revone{``resolved''} sample defined by Eq.~\ref{eq:RPAH_vs_ip}. Histograms on panels (b, d) show how the logarithmic residuals after subtracting this linear trend depend on metallicity. Color on panels (a, c) denotes gas-phase oxygen abundance; red contours show the probability density of the high-metallicity ($\rm 12+\log(O/H) > 8.2$) points (levels correspond to 50, 65, 80, 95 and 99 percentile intervals). The Spearman correlation coefficient ($\rho$) and RMS scatter around the linear fit are reported on panels (a, c). }\label{fig:RPAH_vs_ip}
\end{figure*}

Fig.~\ref{fig:RPAH_vs_ip} shows how ${\rm R_{PAH}}$ changes with $\log({\rm [SIII]/[SII]})$ for a sample of \numHII\ \HII\ regions (panel a). This figure shows that the dependence of PAH fraction on the ionization parameter found by \citet{Egorov2023} is observed across a ten-fold larger sample of \HII\ regions than used in their work.  Fig.~\ref{fig:RPAH_vs_ip}c shows that the same trend is observed with expectedly smaller scatter if we keep only the luminous $\sim 5600$ \HII\ regions \revone{``resolved''} by MUSE (i.e. having sizes exceeding $\rm 2\times FWHM_{PSF}$, according to our selection criteria described in Sec.~\ref{sec:analysis_hii_snrs}). The best-fit linear regression model for the 2D distribution for the luminous \HII\ regions gives the following relation:
\begin{equation}
    {\rm R_{PAH}^*} = (-4.37 \pm 0.52)\times \log({\rm [SIII]/[SII]}) + (1.29\pm0.16).
\label{eq:RPAH_vs_ip}
\end{equation}
Coefficients and associated uncertainties are calculated using the orthogonal distance regression fitting to 500 bootstrapped samples. The same linear regression can also describe the unresolved regions in panel (a), although the scatter around it is higher, and the low-metallicity regions ($\rm 12+\log(O/H) < 8.2$) appear as outliers.  

It is not clear what causes the scatter around the linear regression defined by Eq.~\ref{eq:RPAH_vs_ip}. \citet{Egorov2023} tested several possible physical and observational parameters as potentially responsible for the observed scatter (including $Q^0$ -- the total number of ionizing photons, \OIIIHb\ as the UV hardness tracer, equivalent width of \Ha\ as the age tracer, $\rm F1130W/F770W$ as a tracer of PAH ionization state), and found a secondary dependence of \RPAH\ on metallicity (see their Fig.~3). Here we revisit this conclusion by reproducing the same analysis for a much larger sample of \HII\ regions, especially in the lower metallicity range. We subtract the trend described by Eq.~\ref{eq:RPAH_vs_ip} from the data in Fig.~\ref{fig:RPAH_vs_ip} and demonstrate how the logarithmic residuals depend on the metallicity in panels b and d (for all and resolved only \HII\ regions, respectively). The distribution of the residuals with metallicity is scattered around zero, but no obvious secondary dependence on metallicity is seen in the moderate-to-high metallicity regime in our data. The systematic deviation of the residuals from zero becomes prominent at low metallicities $\rm 12+\log(O/H) < 8.2$. 
%Therefore, thanks to improved statistics in the intermediate range of metallicities $\rm 12+\log(O/H) \sim 8.3-8.45$, we can revisit the conclusion of \citet{Egorov2023} and reject the existence of a  potential secondary dependence of \RPAH\ on metallicity in the moderate-to-high metallicity regime. We will consider how metallicity affects the observed \RPAH\ vs \SIII/\SII\ relation in the low metallicity regime later in Sec.~\ref{sec:metallicity}.

Figure~\ref{fig:RPAH_vs_ip} demonstrates a purely observational result. Physically, the decrease of \RPAH\ \revone{toward} \HII\ regions is consistent with scenarios of destruction or removal of PAH molecules. The fairly strong dependence of ${\rm R_{PAH}}$ on the ionization parameter is indicative of the major role of the ionized gas in the evolution of PAH molecules in star-forming regions. Understanding the physical reasons (if any) behind such a relationship can provide clues about the mechanisms of PAH destruction. We consider in Sec.~\ref{sec:discussion:ip} several possible physical and observational effects that could be responsible for the observed dependence between the \RPAH\ and the ionization parameter.

\section{Discussion}
\label{sec:discussion}

\subsection{Low- and high-metallicity regimes in the PAH life-cycle}
\label{sec:discussion:oh}

\begin{figure}
    \centering
    \includegraphics[width=\linewidth]{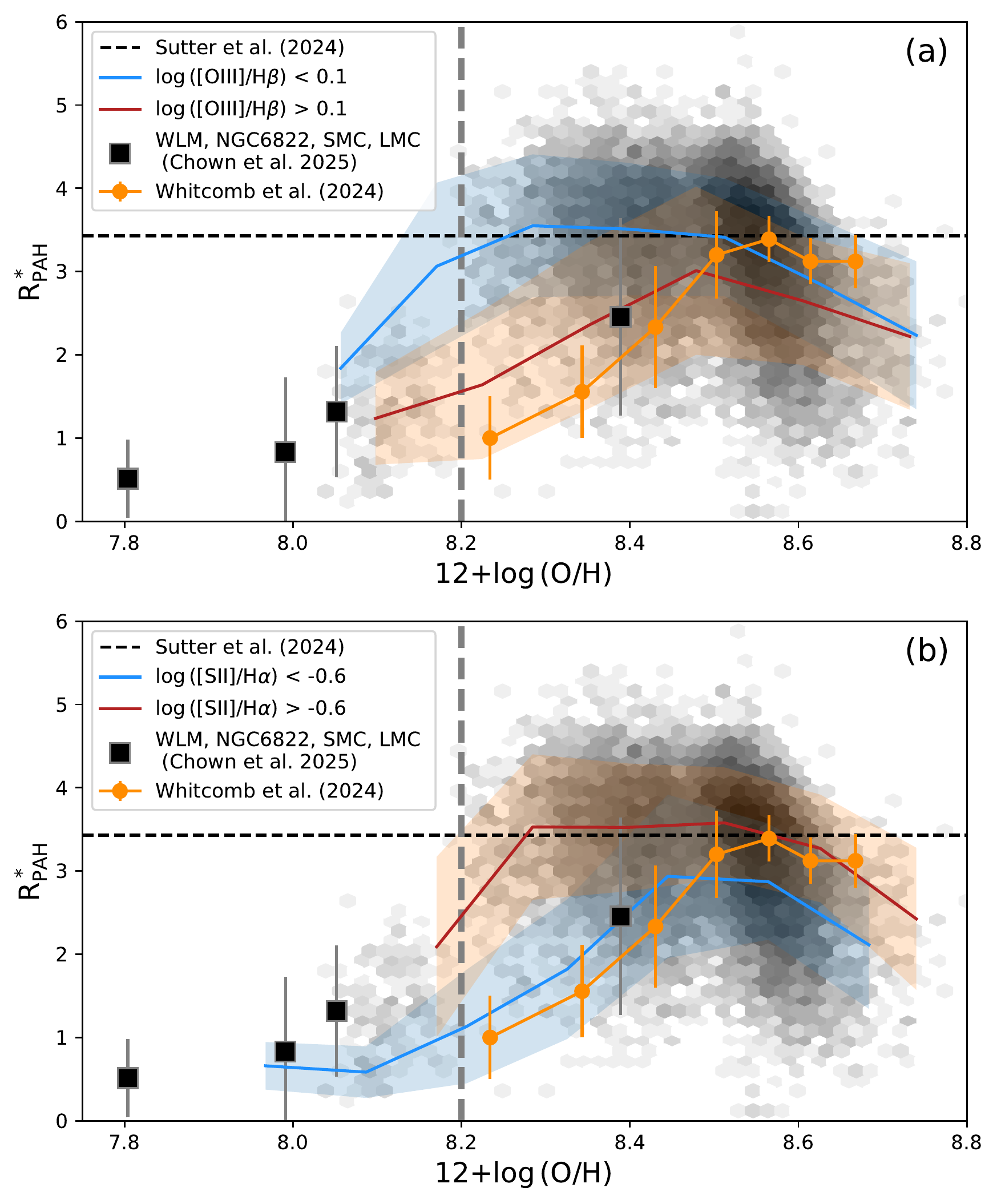}
    \caption{Same as Fig.~\ref{fig:rpah_met_bins}, but with the blue and red curves corresponding to different bins of low and high \OIIIHb\ (panel a) or \SIIHa\ (panel b), respectively. Orange curve and points are translated from average values per metallicity bin in fig.~4 of \citet{Whitcomb2024}.}
    \label{fig:rpah_met_bins_whitcomb}
\end{figure}

Our observations clearly reveal dependence of the PAH fraction on metallicity (Figs.~\ref{fig:RPAH_vs_met}--\ref{fig:diff_vs_hii}), which is consistent with previous resolved and unresolved studies of significantly smaller samples of star-forming regions (see Section~\ref{sec:intro}). Several physical mechanisms can be responsible for the lower PAH fraction at low metallicities. Hard ionizing radiation at low metallicity could lead to more efficient destruction of PAHs, so they rarely survive in \HII\ regions and inner layers of PDRs covered by our apertures. Alternatively, PAHs could form less efficiently (or have smaller sizes; \citealt{Sandstrom2012, Whitcomb2024}) in a low metallicity environment. This is supported by significantly lower \RPAHst\ (compared to high-metallicity bins, see Fig.~\ref{fig:diff_vs_hii}) also in the diffuse ISM, where the processes of their photo-destruction are less relevant.  
Another possible explanation is that PAHs can be destroyed by supernovae shocks more efficiently in low-metallicity environment because of the enhanced SNe rate in the low-mass star-forming galaxies \citep[e.g.][]{OHalloran2006, Jackson2006, Seok2014}. We address the role of shocks in selective PAH destruction in Section~\ref{sec:shocks}. 

The highest metallicity regions in our sample ($\rm 12+\log(O/H) > 8.55$) also reveal a decrease in the PAH fraction (two last bins in Fig.~\ref{fig:RPAH_vs_met}), which cannot be explained by  high \SIIISII\ alone (Fig.~\ref{fig:rpah_met_bins}). Such regions are located in the highest-mass galaxies from our sample. Although we excluded the galactic center regions from consideration, these high-metallicity regions are still associated with bars and other areas with much higher interstellar radiation field (from the old stellar population) than in spirals arms, which could result in decreased mid-IR emission, since the PAHs are being heated less effectively (\citealt{Draine2014, Whitcomb2024, Baron2025}; Pathak et al. in prep). %a lower PAH fraction \citep{Pathak2024}. 

At intermediate metallicities, from $\rm 12+\log(O/H) \sim 8.2$ to $\sim 8.6$, our observations find a constant average \RPAHst, with values for particular regions highly correlated with the ionization parameter. This conclusion is consistent with the results obtained by \citet{Egorov2023}, who found only a weak secondary dependence of \RPAH\ on metallicity for \HII\ regions from four PHANGS-JWST galaxies spanning a range of $12+\log({\rm O/H}) = 8.3-8.7$ (its presence is not confirmed here with the larger sample), and \citet{Sutter2024}, who found a remarkably constant \RPAH\ across the diffuse ISM in the same metallicity range. The same behavior of PAH fraction with metallicity was reproduced by models of \citet{Hirashita2020} considering PAH and dust destruction affecting the dust size distribution in the ISM.  Meanwhile, \citet{Whitcomb2024} derived a higher threshold from the Spitzer-IRS spectral observations of the radial strips in three nearby galaxies: they found that the PAH fraction does not change significantly for $\rm 12+\log(O/H) > 8.5$, below which it steeply drops. Although our results are in very good qualitative agreement with \citet{Whitcomb2024} findings, we try to understand why the quantitative measurements differ. 

One of the explanations for such a discrepancy could be a systematic offset in the metallicity owing to the difference in the methodology. \citet{Whitcomb2024} estimated metallicity relying only on the oxygen abundance gradients and position of the region in the galaxy, while in our study we measure the oxygen abundances for each \HII\ region individually (by using a strong emission line calibration). For the only overlapping galaxy in our samples -- NGC~628 -- there are differences in both central oxygen abundances and metallicity gradients assumed in their work ($12+\log({\rm O/H}) = 8.69 - 0.09 \times r/r_{e}$ measured by \citealt{Rogers2021} using \revone{``direct''} $T_e$-method) and derived by \citet{Groves2023} using the same oxygen abundance measurements as in our work ($12+\log({\rm O/H}) = 8.533 - 0.054\times r/r_{e}$). Such differences, if they result in a systematic bias, can easily explain $\sim 0.1$~dex of the overall discrepancy in the metallicity where PAH fraction drops sharply. The discrepancies between empirical strong-line and \revone{``direct''} methods are not surprising and are a long-standing problem \citep[see, e.g.,][]{Kewley2008}. However, Scal and $T_e$ calibration typically show reasonable agreement \citep[e.g.][]{Brazzini2024}, and therefore differences in the methods alone are unlikely to explain $\sim0.3$~dex systematic shift in the metallicity of the threshold.

%To further investigate possible reasons for the discrepancies in the location of the border between the low and high metallicity regimes between \citet{Whitcomb2024} and our work, we converted their measurements to \RPAHst\ and show them as an orange line overlaid on our measurements in Fig.~\ref{fig:rpah_met_bins_whitcomb}. Based on spectral observations, \citet{Whitcomb2024} compared all PAH emission to the total IR brightness ($\Sigma$PAH/TIR), which correlates linearly with the PAH mass fraction. Therefore, their $\Sigma$PAH/TIR can be scaled to \RPAHst. Estimating the average $\Sigma$PAH/TIR and \RPAHst\ for the NGC~628 in both datasets, we find \RPAHst$ \simeq 0.22 \times \Sigma$PAH/TIR. Using this scaling factor, we converted \citet{Whitcomb2024} measurements (average values in bins in their fig.~4) to \RPAHst. 
Our photometric approach to calculating the PAH fraction utilizes information about only a small number of PAH features, and does not allow precise continuum estimates, and thus can have some bias compared to spectral measurements. 
\citet{Whitcomb2024} is based on spectral observations, 
and compares all PAH emission to the total IR brightness ($\Sigma$PAH/TIR).  As this correlates linearly with the PAH mass fraction, their $\Sigma$PAH/TIR can be scaled to \RPAHst, to allow a more direct comparison. Estimating the average $\Sigma$PAH/TIR and \RPAHst\ for the NGC~628 regions in both datasets, we find \RPAHst$ \simeq 0.22 \times \Sigma$PAH/TIR. Using this scaling factor, we converted the results from \citet{Whitcomb2024} (using the average values in bins in their fig.~4) to \RPAHst, and show them as an orange line overlaid on our measurements in Fig.~\ref{fig:rpah_met_bins_whitcomb}. 

%Comparison of \citet{Whitcomb2024} and our measurements in Fig.~\ref{fig:rpah_met_bins_whitcomb} 
This comparison indicates another possible origin for the discrepancy in the threshold between the high- and low-metallicity regimes. Splitting our data into two subsamples with high and low \OIIIHb\ or \SIIHa\ (similarly to Fig.~\ref{fig:rpah_met_bins} for \SIIISII), we find that considering only hard and soft ionization regions (that is, with high \OIIIHb\ and low \SIIHa, and vice versa) can lead to systematic differences in the position of the metallicity threshold: harder ionization regions exhibit a higher metallicity turnover.  Interestingly, the low-metallicity end of the \citet{Whitcomb2024} curve coincides very well with our measurements when only regions with relatively hard ionizing spectra are considered. However, since their data in this tail correspond to kpc-sized bins that include diffuse ISM and \HII\ regions, such ionization bias is very unlikely, and systematics in metallicity estimates remain the most probable explanation.

\subsection{Is the PAH fraction physically linked with the properties of ionized gas?}
\label{sec:discussion:ip}

By construction, each point in Fig.~\ref{fig:RPAH_vs_ip} represents an \HII\ region. High physical resolution observations of a small subset of Galactic nebulae reveal sharp borders in the PAH emission \citep[e.g.][]{Hernandez-Vera2023, Abergel2024, Chown2024}. Given this, one can expect that efficient PAH destruction would produce much lower values of \RPAHst\ than what is observed in Fig.~\ref{fig:RPAH_vs_ip}. The PAH fraction should be around zero in all (especially the brightest) \HII\ regions, which approximately corresponds to \RPAHst$\sim 0.5$ (see Fig. 7 in \citealt{Sutter2024}). However, such sharp borders between PAH emission and \HII\ regions cannot be observed in relatively distant extragalactic objects even if the ionized gas is traced by high-resolution Pa$\alpha$ JWST images \citep{Pedrini2024}. Instead, we see a gradual decrease with the ionization parameter even for the brightest \revone{``resolved''} \HII\ regions (Fig.~\ref{fig:RPAH_vs_ip}c), with non-zero PAH fraction at any measured ratio of \SIII/\SII\ (with the exception of some of the lowest metallicity regions). Here we consider physical and observational effects that can be responsible for the measured non-zero PAH fractions in \HII\ regions and setting such a uniform observational trend with the \SIII/\SII\ present across all galaxies in our sample. \revone{For all tests in the rest of this Section, we rely on \RPAH\ (not \RPAHst) measurements, which include the 7.7 $\mu$m and 11.3 $\mu$m PAH bands and are not dependent on the luminosity-sensitive conversion between \RPAHst\ and \RPAH. However, the results remain unchanged when using \RPAHst.}

\paragraph{\em Physical connection between the PAH fraction and the number of EUV photons.} By definition, the ionization parameter is proportional to the number of hydrogen-ionizing photons relative to the number of hydrogen atoms. The more such extreme UV (EUV) photons are present in a region, the higher the ionization parameter, and the more PAH molecules are exposed by such harsh ionizing radiation. Laboratory studies find that large PAHs are mostly photoionized when exposed to energies above 13.6~eV, and start experiencing photodissociation at energies below 20~eV \citep{Wenzel2020}. Because most EUV photons have the potential to destroy PAHs, it is primarily the number density of EUV photons and not the hardness of the UV spectrum that determines the efficiency of PAH destruction. The latter explains the absence of the secondary dependence of \RPAH\ on \OIIIHb\ and metallicity for the high metallicity range (Figs.~\ref{fig:RPAH_vs_ip}, \ref{fig:rpah_met_bins_whitcomb}, see also \citealt{Egorov2023}). From modeling, \citet{Murga2019} found that the characteristic timescale for the photodestruction of PAH-like grains decreases significantly with increasing radiation field intensity, providing another physical basis for the observed anti-correlation between \RPAH\ and the ionization parameter. Overall, the scenario of ionization of PAHs and their subsequent destruction (through photodestruction or sputtering and fragmentation) regulated by the intensity of EUV ionizing radiation (and hence -- ionization parameter) is consistent with the observed decrease in the \RPAH\ in the \HII\ regions compared to diffuse ISM (Fig.~\ref{fig:diff_vs_hii}, see also \citealt{Sutter2024, Pedrini2024}).

\paragraph{\em PDRs dominate the \RPAH\ measurements due to limited resolution.}
As described in Sec.~\ref{sec:analysis_hii_snrs}, we identified the borders of the \HII\ regions based on the MUSE data, which have angular (and spatial) resolution lower than that of the JWST images. At the distance of our sample, %The sizes derived from the MUSE data tend to be overestimated, and therefore, we expect that 
our spatial masks defining borders of \HII\ regions are not perfectly isolating them from the PDRs where the emission of PAHs dominates \citep[e.g.][]{Anderson2014}. Even if the small number of \HII\ regions in our sample are truly resolved, we still expect to see a contribution from PDRs and the diffuse ISM along the line-of-sight. Therefore, the non-zero values of \RPAH\ can be naturally explained if the \RPAH\ values reflect a combination of the surrounding diffuse ISM, PDRs, and the \HII\ region within the nebular mask boundary. The observed anti-correlation could also reflect the strong link between the properties of the surrounding ISM (including PDRs) and ionized gas in \HII\ regions. In particular, due to the leaky nature of \HII\ regions \citep[e.g.][]{Oey2007, DellaBruna2021, Ramambason2020, Teh2023}, EUV radiation can penetrate into surrounding ISM and ionize or destroy PAHs there, as described above. Therefore, the results in Fig.~\ref{fig:RPAH_vs_ip} can reflect the processes occurring in the zones of influence of \HII\ regions rather than in \HII\ regions themselves. 

\begin{figure}%[!htbp]
\centering
    \includegraphics[width=0.745\linewidth]{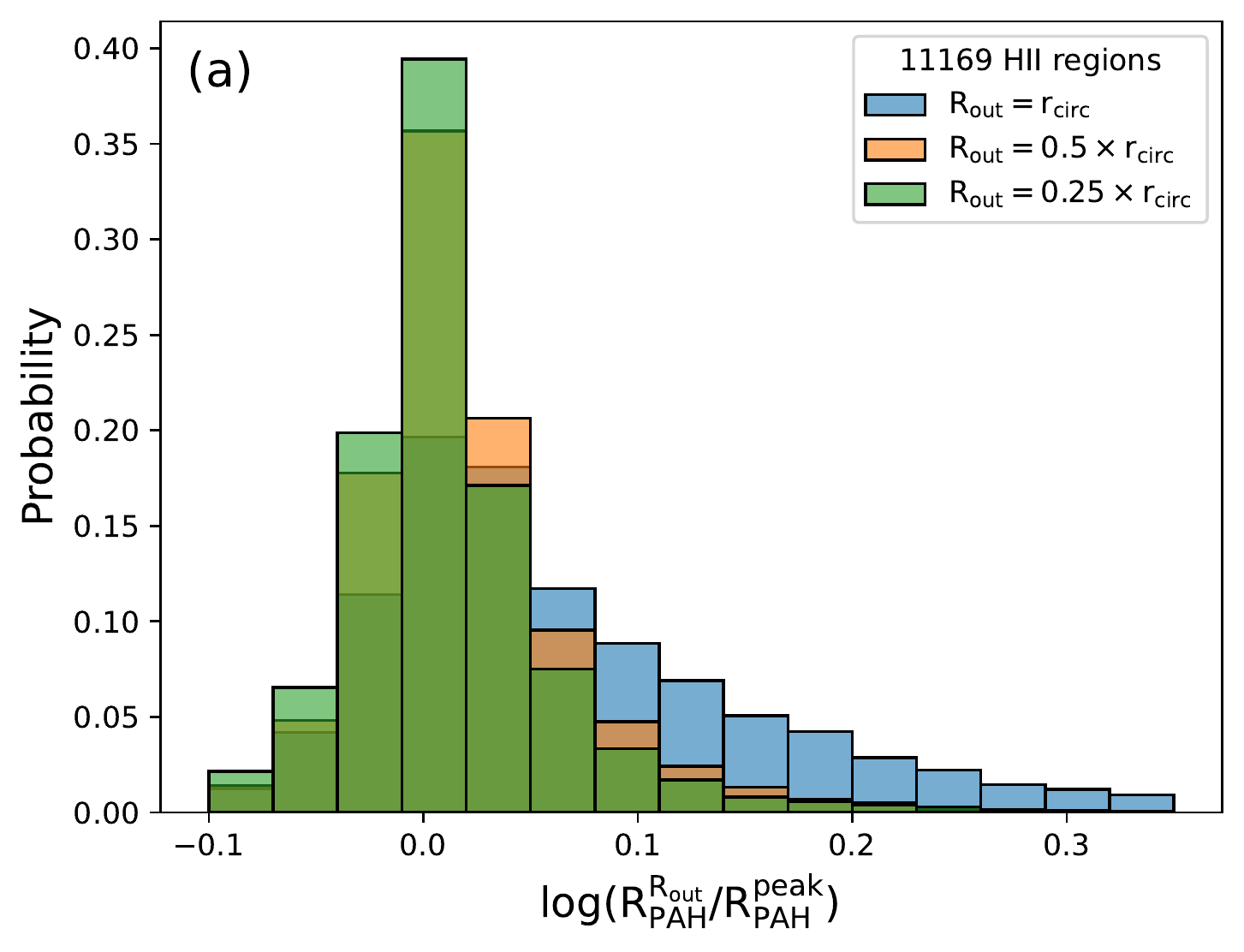}

    \includegraphics[width=0.765\linewidth]{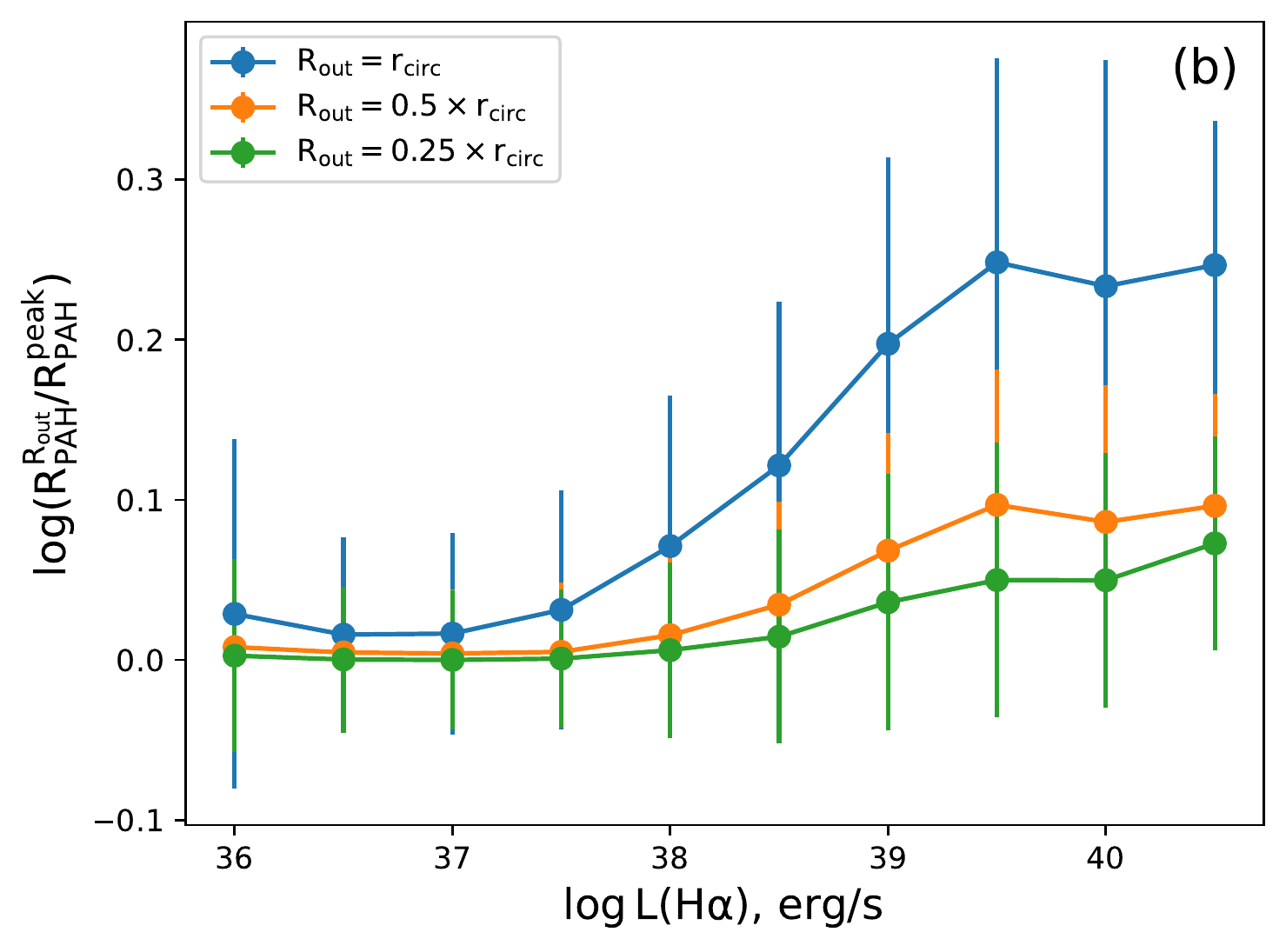}
    \caption{About 45\% of the \RPAH\ measurements are overestimated by more than 10\% when measured in the MUSE-based apertures for \HII\ regions compared to the values in their centers. Panel (a) shows the probability distribution of \RPAH\ measured in different circular apertures for the same sample of \HII\ regions related to the value of $\rm R_{PAH}$ measured at the local peaks of \Ha\ brightness within the \HII\ region. Panel b demonstrate how such ratios change for different \Ha\ luminosity bins. Different colors correspond to different outer radii of the circular aperture.  }\label{fig:RPAH_hist_diffap}
\end{figure}

To test whether our measurements of \RPAH\ are indeed significantly affected by the overestimated sizes of \HII\ spatial masks, we compare the values of \RPAH\ measured in circular apertures with radii equal to $r_{\rm circ}$, $0.5 \times r_{\rm circ}$, $0.25 \times r_{\rm circ}$, where $r_{\rm circ}$ is circularized radius of the \HII\ region derived from MUSE data, to the value recovered at the location of a local peak in \Ha\ distribution within the \HII\ region mask. %If our measurements are not affected by aperture size and if a sharp border between PAHs and ionized gas is present in all \HII\ regions, we expect to see no differences between the measurements in different apertures. 
The results of our test are shown in Fig.~\ref{fig:RPAH_hist_diffap}, where different colors correspond to distribution of the \RPAH\ measured in the chosen aperture relative to that at the location of \Ha\ peak, where $\rm {\rm R_{PAH}^{peak}}$ \revone{is supposed} to be minimal.

\begin{figure*}%[!htbp]
\centering
    \includegraphics[width=0.68\linewidth]{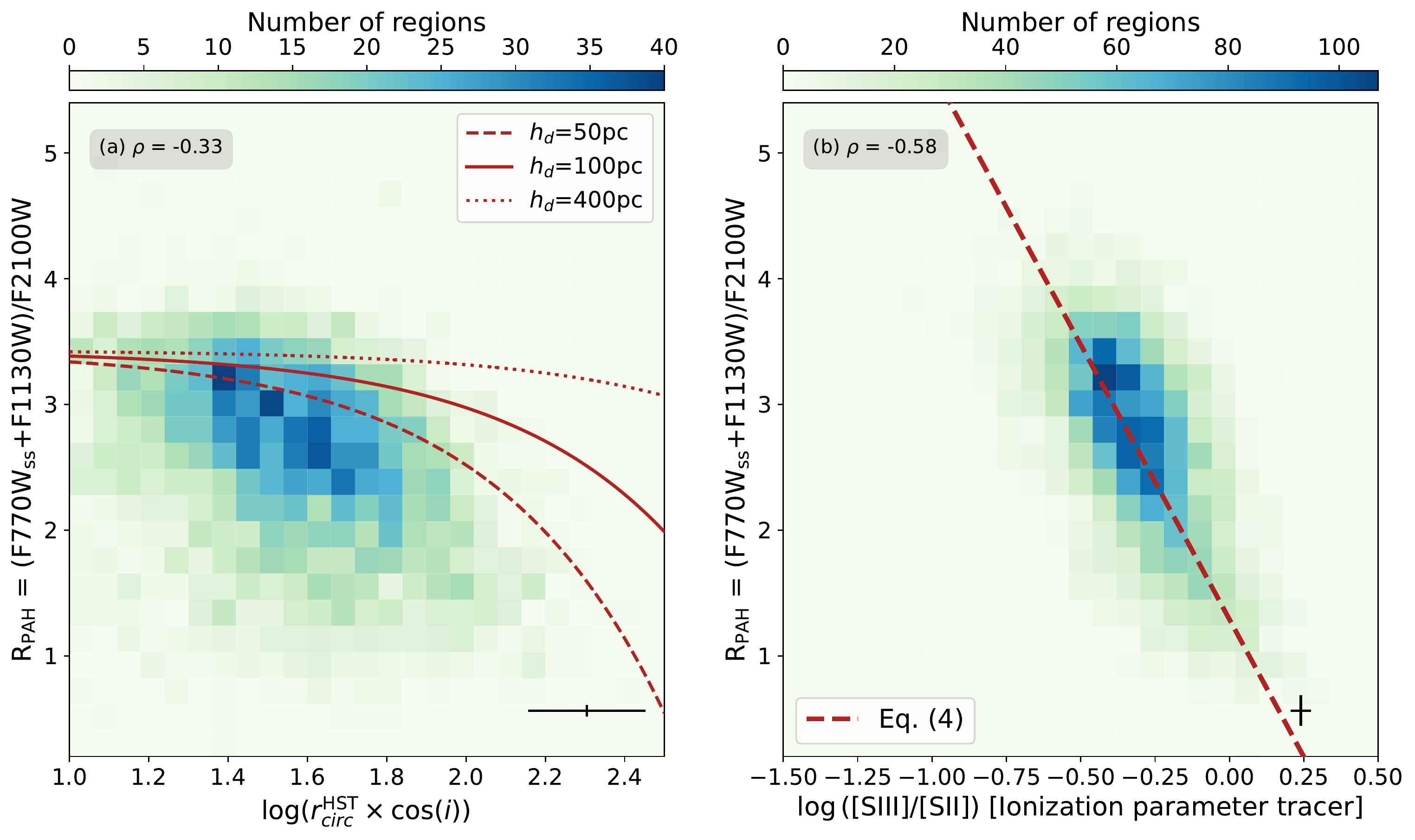}
    \caption{Histograms showing distribution of \RPAH\ vs the logarithm of the circularized radii of the \HII\ regions derived from HST data (from Barnes et al., submitted; panel (a)), and vs the ionization parameter tracer $\rm \log([SIII]/[SII])$ (panel (b)). Colors denote a number of the regions in each bin. Curved dashed, solid and dotted lines on the panel (a) show models defined by Eq.~\ref{eq:RPAH_vs_hd} with the \revone{disk} scale height $h_d=50$, 100 and 400~pc, respectively. Red dashed line on the panel (b) corresponds to Eq.~\ref{eq:RPAH_vs_ip}. Anti-correlation between \RPAH\ and size of the \HII\ region is much weaker than that with the ionization parameter and can be partially explained by the contribution of the PAHs from the diffuse ISM along the line of sight.}\label{fig:RPAH_vs_Reff}
\end{figure*}

We do not see significant differences between the measurements made in the two reduced apertures compared to $\rm R_{PAH}^{peak}$ and to each other. However, about 45\% of regions show higher \RPAH\ by at least 10\% when measured in the largest aperture corresponding to the adopted size of the \HII\ regions in our analysis. As follows from Fig.~\ref{fig:RPAH_hist_diffap}b, the difference is greater in brighter regions, which likely indicates that we cannot isolate diffuse ISM from \HII\ regions at all in the faintest areas using the JWST MIRI/F2100W resolution in our sample.

From this test, we can conclude that our integrated measurements in Fig.~\ref{fig:RPAH_vs_ip} are contaminated by a contribution from the PDR and surrounding diffuse ISM due to insufficient resolution to isolate \HII\ regions. To test if this fact alone is responsible for the observed anti-correlation between \RPAH\ and the ionization parameter, we tested also how \RPAHst\ measured at the \Ha\ peaks of the \revone{``resolved''} \HII\ regions changes with \SIIISII\ and did not find significant differences with the Fig.~\ref{fig:RPAH_vs_ip}c -- the anti-correlation remains strong and can be fitted by Eq.~\ref{eq:RPAH_vs_ip}. Therefore, we can rule out that this relation is the consequence of the aperture effect alone.   %changes with  in Fig.~\ref{fig:RPAH_ip_diffap} the version of Figs.~\ref{fig:RPAH_vs_ip}c, but with \RPAH\ measured in the \Ha\ peaks.
% \begin{figure*}[!htbp]
% \centering
%     \includegraphics[width=0.49\linewidth]{figs/RPAH_vs_ip_v20_resolved_peak.pdf}
%     \includegraphics[width=0.49\linewidth]{figs/RPAH_vs_ip_v20_bgrsub_resolved_peak.pdf}
%     \caption{Same as in Fig.~\ref{fig:RPAH_vs_ip} (b) but measured at the location of \Ha\ peaks for each \HII\ region. Red dashed line show the relation defined by Eq.~\ref{eq:RPAH_vs_ip}. The anti-correlation between \RPAH\ and \SIII/\SII\ remains the same for the smallest apertures probing the interiors of \HII\ regions, although the absolute values of \RPAH\  decrease.}\label{fig:RPAH_ip_diffap}
% % \end{figure*}
% Although the systematic offset from the relation defined by Eq.~\ref{eq:RPAH_vs_ip} is prominent, the anti-correlation remains strong. Therefore, we can rule out that this relation is the consequence of the aperture effect alone. 

We note that the same effect as seen in Fig.~\ref{fig:RPAH_hist_diffap} can be in principle observed even if our data would allow perfect isolation of \HII\ regions provided that PAHs are not completely destroyed or removed at the edges of \HII\ regions. Recent JWST observations of the Horsehead nebula \citep{Abergel2024} revealed photoevaporative flows from PDRs to the interior of the \HII\ region visible in F770W and F335M filters centered on PAHs. The characteristic length of such features ($\sim 0.05$~pc) is much smaller than can be resolved in similar extragalactic regions. However, as was demonstrated by recent SDSS-V/LVM IFU observations, the central region of Orion (which includes the Horsehead nebula) is not representative of the population of extragalactic \HII\ regions \citep{Kreckel2024}. Therefore, in principle, such an effect could be more prominent in the regions surrounding more energetic sources like what is seen in the PHANGS sample of \HII\ regions. Furthermore, \citet{Bolatto2024} observed PAH emission closely resembling the ionized gas in outflow of M82 and argued that part of it can come from the surviving PAHs at the interface between the ionized and neutral ISM. Future statistical morphological analysis of a large sample of regions in the Local Group galaxies with very high resolution for both ionized gas (e.g. with the Pa$\alpha$ emission line) and PAHs (using 3.35~$\mu$m feature) with JWST %, as prototyped by \cite{Pedrini2024}, 
can provide clues about how steep the decrease in PAH abundance is in normal \HII\ regions. 

\paragraph{\em Time evolution of PAHs.}
Independently of whether the result presented in Fig.~\ref{fig:RPAH_vs_ip} reflects the processes in \HII\ regions or in the PDRs, the general behavior of this plot can be attributed also to the time evolution of the \HII\ regions. Theoretical models and observational works demonstrate that at fixed \HII\ region luminosity, the ionization parameter decreases as a function of time \citep[e.g.][]{Dopita2006, Scheuermann2023}. From photoionization models, it follows that the observed line ratio \SIIISII\ also decreases with the age of the ionizing star cluster even at the fixed value of ionization parameter (see Fig.~\ref{fig:cloudy} in Appendix~\ref{sec:app:cloudy}). Therefore, the anti-correlation between \RPAH\ and the \SIII/\SII\ observed in Fig.~\ref{fig:RPAH_vs_ip} could reflect a time evolution of the PAH fraction in the ISM illuminated by EUV radiation, where each bin along the x-axis could be interpreted as a time stamp. However, a lack of correlation between \RPAH\ and EW(H$\alpha$), another good tracer of the age of \HII\ regions, makes this interpretation unlikely \citep{Egorov2023}.

\paragraph{\em Line-of-sight projection of the diffuse ISM onto \HII\ regions}
Star-forming regions are embedded in the large reservoirs of the diffuse ISM. Therefore, some contamination of the measured fluxes in optical emission lines (from DIG) and in mid-IR bands (from neutral diffuse ISM) is expected. The absolute brightness of \SII\ and especially \SIII\ is typically much lower in DIG than in \HII\ regions, and thus its contribution along the line of sight should not significantly affect the observed \SIII/\SII\ and the ionization parameter for most of the regions. \citet{Egorov2023} demonstrated that the anti-correlation between \RPAH\ and the ionization parameter remains unchanged when only \HII\ regions with high contrast over DIG are considered. The contamination of \RPAH\ measurements by line of sight emission from the neutral ISM can be however much higher as the fluxes in mid-IR bands are typically high outside the star-forming regions \citep{Leroy2023, Sandstrom2023, Pathak2024}. Below we consider if such contamination alone can explain the observed relation in Fig.~\ref{fig:RPAH_vs_ip}.

To test this, we consider here a toy model of a spherical \HII\ region with radius $r_{\rm HII}$ located at the mid-plane of the galactic \revone{disk}. %, which is oriented face on. 
We assume that the vertical density distribution of the dust in the \revone{disk} is mimicking the gas density distribution and can be described by a Gaussian with the \revone{disk} scale height $h_d$. Also, we assume that the \HII\ region is completely devoid of PAHs, and outside the \HII\ region their volumetric density is distributed according to the same law as for dust grains. The latter assumption is in agreement with the findings by \citet{Sutter2024}, who measured fairly constant distribution of \RPAH$\sim 3.43$ across the diffuse ISM of 19 PHANGS galaxies. We therefore assume in our toy model that outside the \HII\ region, ${\rm R_{PAH}} = {\rm R_{PAH}^0} = 3.43$. Then we can estimate how \RPAH\ should change with the size of \HII\ region assuming that all PAHs that we observe \revone{toward} \HII\ region are from the diffuse ISM along the line of sight:
\begin{multline}
        {\rm R_{PAH}} \sim \frac{M_{\rm PAH}}{M_{\rm dust}} \simeq \frac{\int_{{diff}}\rho_{PAH}dV}{\int_{{diff+HII}}\rho_{dust}dV} \simeq \\
        \frac{\rho_{PAH}^0(S_{HII}\int_{diff+HII}e^{-z^2\cos^2(i)/2 h_d^2}dz - V_{HII})}{\rho_{dust}^0S_{HII}\int_{diff+HII}e^{-z^2\cos^2(i)/2h_d^2}dz} \simeq \\
        {\rm R_{PAH}^0}\times(1-\frac{r_{\rm HII}\cos(i)}{3 \sqrt{2\pi} h_d}), 
\label{eq:RPAH_vs_hd}\end{multline}
where $M_{PAH}$ and $M_{dust}$ are total masses of the PAHs and small grains along the line-of-sight in the aperture limited by the \HII\ region footprint, $\rho_{PAH}$, $\rho_{dust}$, $\rho_{PAH}^0$ and $\rho_{dust}^0$ are volumetric densities of the PAHs and small dust grains in the particular volume $dV$ and in the midplane of galactic \revone{disk} outside \HII\ region, respectively; $z$ is the distance along the line-of-sight from the midplane;  $i$ is inclination of the galactic \revone{disk}; $S_{HII}$ and $V_{HII}$ are the area in the galactic plane and the volume occupied by the \HII\ region, respectively. We note as one of the caveats that this toy model does not account for the probable increase of PAH and dust masses and luminosities around the \HII\ regions. 

By definition, the ionization parameter is proportional to $1/r_{\rm HII}^{2}$, which means that it decreases with the expansion of the \HII\ region (assuming constant density and ionizing flux). Photoionization models also reveal such a trend for \SIII/\SII\ line ratio: while \SIII/\SII\ increases with the ionizing photons rate $Q^0$, it decreases with the radius of the region for each constant value of $Q^0$ (see Appendix~\ref{sec:app:cloudy}). Therefore, to some extent, the \SIII/\SII\ ratio should correlate with the size of the regions, and thus it can be assumed that the observed trend between \RPAH\ and \SIII/\SII\ can result from the \RPAH\ vs. $r_{\rm HII}$ relation, which in turn is expected from our toy model according to Eq.~\ref{eq:RPAH_vs_hd}.

In Fig.~\ref{fig:RPAH_vs_Reff}a, we check if \revone{$\rm R_{PAH}$} correlates with the \HII\ size and if this correlation can be explained by our toy model. As a proxy of $r_{\rm HII}$, we consider circularized radii of \HII\ regions $r_{\rm circ}^{\rm HST}$ derived by Barnes et al. (submitted) from HST \Ha\ narrow-band images \citep{Chandar2025} for about 5000 \revone{\HII\ regions} in 19 PHANGS-JWST Cycle~1 galaxies. Unlike MUSE \Ha\ images, HST observations resolve \HII\ regions and provide more reliable measurements of their sizes. 
We indeed see the anti-correlation between sizes and \RPAH. However, the scatter is much higher and the significance of the correlation is much weaker than for the \RPAH\ vs \SIII/\SII\ shown on panel (b) for the same subsample of regions and in the same 2D histogram representation for comparison. Note that the sizes in panel (a) are multiplied by $\cos(i)$ to take into account the galaxy inclination in the same way as in Eq.~\ref{eq:RPAH_vs_hd} for further comparison with the toy model. However, the scatter and the correlation coefficient improve insignificantly for the uncorrected values.

% Given the weakness of the  correlation of \RPAH\ vs $\log(r_{circ}^{\rm HST})$, it probably just reflects the more universal dependence of \RPAH\ on the ionization parameter. Nevertheless, we can try to reproduce this relation assuming its purely observational origin due to the contamination by the diffuse ISM along the line of sight. 

\begin{figure}
    \centering
    \includegraphics[width=0.8\linewidth]{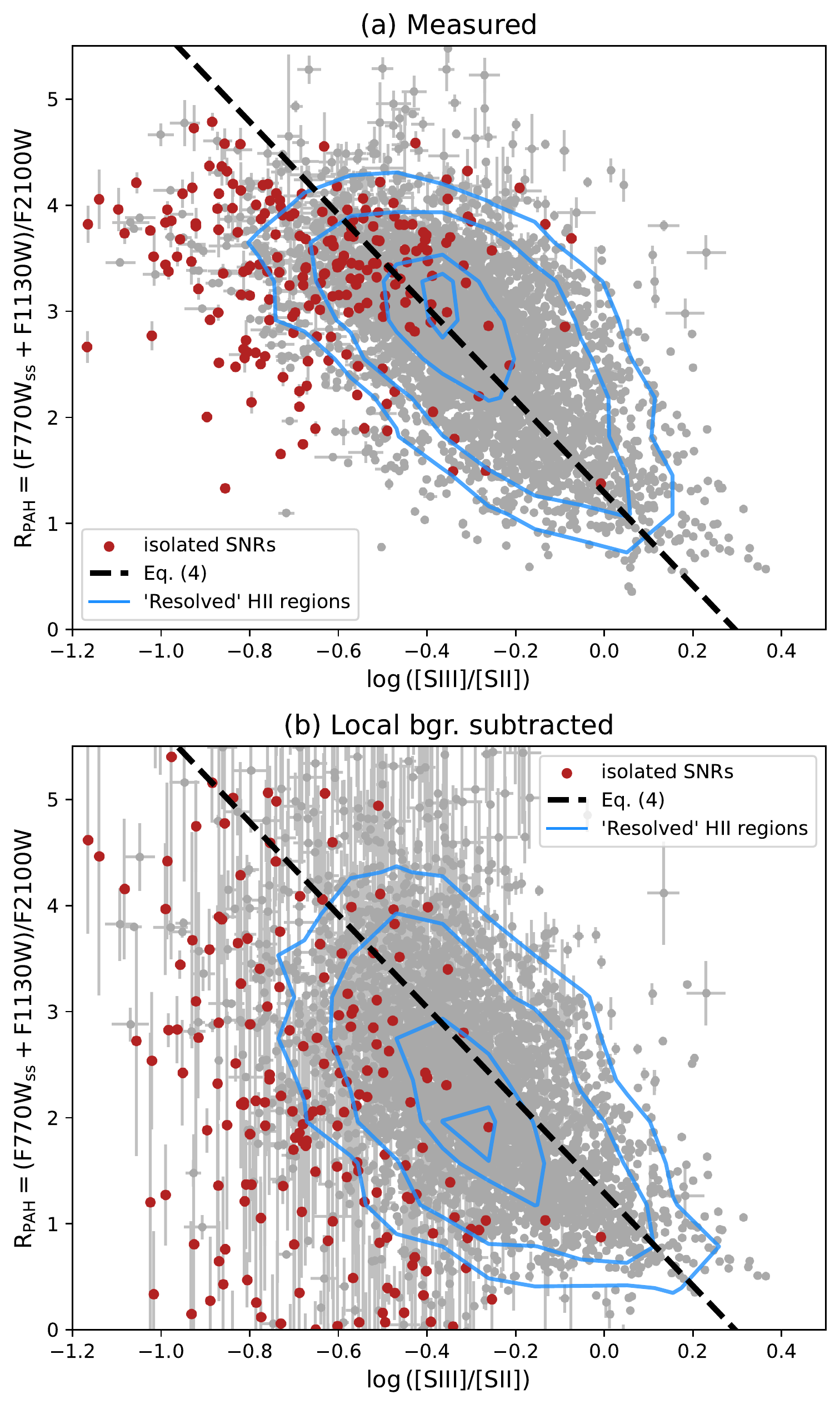}
    \caption{Distribution of \RPAH\ vs $\rm \log([SIII]/[SII])$ for the resolved \HII\ regions (gray points; same as in Fig.~\ref{fig:RPAH_vs_ip}c) and isolated SNRs (red points). Panels (a) and (b) correspond to \RPAH\ measurements from the observed and the local background subtracted fluxes. The black dashed line corresponds to the relation defined by Eq.~(\ref{eq:RPAH_vs_ip}). Cyan contours show probability density distribution for \HII\ regions (levels correspond to 65, 80, 95 and 99 percentile intervals).} %SNRs show in general higher \RPAH, but also lower $\rm \log([SIII]/[SII])$ compared to the \HII\ regions.}
    \label{fig:RPAH_vs_ip_snrs}
\end{figure}

\begin{figure*}%[!htbp]
    \centering
    \includegraphics[width=0.46\linewidth]{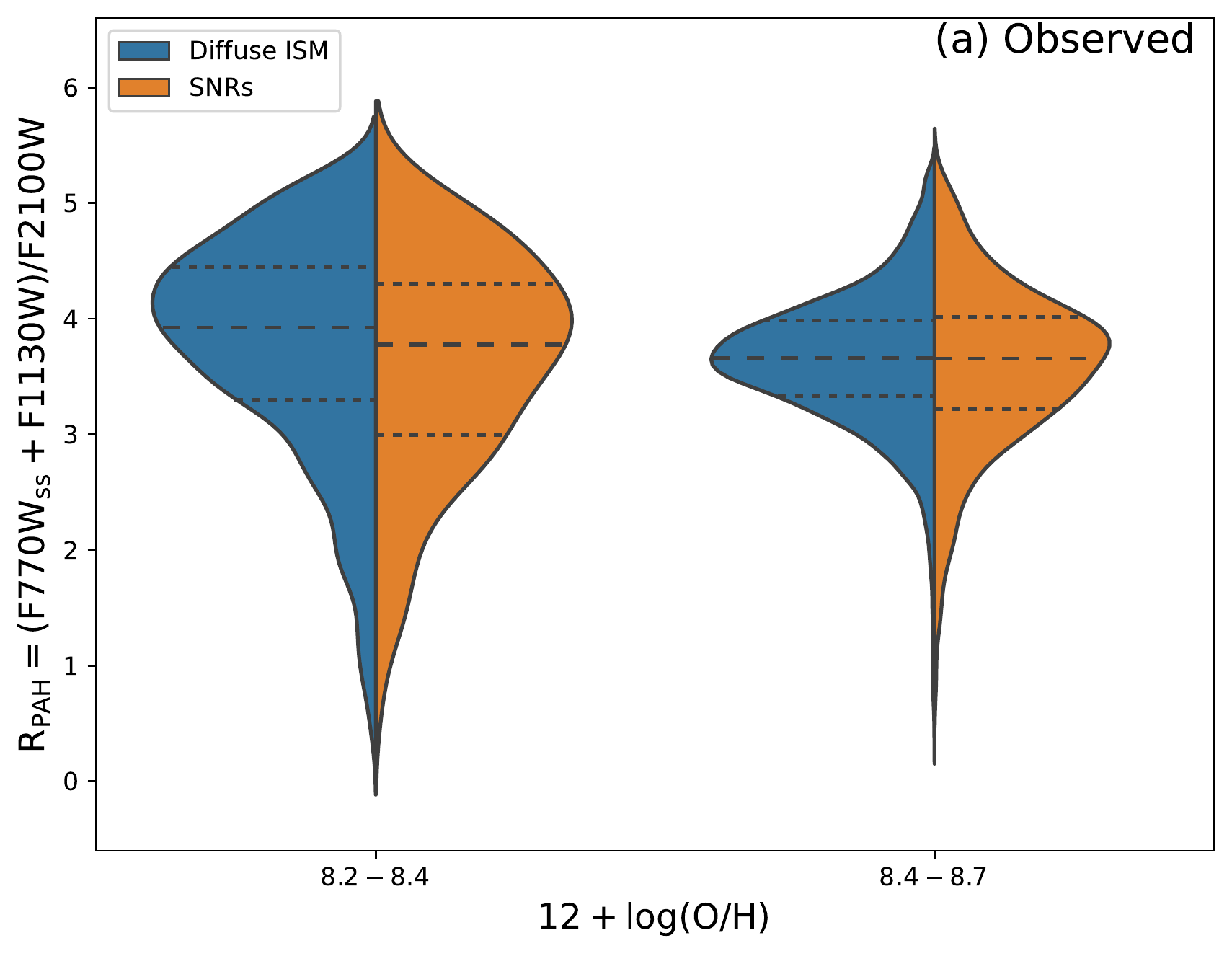}
    \includegraphics[width=0.46\linewidth]{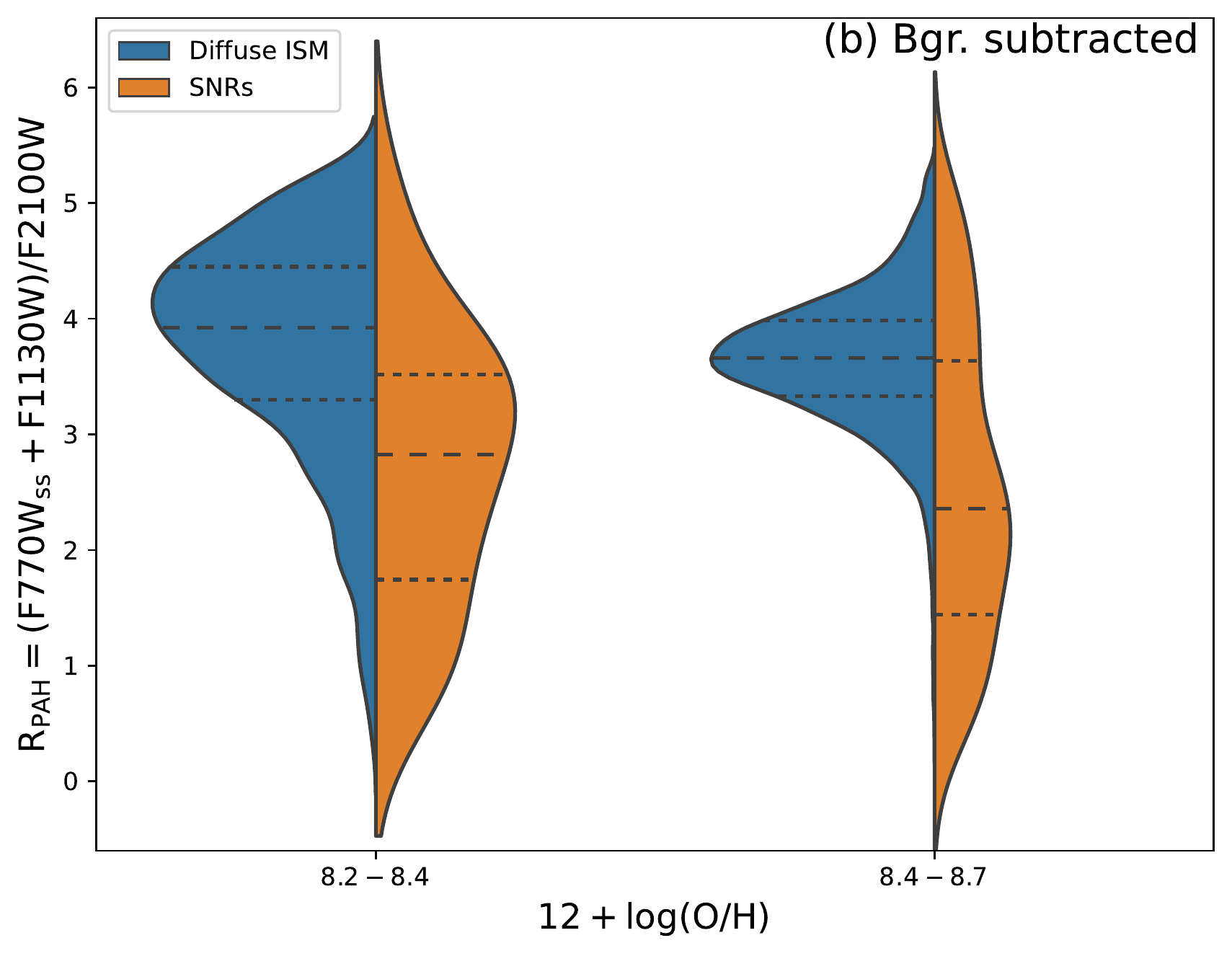}
    \caption{Same as in Fig.~\ref{fig:diff_vs_hii}, but for SNRs (orange) compared to the diffuse ISM (blue). Panels (a) and (b) correspond to \RPAH\ measurements from the observed and the local background subtracted fluxes. SNRs do not reveal a decrease in \RPAH\ measured from the observed fluxes indicating that SN shocks are unlikely major contributors to the selective PAH destruction at 50pc scale. Meanwhile, more localized effect is evident from the background-subtracted measurements.}
    \label{fig:diff_vs_snrs}
\end{figure*}

In Fig.~\ref{fig:RPAH_vs_Reff}a, we overlaid the distribution of \RPAH\ according to Eq.~(\ref{eq:RPAH_vs_hd}) for three adopted values of the PAH scale height: $h_d = 50, 100$ and 400~pc. Our toy model generally shows reasonable agreement with the observations assuming $h_d = 50$~pc, which is measured for molecular gas \citep[][]{Jeffreson2022}, or $h_d = 100$~pc, which is a fiducial scale height for the star-forming \revone{disks} \citep[][]{Schinnerer2024} in spiral galaxies. However, a single value of $h_d$ cannot reproduce the whole range of \RPAH\ measurements, and our toy model overestimates \RPAH\ for the largest regions. The latter can be in principle explained by the violation of the assumption of spherical symmetry for the large \HII\ regions that have a size comparable to the \revone{disk} scale height. The model significantly overestimates \RPAH\ if the PAHs are assumed to be distributed in a \revone{disk} with $h_d \sim 200-400$~pc, which is typically estimated for \HI\ in the central parts of galaxies \citep[e.g.][]{Yim2014, Bacchini2019, Patra2020}. Overall, observations show a steeper decrease of \RPAH\ with growing $r_{\rm circ}^{\rm HST}$ than is seen in the model. Furthermore, if $r_{\rm circ}^{\rm HST}$ still overestimates the sizes of the \HII\ regions, the observational data would shift further to the left of the model lines. Barnes et al. (submitted) showed that the sizes indeed appear to be smaller than $r_{\rm circ}^{\rm HST}$ by $\sim0.3$ dex when computed from the second spatial moments of the emission within the
source boundary. Therefore, we can conclude that within the assumption in our toy model, the effect of the line-of-sight projection can only partially explain \RPAH\ vs $r_{\rm circ}^{\rm HST}$ (and to some extent - vs \SIII/\SII) if PAHs are distributed in the thin star-forming \revone{disk} with scale height comparable to what is typically measured for molecular gas. Note, however, that the observations show that PAHs are also mixed with atomic gas in our Galaxy \citep[e.g.][]{Boulanger1985, Puget1985, Dwek1997}, which means that they are not necessarily locked in such a thin \revone{disk}.

To summarize this section, we conclude that most of the mechanisms considered here can contribute to the observed relation between \RPAH\ and the ionization parameter. Contamination by diffuse ISM along the line of sight can lead to a higher \RPAH\ at lower \SIII/\SII\ ratio due to the secondary dependence of this ratio on the size of the \HII\ region, although such projection effect alone is unlikely to completely explain the observed tight anti-correlation. The limited spatial resolution does not allow us to perfectly isolate the interiors of the \HII\ regions from the PDRs, and therefore this relation probably reflects a link between the processes in the PDRs and the properties of the ionized gas. These processes lead to a decrease in \RPAH\ in regions with a high ionization parameter (that is, a high density of ionizing photons), while the hardness of the ionizing radiation is not important in the moderate-to-high metallicity regime. Regardless of whether these processes occur in PDRs or in the interface between ionized and neutral gases in the \HII\ regions, the overall picture is consistent with the scenario in which PAHs are ionized and destroyed by hydrogen-ionizing photons. \revone{In the meantime, the particular physical mechanisms regulating PAH destruction remain uncertain, and the analysis of other spectral features sensitive to the physical properties of PAHs is thus required. In particular, upcoming 3.3 $\mu$m maps for PHANGS galaxies (Koizol et al., in preparation; \citealt{Sandstrom2023}), combined with 7.7$\mu$m and 11.3$\mu$m data presented here, will provide diagnostics of whether PAHs are destroyed primarily by ionization or by changes in their sizes \citep[e.g.,][]{Draine2021, Baron2024a, Dale2025}.}

\subsection{PAH fraction is insensitive to SNe shocks at 50~pc scale}
\label{sec:shocks}

Fig.~\ref{fig:RPAH_vs_ip_snrs}a shows the similar correlation between \RPAH\ and the ionization parameter as in Fig.~\ref{fig:RPAH_vs_ip}b with overlaid measurements for SNRs.

In previous sections, we focused mainly on the observational signatures of PAH destruction in the \HII\ regions caused by the influence of EUV radiation. Meanwhile, theoretical works and observations suggest that shocks could also destroy PAHs \citep[e.g.][]{OHalloran2006, Micelotta2010, Zhang2022}. In particular, more efficient PAH destruction by SNe shocks in low-metallicity galaxies was considered as one possible explanation of the metallicity dependence of the PAH fraction. PHANGS galaxies provide an excellent opportunity to test the role of supernova shocks in selective PAH destruction. In a recent paper, \citet{Li2024} presented a catalog of 964 isolated SNRs carefully selected with five different criteria (see details in Section~\ref{sec:analysis_hii_snrs}). Here we analyze how \RPAH\ averaged in 50~pc circular apertures around these isolated SNRs differs from what is derived in \HII\ regions and the diffuse ISM and search for indications of the more localized signatures of PAH selective destruction.

As seen from this plot, SNRs typically have a smaller range of \SIII/\SII, which is expected as they have different ionization mechanisms compared to \HII\ regions. Most SNRs are concentrated in the upper left part of the \RPAH\ vs \SIII/\SII\ relation defined by Eq.~(\ref{eq:RPAH_vs_ip}), although \RPAH\ for SNRs does not seem to depend strongly on \SIII/\SII\ in the same way as it does for \HII\ regions. However, assuming that Eq.~(\ref{eq:RPAH_vs_ip}) reflects the effects of photoionization and photodestruction of PAHs, not seeing such a clear dependence for SNRs is not surprising. 

In Fig.~\ref{fig:diff_vs_snrs}a, we compare \RPAH\ for SNRs and diffuse ISM for the two metallicity bins covered by the \citet{Li2024} sample in the same way as we did for \HII\ regions in Sec.~\ref{sec:metallicity}. If shocks are important for selective PAH destruction, and if their effect can be measured at 50~pc scales, we would expect to see lower \RPAH\ for SNRs compared to diffuse ISM. However, we do not see any significant differences in Fig.~\ref{fig:RPAH_vs_ip_snrs}a or Fig~\ref{fig:diff_vs_snrs}a. 

The SNRs are often faint in mid-IR emission and show limited contrast with the background. To better isolate the localized effects of SNRs, we repeat the same analysis by measuring \RPAH\ from the local background-subtracted fluxes, as described in Sec.~\ref{sec:analysis_rpah}. About 65\% of the original sample of SNRs reveal the signal in all three considered MIRI bands after the backround removal, thus ensuring that the measured residual signal from SNRs is not a result of random fluctuations. For comparison, similar background subtraction for diffuse ISM regions (which are nearly homogeneous by construction) resulted in 34\% positive residuals, which is close to what is expected from the noise distribution. The resulting values of \RPAH\ are scattered on the \RPAH\ vs \SIIISII\ plot (Fig.~\ref{fig:RPAH_vs_ip_snrs}b) demonstrating that the correlation defined by Eq.~(\ref{eq:RPAH_vs_ip}) is not applicable for SNRs. Meanwhile, we can recover the lower \RPAH\ at the SNRs locations compared to the original measurements for diffuse ISM in Fig.~\ref{fig:diff_vs_snrs}b. 

This test demonstrates that the effects of SNe shocks on the PAH fraction without a diffuse background correction are indistinguishable in unresolved studies, most likely because of the dominant contribution of the IR emission from the diffuse ISM in the aperture. % or because very small grains are also being destroyed by the shocks. 
Therefore, the signatures of the selective PAH destruction by shocks can be identified at the SNR locations, but is not dominant on 50~pc scales as high contamination by the diffuse ISM prevents their detection. This means that selective PAH destruction by SN shocks, suggested earlier to explain the decrease of the PAH fraction at low metallicities, is unlikely to affect a global PAH fraction on galaxy scales.

\section{Summary}
\label{sec:summary}
Based on the JWST MIRI and NIRCam imaging and MUSE integral field spectroscopic observations of 42 nearby galaxies, we investigate how the dust mass fraction of PAHs traced by mid-IR emission (\RPAH) is linked to/regulated by the properties of the ionized gas in star-forming regions. The galaxies in our sample are from the PHANGS sample and cover stellar masses from $1.5\times10^9$ to $1.2\times 10^{11} M_\odot$, with individual \HII\ regions having oxygen abundances mostly from $\rm 12+\log(O/H)$ from 8.0 to 8.8 (i.e. $0.2-1.3Z_\odot$) and located at distances from 5 to 23 Mpc (i.e. allowing spatial resolution of $<100$~pc with MUSE and $<80$~pc with JWST at 21$\mu$m - the longest wavelength considered here). Using the spatial masks (derived from the MUSE \Ha\ images), we isolated the \HII\ regions from the diffuse ISM and measured the \RPAH\ ratio, which traces PAH fraction, for every \HII\ region and a control sample of diffuse ISM regions. For 19 more massive (and metal-rich) galaxies, we also measured \RPAH\ in the circular 50~pc apertures surrounding the sample of SNRs isolated from \HII\ regions. Comparing the measured values of \RPAH\ with the properties of the ionized gas, we obtained the following main results:

\begin{itemize}
    \item The PAH fraction drops at metallicities $\rm 12+\log(O/H) < 8.2$ in both \HII\ regions and diffuse ISM. Above this limit, it is relatively insensitive to metallicity. PAHs could be destroyed more efficiently in the low-metallicity \HII\ regions due to the higher UV hardness. At the same time, the same steep decrease for the diffuse ISM indicates a possible less efficient formation of PAHs in a low-metallicity environment.
    \item The PAH fraction for \HII\ regions is significantly lower than for diffuse ISM at all metallicities considered here. We argue that PAHs are being destroyed in \HII\ regions by intense hydrogen-ionizing radiation. 
    \item We find strong %and universal 
    anticorrelation between \RPAH\ and ionization parameter (traced by the $\rm \log([SIII]9069/[SII]6717+6731)$ emission line ratio)  across our large sample of relatively high metallicity \HII\ regions with $\rm 12+\log(O/H) > 8.2$. %This result confirms previous findings by \citep{Egorov2023} for much smaller sample of the regions from 4 galaxies. %We do not see any significant secondary correlation with metallicity for that range (contrary to previous finding in \citealt{Egorov2023}) , UV hardness, age or other properties of star-forming regions and ionizing clusters considered here.
    \item We argue that the relation between the PAH fraction and the ionization parameter is driven by a physical connection between the properties of the ionized gas and PAHs at the edges of \HII\ regions and PDRs. Assuming that PAHs are distributed in the thin \revone{disk}, effects of line-of-sight projection also contribute significantly to the observed relation but cannot completely explain it.

    \item   We do not see differences in PAH fraction between diffuse ISM and isolated SNRs measuring it from the observed JWST/MIRI band ratios. Only applying careful local background subtraction do we recover evidence for PAH destruction at SNR sites. This indicates that supernova shocks do not contribute significantly to the selective PAH destruction when averaged over 50~pc scales. In particular, the previously suggested explanation of the metallicity dependence of the PAH fraction resulted from higher SN activity in low-metallicity galaxies seems to be unlikely. %On the other hand, signatures of selective PAH destruction are identified at the locations of the SNR peaks.

\end{itemize}

Overall, our results agree with the scenario of PAH ionization and destruction by hydrogen-ionizing photons in star-forming regions, and of less efficient formation of PAHs in the low metallicity environment. The particular physical mechanisms, spatial scales and timescales for PAH destruction remain unclear. Future systematic resolved studies of extragalactic \HII\ regions and their surroundings now possible with JWST by imaging of the ionized gas (via Pa$\alpha$ emission line) and the small PAHs glowing at 3.3$\mu$m, or by resolved spectroscopy with NIRSpec and MIRI-MRS will certanly provide a huge next step in our understanding of the PAH evolution cycle.

\begin{acknowledgements}
This work is based on observations made with the NASA/ESA/CSA JWST. The data were obtained from the Mikulski Archive for Space Telescopes at the Space Telescope Science Institute, which is operated by the Association of Universities for Research in Astronomy, Inc., under NASA contract NAS5-03127. The observations are associated with JWST programs 2107 and 3707. The specific observations analyzed can be accessed via \url{https://doi.org/10.17909/ew88-jt15}.
Based on observations collected at the European Southern Observatory under ESO programmes 0111.C-2109 (PI: Egorov), 0108.B-0249 (PI: Kreckel), 094.C-0623 (PI: Kreckel), 095.C-0473,  098.C-0484 (PI: Blanc), 1100.B-0651 (PHANGS-MUSE; PI: Schinnerer), as well as 094.B-0321 (MAGNUM; PI: Marconi), 096.B-0309, 098.B-0551, 099.B-0242, 0100.B-0116 (MAD; PI: Carollo), 097.B-0640 (TIMER; PI: Gadotti), 0104.B-0404, 109.22VU.001 (PI: Erwin), 094.B-0298 (PI: Walcher), 60.A-9487 (PI: Gadotti), 109.2332.001 (PI: Belfiore), 0102.B-0617 (PI: Fluetsch), 111.24KE.001 (PI: Barnes), 097.D-0408 (PI: Anderson). 

OE acknowledges funding from the Deutsche Forschungsgemeinschaft (DFG, German Research Foundation) -- project-ID 541068876.
MB gratefully acknowledges support from the ANID BASAL project FB210003. This work was supported by the French government through the France 2030 investment plan managed by the National Research Agency (ANR), as part of the Initiative of Excellence of Université Côte d’Azur under reference number ANR-15-IDEX-01.
KK, MCS, EE, MH, and JL acknowledge support from the Deutsche Forschungsgemeinschaft (DFG, German Research Foundation) in the form of an Emmy Noether Research Group (grant number KR4598/2-1, PI Kreckel) and the European Research Council’s starting grant ERC StG-101077573 (“ISM-METALS"). 
HK, KS acknowledge funding from JWST-GO-2107.006-A. HAP acknowledges support from the National Science and Technology Council of Taiwan under grant 113-2112-M-032-014-MY3.
JC acknowledges funding from the Belgian Science Policy Office (BELSPO) through the PRODEX project ``JWST/MIRI Science exploitation'' (C4000142239).
\end{acknowledgements}


\bibliographystyle{aa}
\bibliography{PHANGS_PAHs}

\appendix

\section{Testing effects of background and stellar continuum subtraction}
\label{sec:app:fluxes}

\begin{figure}[!htbp]
    \includegraphics[width=0.49\linewidth]{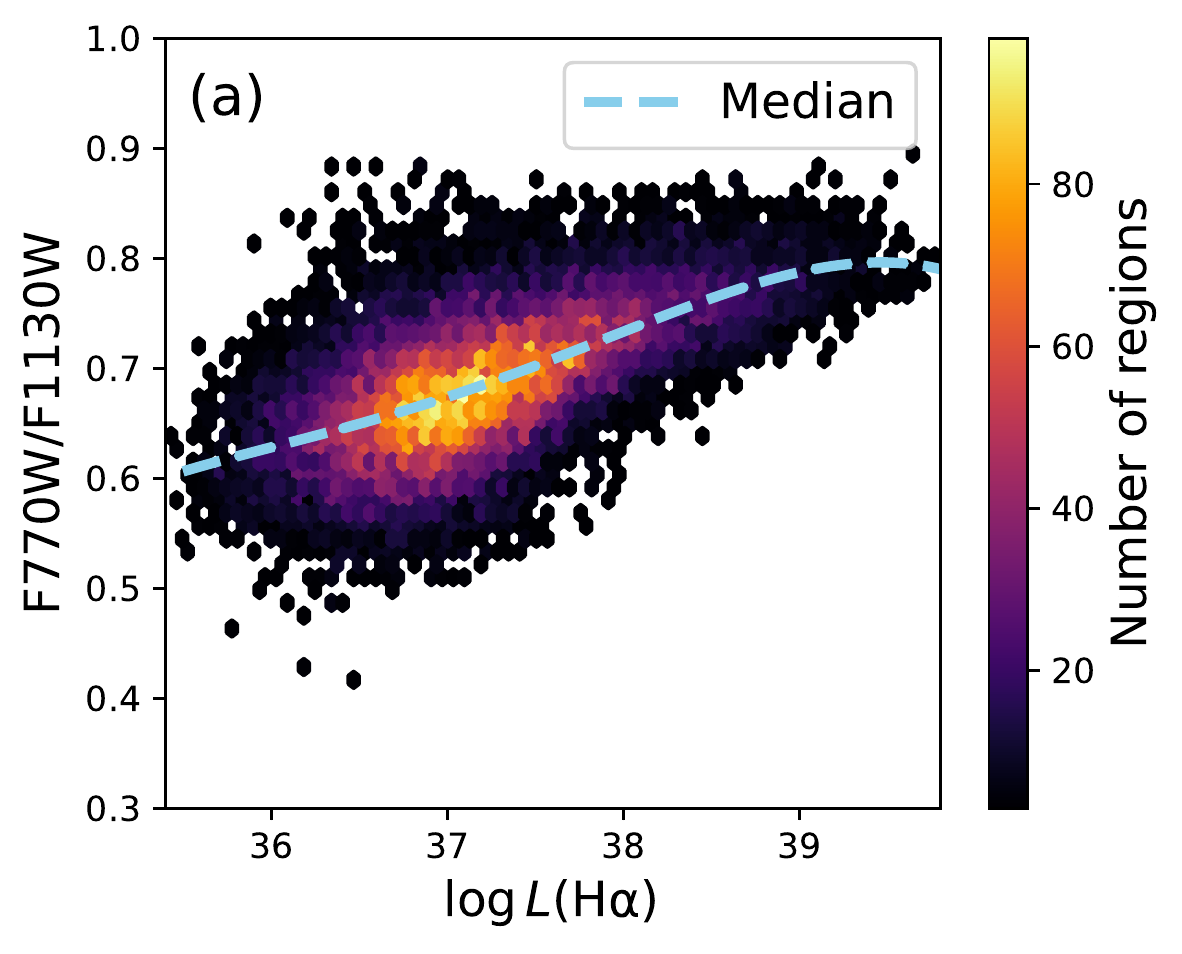}~
    \includegraphics[width=0.49\linewidth]{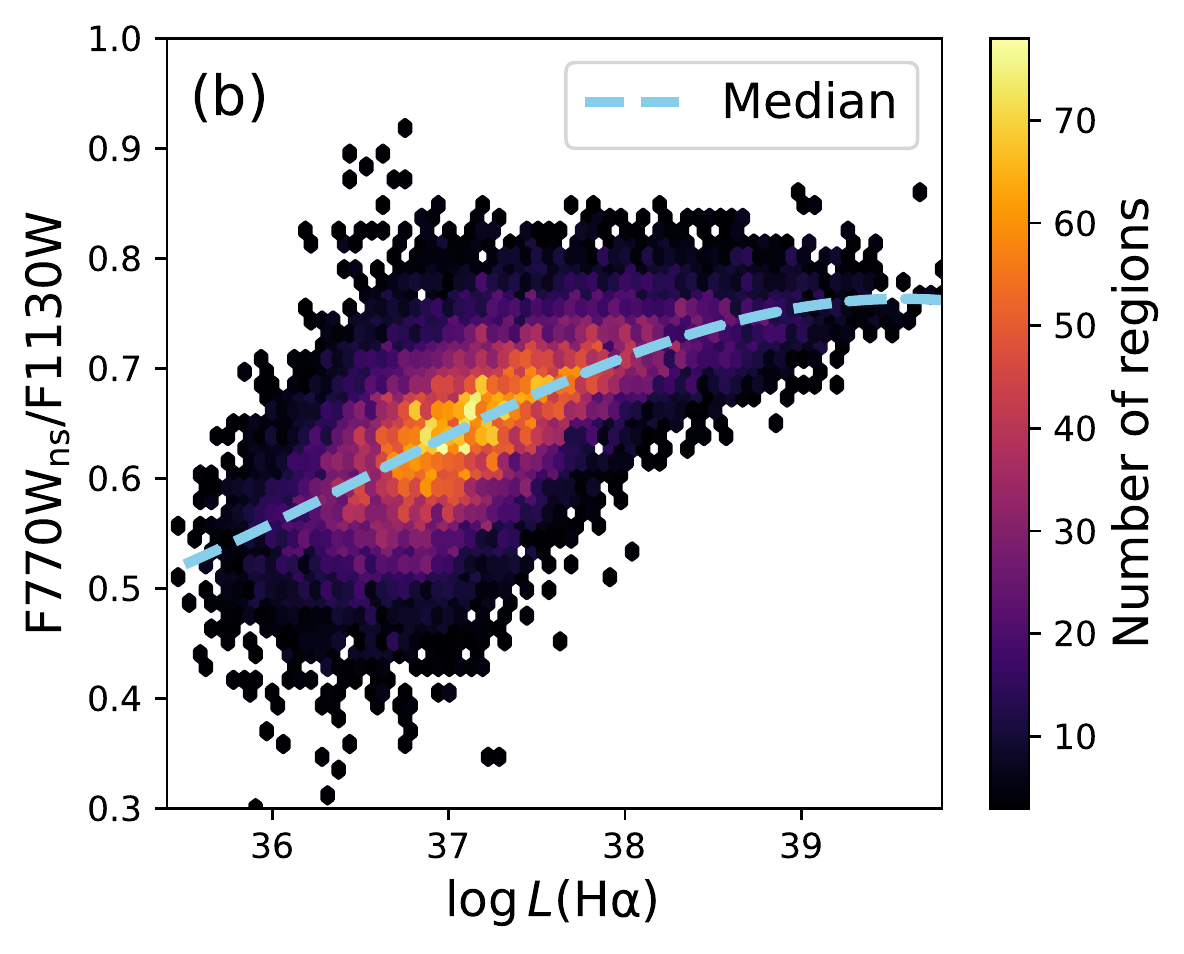}

    \includegraphics[width=0.49\linewidth]{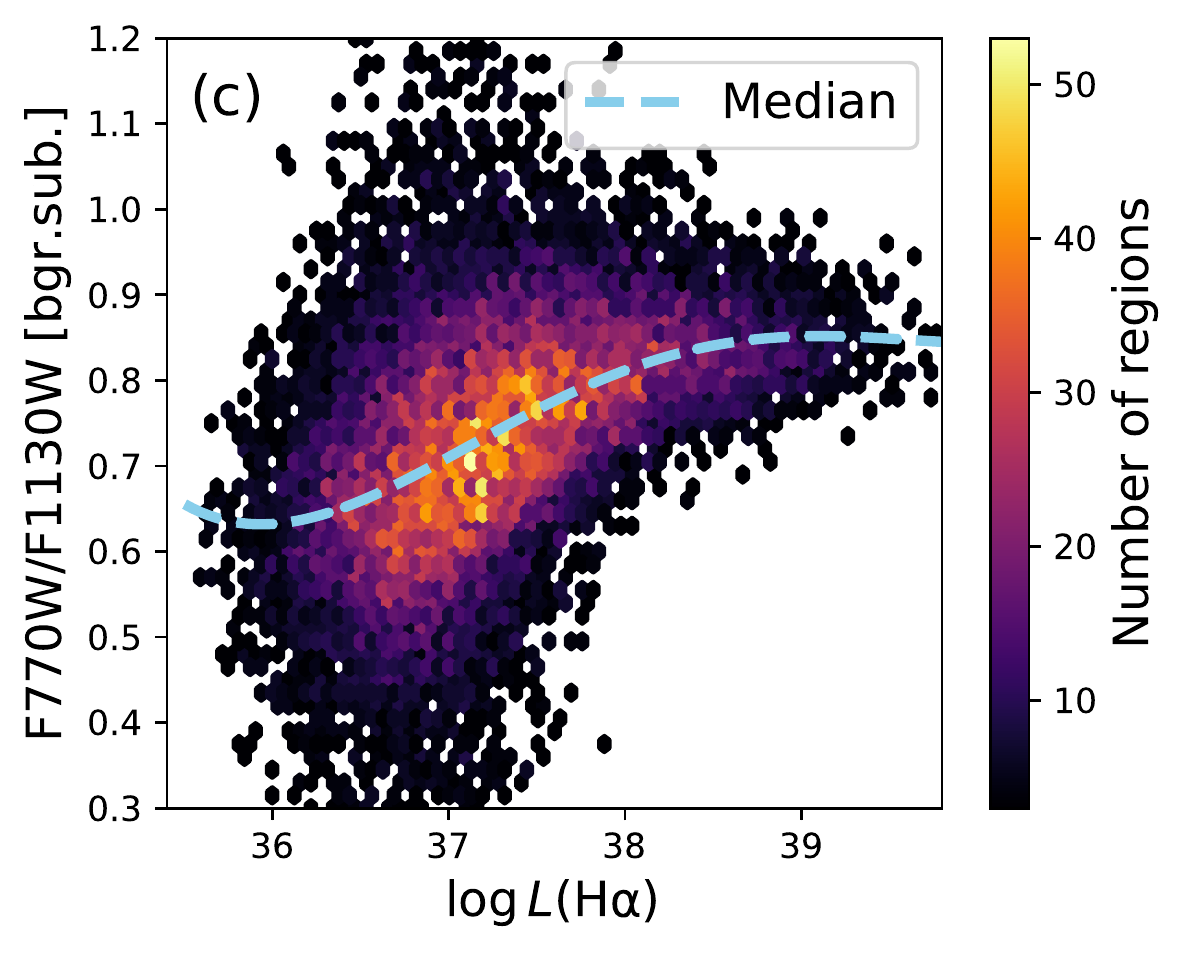}~
    \includegraphics[width=0.49\linewidth]{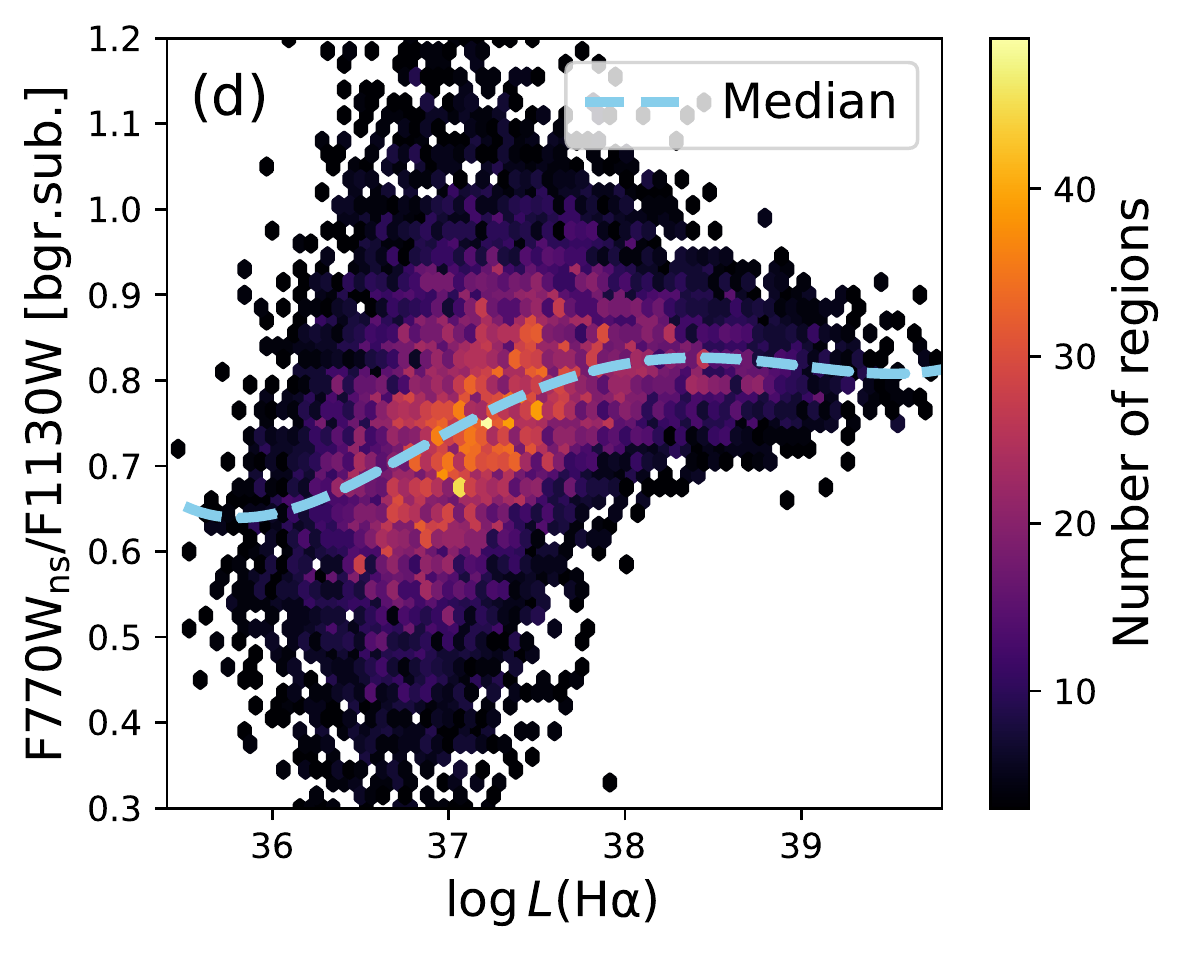}
    \caption{Band ratio $F770W$/$F2100W$ grows with the \Ha\ luminosity of \HII\ regions. This effect is seen for both observed (panel a) and stellar continuum subtracted (panel b) data in F770W. Panels c and d show the same line ratios after local background subtraction. Color denotes the number density of the regions, and the cyan curve represent a smoothed running median across the \Ha\ luminosities.}
    \label{fig:F770_F1130_vs_Ha}
\end{figure}

The present study relies on measurements of the brightness in mid-IR photometric bands that contain not only emission from PAHs or very small grains but also emission from the considered regions but also contaminated by stellar continuum and background emission from diffuse ISM. The effects of subtraction of stellar populations were analyzed in detail by \citet{Sutter2024, Baron2025}, and discussed in Sec.~\ref{sec:analysis_rpah}. Here we additionally demonstrate that it does not significantly affect $F770W/F1130W$ except causing some non-linear dependence on $\log L({\rm H\alpha})$ for the faintest \HII\ regions ($\log L({\rm H\alpha}) < 37$; see Fig.~\ref{fig:F770_F1130_vs_Ha}a,b). In extreme cases, this can give up to 0.05~dex difference \RPAH\ and thus do not affect our results.

\begin{figure}[!htbp]
\includegraphics[width=\linewidth]{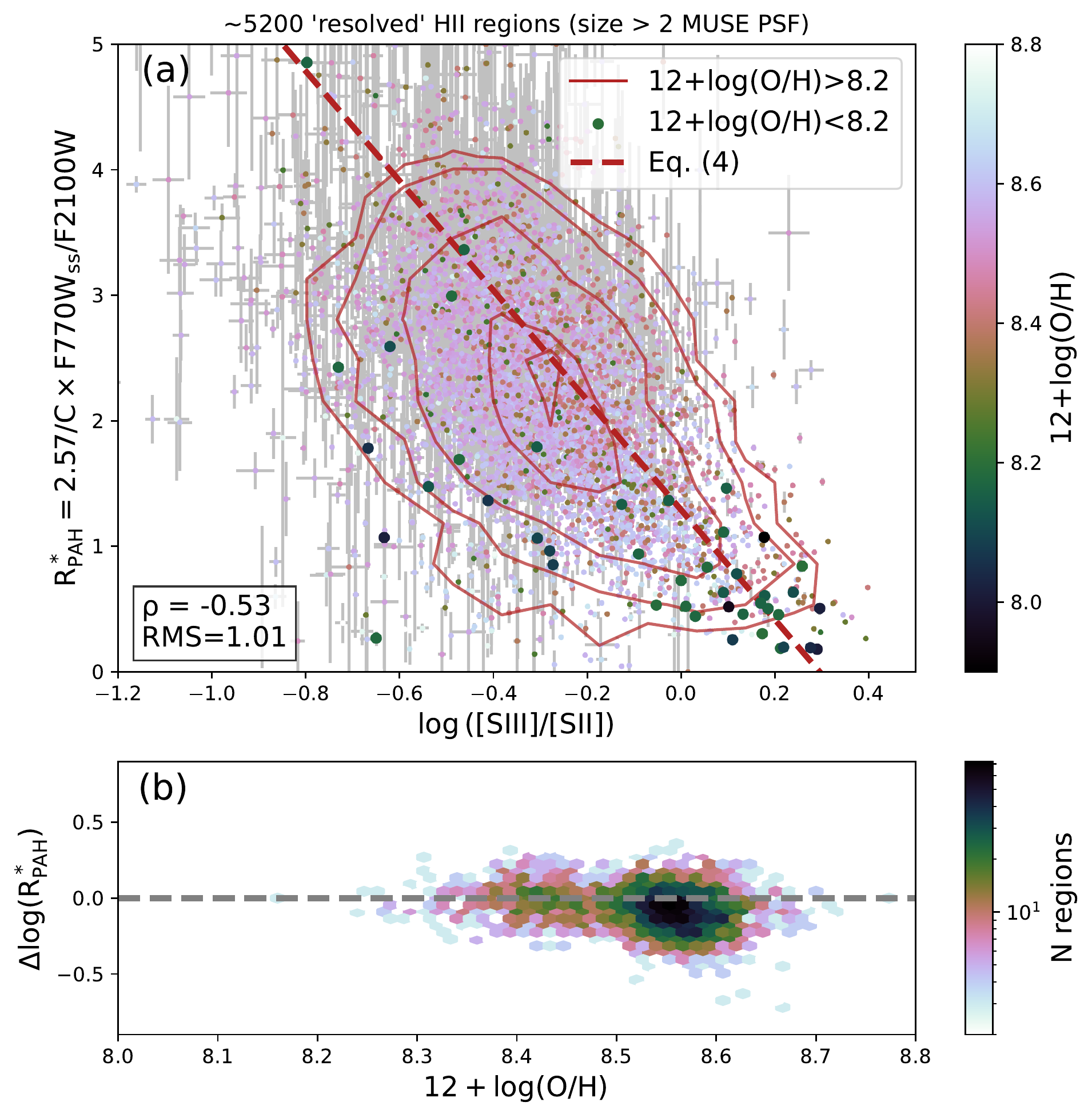}
\caption{Same as Fig.~\ref{fig:RPAH_vs_ip}c,d but for \RPAHst\ derived from the local background subtracted measurements. The background subtraction does not affect the conclusions made for \HII\ regions. }
\label{fig:RPAH_vs_ip_bgrsub}
\end{figure}

The brightness in all mid-IR bands considered here shows a tight semilinear correlation with the extinction-corrected \Ha\ brightness \citep{Leroy2023}, which we also see in the integrated measurements for the \HII\
regions. We find that subtraction of the local background does not affect this linear correlation but only leads to a systematic offset and a slight change in the slope of that correlation. Meanwhile, this removing background produces noticeable changes in $F770W/F1130W$ and $F770W_{ss}/F1130W$ ratios: their dependence on $\log({\rm H\alpha)}$ has large scatter and becomes non-linear (Fig.~\ref{fig:F770_F1130_vs_Ha}c,d), in contrast with those based on original flux measurements (panels a,b). In turn, this can result in stronger disagreement between the measurements \RPAH\ and \RPAHst, and therefore we provide a separate calibration for the background-subtracted version of \RPAHst{} (see Appendix~\ref{sec:app:calibration}). Meanwhile, background subtraction does not significantly affect the measurements of \RPAH\ itself (and thus recalibrated version of \RPAHst). Thus, Fig.~\ref{fig:RPAH_vs_ip_bgrsub} demonstrate that \RPAHst\ after background subtraction still shows the same strong correlation with $\rm \log ([SIII]/[SII])$ as in Fig.~\ref{fig:RPAH_vs_ip}, with a slightly larger scatter and small systematic offset ($<0.1$~dex) relative to the linear trend defined by Eq.~(\ref{eq:RPAH_vs_ip}). Therefore, mostly to prevent an increase in the scatter, we use original measurements without background subtraction. Note, however, that the local background subtraction is crucial for interpretation of the comapact sources, such as SNRs (see Sec.~\ref{sec:shocks}).

\section{Calibration of ${\rm R_{PAH}^*}$ prescription}
\label{sec:app:calibration}

\begin{figure}[!htbp]
    \includegraphics[width=\linewidth]{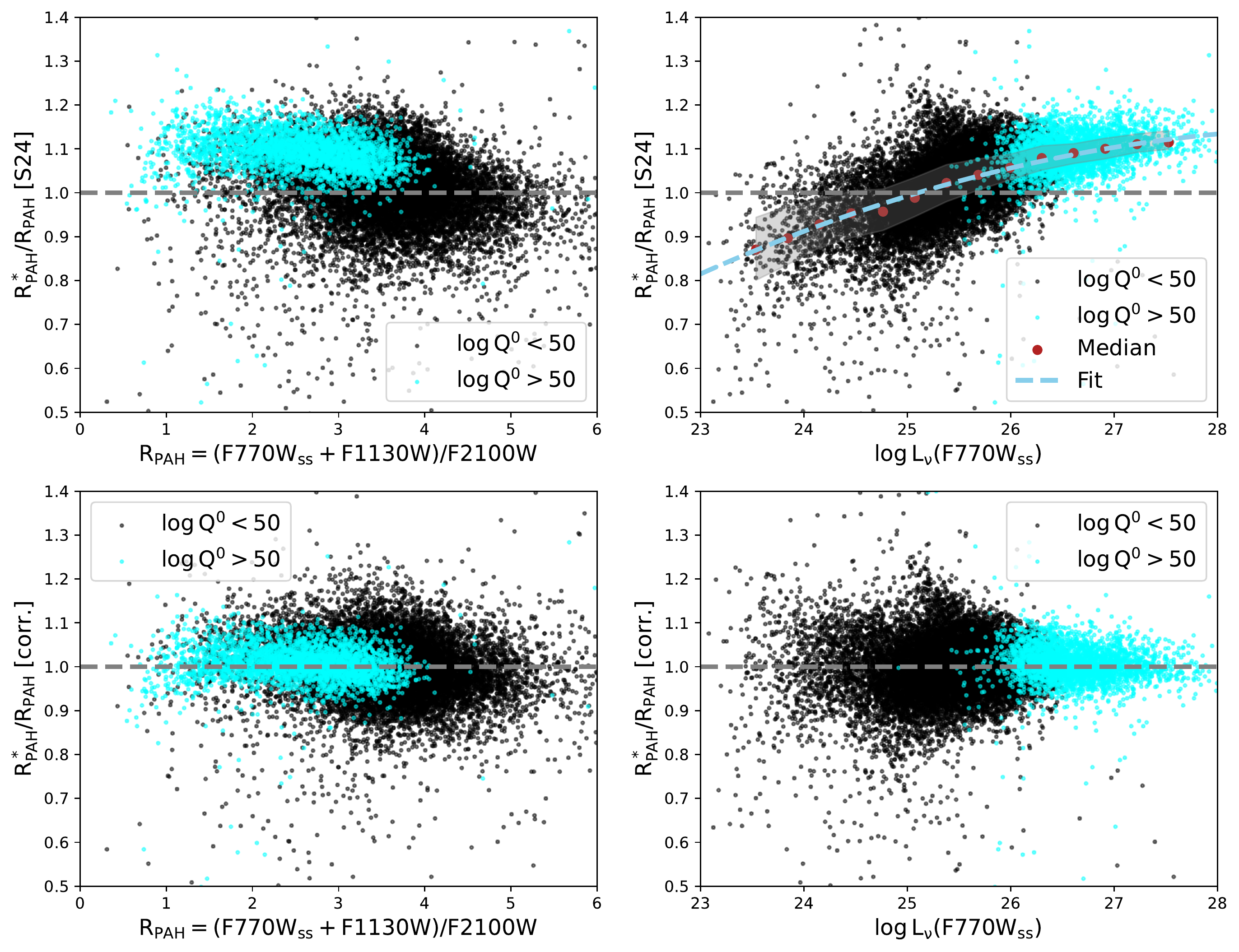}
    \caption{The ratio of two tracers of PAH mass fraction (\RPAH\ and \RPAHst) for $\sim 12000$ \HII\ regions from 19 PHANGS galaxies (top row) depending on \RPAH\ (left panel) and monochromatic luminosity $\rm L_{\nu}(F770W_{\rm ss})$ (right panel). The cyan and black colors denote regions with hydrogen ionizing photons rate $Q^0 > 10^{50}$~s$^{-1}$ and $Q^0 < 10^{50}$~s$^{-1}$, respectively. \RPAHst\ and \RPAH\ are consistent with each other for most of the regions, but \RPAHst\ systematically overestimates PAH fraction for the luminous \HII\ regions. Bottom row demonstrate the same ratio with \RPAHst\ calculated after applying correction described in the text.}
    \label{fig:RPAH_vs_RPAH*}
\end{figure}

\begin{figure}[!htbp]
    \includegraphics[width=\linewidth]{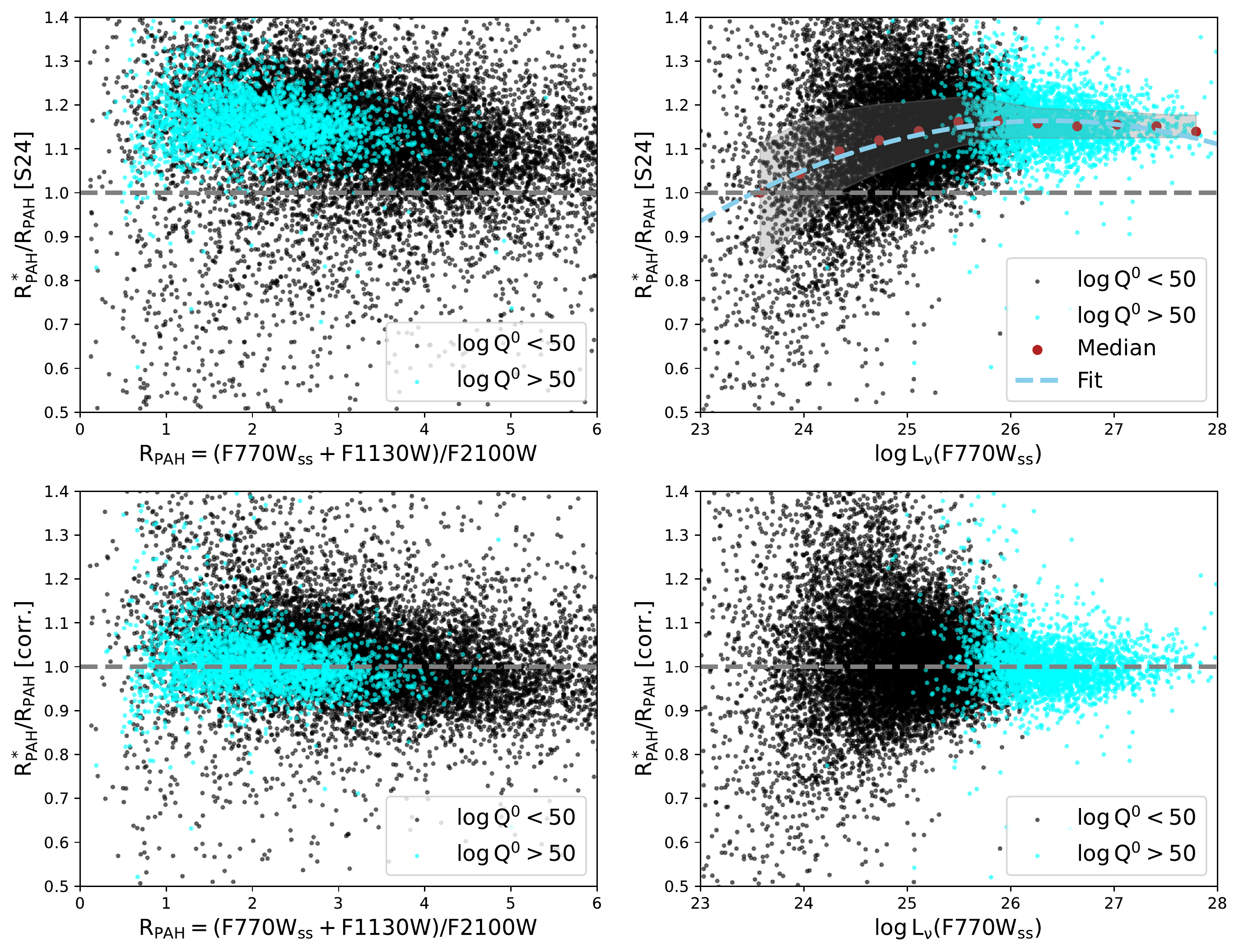}
    \caption{Same as Fig.~\ref{fig:RPAH_vs_RPAH*} but  after removing the local background from the flux measurements in all involved JWST bands.}
    \label{fig:RPAH_vs_RPAH*_bgrsub}
\end{figure}

Since $F1130W$-band images necessary for calculating \RPAH\ are not available for most galaxies in our sample, we rely on \RPAHst$\sim F770W_{\rm ss}/F2100W$ as a tracer of the PAH fraction. \citet{Sutter2024} demonstrated that \RPAHst$= 2.57 \times F770W_{\rm ss}/F2100W$ provide a very good agreement with \RPAH\ measurements when applied to diffuse ISM. Fig.~\ref{fig:F770_F1130_vs_Ha} shows that $F770W/F1130W$ is not constant in the range of \Ha\ luminosities considered here. Thus, here we refine \RPAHst\ prescription from \citet{Sutter2024} against \RPAH\ measurements to make it applicable for nebulae of different luminosities. 

The top panels of Fig.~\ref{fig:RPAH_vs_RPAH*} shows how the ratio between original \RPAHst\ (from \citealt{Sutter2024} approximation) and \RPAH\ changes with \RPAH\ and $F770W$ derived for our sample of \HII\ regions from 19 PHANGS galaxies. In agreement with \citet{Sutter2024} findings for diffuse ISM, \RPAHst\ yields very consistent values with \RPAH\ when measured in relatively faint \HII\ regions, and the typical scatter around the 1-to-1 relation is within 0.05 dex ($<12$\%). However, \RPAHst\ is systematically higher (by $\sim 7-17$\%) than \RPAH\ in bright star-forming regions \HII\ regions. For example, cyan \revone{color} in Fig.~\ref{fig:RPAH_vs_RPAH*} shows the regions more luminous than $L({\rm H\alpha}) \sim 1-3\times10^{38}\ {\rm erg\ s^{-1}}$, which roughly corresponds to $Q^0 \sim 10^{50}$~s$^{-1}$, or an \HII\ region ionized by 4-5 OV5 stars assuming no escape of ionizing photons. %This value is the same as the threshold that separates two different branches of linear correlation between \RPAH\ and the ionization fraction of the region, as was found by \citet{Egorov2023}. 
Most probably, this observed trend reflects the different band $F1130W/F770W$ ratio (depending on the PAHs ionization or the radiation field spectrum) toward the bright and faint \HII\ regions. Fig.~\ref{fig:F770_F1130_vs_Ha} (see also \citealt{Egorov2023}) shows that low $F770W/F1130W$ tends to be observed preferentially in faint \HII\ regions. \citet{Baron2025} revealed strong correlation between $F1130W/F770W$ and \SIIHa, which also implies that higher values of $F770W/F1130W$ are expected from the diffuse ISM (in general showing higher \SIIHa\ compared to the \HII\ regions, \citealt{Belfiore2022}). Finally, \citet{Baron2025} found that some regions with anomalously low $F770W/F1130W$ ratios also have elevated stellar-to-PAH (i.e. $F200W/F770W$) ratios different from adopted by \citet{Sutter2024} for calculating $F770W_{\rm ss}$. 

The top right panel of Fig.~\ref{fig:RPAH_vs_RPAH*} demonstrates that \RPAHst/\RPAH\ ratio correlates with a logarithm of monochromatic luminosity $\log{\rm L_\nu (F770W_{ss})}$ and can be well approximated by the second-order polynomial: 

\begin{equation}
\label{eq:app_rpah_star}
    {\rm R^*_{PAH}/R_{PAH}} = a_2\log(\rm L_{\nu}(F770W_{\rm ss}))^2 +a_1\log(\rm L_{\nu}(F770W_{\rm ss})) + a_0,
\end{equation}
\revone{with the best-fit result obtained with
$a_2=-0.00832\pm0.00081$, $a_1=0.4881\pm0.0403$, and $a_0=-6.01\pm0.52$.} %$a_2=-0.01697\pm0.00197$, $a_1=0.9063\pm0.0981$, and $a_0=-10.98\pm1.22$.
% $a_2=-0.021\pm0.002$, $a_1=0.208\pm0.013$, and $a_0=0.61\pm0.02$. 
The uncertainties are obtained as the standard deviation of the measured coefficients across 500 randomly bootstrapped samples, each comprising \revone{2/3} of the original dataset. Dividing the \RPAHst\ measurements by the $\rm L_{\nu}(F770W_{\rm ss})$-dependent multiplicative factor defined by this equation, we put them into agreement with \RPAH\ across the entire range of {\RPAH} and \rm $L_{\nu}(F770W_{\rm ss})$ values and \Ha\ luminosities (see bottom panels of Fig.~\ref{fig:RPAH_vs_RPAH*}. Therefore, in this study we use the \RPAHst\ refined in such a way (and thus derived with Eqs.~(\ref{eq:rpah_star}, \ref{eq:rpah_corr}).

Removing the local background changes the $F770W_{\rm ss}$ (and thus $\rm L_{\nu}(F770W_{\rm ss})$) values and thus such a non-linear correction can be invalid when applied to the background-subtracted measurements. Indeed, Fig.~\ref{fig:RPAH_vs_RPAH*_bgrsub} shows the same distributions as in Fig.~\ref{fig:RPAH_vs_RPAH*}, but for the measurements made after removing the local background. Unlike previously discussed, \RPAHst\ prescription from \citet{Sutter2024} and \RPAH\ measurements have a nearly constant multiplicative offset by a factor of $\sim 1.15$ for the entire sample of \HII\ regions. For consistency, we approximate it with the same second-order polynomial (Eq.~\ref{eq:app_rpah_star}), but with different \revone{best-fit coefficients: $a_2=-0.02009\pm0.00141$, $a_1=1.0597\pm0.0736$, and $a_0=-12.81\pm0.96$.} 
%$a_2=-0.01954\pm0.00335$, $a_1=0.9905\pm0.1610$, and $a_0=-11.39\pm1.97$
%$a_2=-0.011\pm0.006$, $a_1=0.090\pm0.035$, and $a_0=0.98\pm0.04$. %The recovered trend has a noticeable linear gradient with $F770W_{\rm ss}$, while the second-order term is rather negligible. 
All \RPAHst\ values relying on the background-subtracted measurements are calculated using these coefficients (namely, with Eqs.~(\ref{eq:rpah_star}, \ref{eq:rpah_corr_bgrsub}).

\section{Linking observational properties of \HII\ regions with Cloudy models}
\label{sec:app:cloudy}

\begin{figure*}
    \centering
    \includegraphics[width=0.32\linewidth]{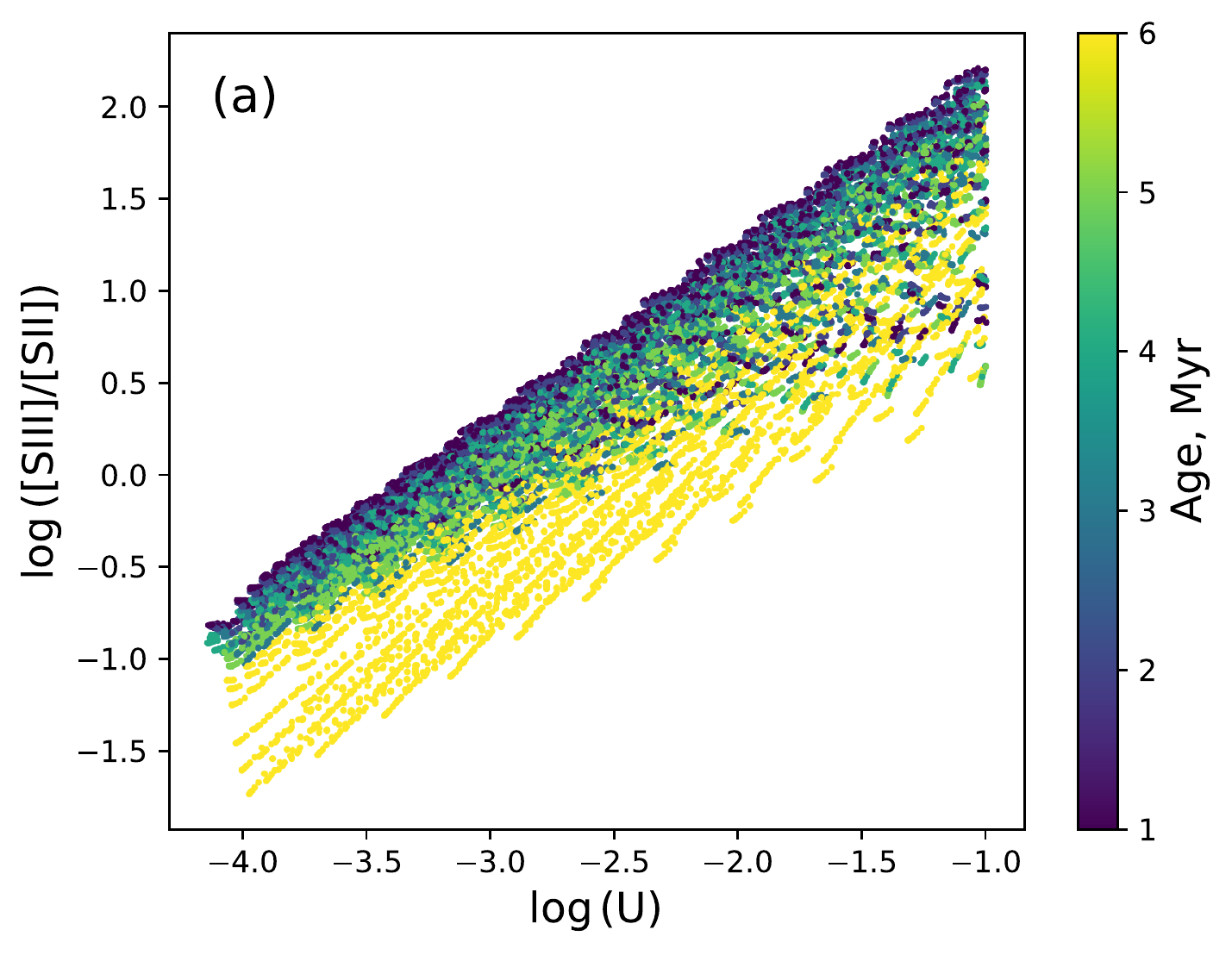}
    \includegraphics[width=0.32\linewidth]{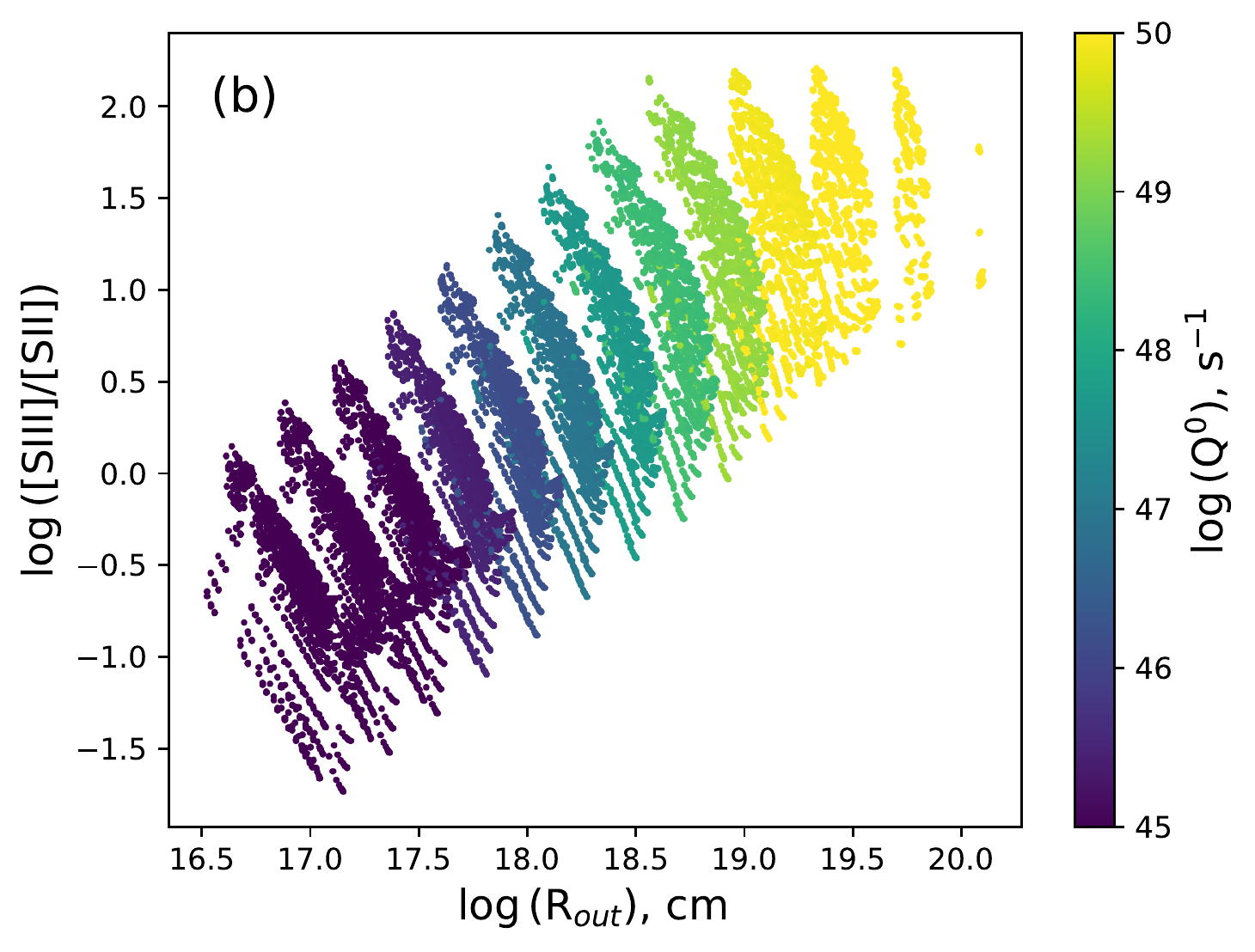}
    \includegraphics[width=0.32\linewidth]{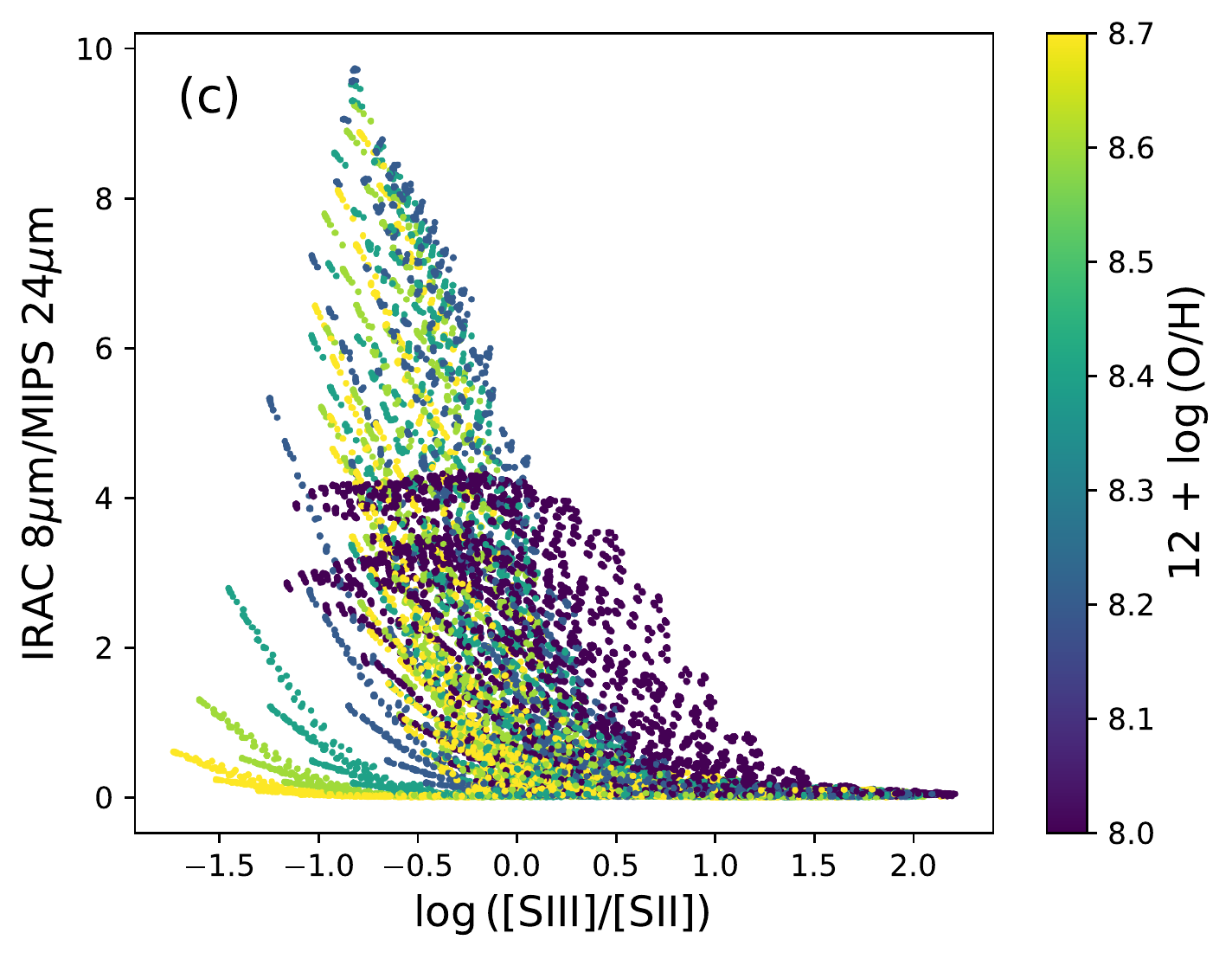}
    \caption{Dependence between different observational and physical parameters derived from the BOND models \citep{ValeAsari2016} produced with the Cloudy photoionization code \citep{Ferland2017}. Panel a: $\rm\log([SIII]/[SII])$ vs the logarithm of the ionization parameter, $\rm \log(U)$, color-coded with the age of the ionizing star cluster. $\rm\log([SIII]/[SII])$ strongly correlates with the ionization parameter, with the secondary dependence on the age. Panel b: $\rm\log([SIII]/[SII])$ vs logarithm of the outer size of the \HII\ region, $\rm \log(R_{out})$, color-coded by the logarithm of the hydrogen ionizing photons rate, $\rm \log(Q^0)$. At the fixed value of the ionizing flux, the observed $\rm\log([SIII]/[SII])$ depends on the size of the region. Panel c: Dependence of $F_{8\mu m}/F_{24\mu m}$, similar to \RPAHst, on the $\rm\log([SIII]/[SII])$. The existing Cloudy models can only partially reproduce the trend seen in Fig.~\ref{fig:RPAH_vs_ip}.}
    \label{fig:cloudy}
\end{figure*}

Similarly to \cite{Egorov2023}, this study relies on the \SIII/\SII\ emission lines ratio as a tracer of the ionization parameter \citep[see, e.g.][]{Diaz1991, Kewley2019}. Based on photoionization models, here we demonstrate this dependence, as well as a secondary relation of this ratio on other physical properties considered in this study.

For simplicity, we considered existing \revone{``BOND''} models \citep{ValeAsari2016} accessible via the 3MdB database \citep{Morisset2015}. The photoionization model grids were produced using the Cloudy v17.01 code \citep{Ferland2017} by varying the parameters of the cloud (its size and geometry, oxygen and nitrogen abundances) and ionizing source (age of the cluster and input ionization parameter) as described in \citet{ValeAsari2016}. We only consider models with oxygen abundance $\rm 12+\log(O/H)=7.6 \dots 9.0$, ionization parameter $\log(U) = -5 \dots -1$, and hydrogen density $\log(n_{\rm H}/cm^{-3}) = 0\dots2.5$ spanning the properties of the \HII\ regions in this study. 

Fig.~\ref{fig:cloudy}a shows that $\rm \log([SIII]/[SII])$ indeed linearly depends on the logarithm of the ionization parameter, although it also exhibits a clear secondary dependence on the age of the ionizing star cluster: the \SIII/\SII\ ratio is typically lower for older regions of the same ionization parameter. As discussed in Sec.~\ref{sec:HIIregs} and in \citet{Egorov2023}, our interpretation of the observational dependence of the PAH fraction on the ionization parameter is unlikely to be biased by this secondary dependence on the age as we do not observe a correlation with other age-sensitive tracers like EW(\Ha). 

By definition, $\log U$ linearly depends on the logarithm of the hydrogen ionizing photons rate, $\rm \log(Q^0)$, and thus $\rm \log([SIII]/[SII])$ must do. This is indeed seen in Fig.~\ref{fig:cloudy}b. This plot also demonstrates that at constant ionizing photon rate (and thus mass and age of the ionizing cluster), larger \HII\ regions exhibit lower \SIII/\SII\ line ratio. This secondary dependence is a reason why the line-of-sight projection effect can be partially responsible for the observed correlation between \RPAH\ and $\rm \log([SIII]/[SII])$ (see discussion in Sec.~\ref{sec:HIIregs}). 

Cloudy code does not calculate all the chemical reaction taking place in the PDRs and depends on the input from the dedicated sub-grid models. By default, it considers simplified, but reasonable process of PAH destruction in \HII\ regions: PAH fraction $q_{\rm PAH}$ linearly decreases with the fraction of the ionized hydrogen, $\rm H^+/(H^+ + H^0)$, at the particular radius. In turn, the ionization fraction directly depends on the number of the hydrogen ionizing photons available to ionize the ISM. PAH fraction in PDRs unaffected by the ionizing radiation is defined by the net chemical reactions depending on the dust content and size distribution, and thus related to the processes of the PAH formation rather then of their destruction. In principle, such a simplified picture is in qualitative agreement with our proposed interpretation of the observational results. Therefore, we expect to see in integrated Cloudy models a dependence between \RPAH\ and \SIII/\SII\ similar (although, not necessary identical) to what is presented in Figs.~\ref{fig:RPAH_vs_ip}. Fig.~\ref{fig:cloudy}c such dependence, where $F_{8\mu m}/F_{24\mu m}$ (corresponding to the modeled brightness at Spitzer IRAC4/MIPS24 bands) is given as \RPAH\ tracer. This tracer must produce qualitatively similar results to those obtained with \RPAHst. 

In Fig.~\ref{fig:cloudy}c, we indeed observe a similar decrease of IRAC4/MIPS24 with \SIII/\SII\ to what is seen for the observed \HII\ regions. However, there are several noticeable differences between these simple Cloudy models and the observations. The models exhibit much larger scatter of IRAC4/MIPS24 in the high-metallicity regime than \RPAH\ or \RPAHst\ in the observations, with some region having almost no surviving PAHs even for low values of the \SIII/\SII\ ratio. Furthermore, the metallicity dependence is noticeable in this plot, while it is absent in the observational results. Similarly to the observations, the Cloudy models reproduce a different behavior of the PAH fraction tracer with \SIII/\SII\ in the low-metallicity regime ($\rm 12+\log(O/H)<8.2$), however, the resulting PAH fraction in the models is significantly higher than in the observations (especially for high \SIII/\SII\ ratio). Therefore, despite the qualitative agreement, the real evolution of PAHs around \HII\ regions is more complex and PAH destruction is unlikely just linearly dependent on the ionization fraction as assumed in the Cloudy models. 

\section{\revone{Online catalog}}
\label{app:catalog}

\revone{We report most of the measurements used or derived in this work in the online table accessible at the CDS. A list of the columns included in the catalog is provided in Table~\ref{tab:catalog}. The table contains information on \numHII\ \HII\ regions and \numSNRs\ SNRs selected according to the S/N and environmental criteria presented in Sections~\ref{sec:analysis_hii_snrs} and \ref{sec:analysis_snrs}. The parent samples of \HII\ regions and SNRs for the 19 PHANGS-MUSE galaxies are taken from the catalogs of \citet{Groves2023} and \citet{Li2024}, respectively. The full catalog of \HII\ regions for 23 additional galaxies will be published in future work (Egorov et al., in prep.). The exact identification of \HII\ regions in that work may differ from what is presented here. 
}

\begin{table*}
    \caption{\revone{Columns in the catalog accessible at the CDS.}}
    \label{tab:catalog}
    \centering
    \begin{small}
    \begin{tabularx}{\linewidth}{llX}
    \hline
        Column & Unit & Description \\
    \hline
        ID & & Region ID. Equal to the IDs from \citet{Groves2023} or \citet{Li2024} catalogs for 19 PHANGS-MUSE galaxies. \\
        GalName & & Galaxy name \\
        RA & deg & RA (J2000) of the region center \\
        Dec & deg & Dec (J2000) of the region center \\
        neb\_type & & Type of nebula (\HII\ region or SNR) \\
        r\_circ\_ang & arcsec & Circularized radius of the region based on the MUSE data$^a$ \\
        r\_circ\_hst & arcsec & Circularized radius of the region based on the HST-\Ha\ images (available for 19 galaxies)$^b$ \\
        r\_circ\_muse\_psf & & Ratio of r\_circ\_ang to the average PSF of the MUSE data$^c$ \\
        phys\_area & pc$^2$ & Physical area covered by the region, based on its borders derived from the MUSE data$^a$ \\ 
        phys\_area\_hst & pc$^2$ & Physical area covered by the region, based on its borders derived from the HST-\Ha\ images$^b$ \\
        \textsc{band}$^*$ & $\rm MJy\ sr^{-1}$ & Brightness in the specified JWST/MIRI band, averaged over the area covered by the region. \\
        \textsc{band}\_err$^*$ & MJy sr$^{-1}$ & Uncertainty of the \textsc{band} measurements \\
        F\_\textsc{line}$^\dag$ & $\rm 10^{-17} erg\ s^{-1}\ cm^{-2}$ & Total flux (corrected for reddening) in the specified optical emission line, integrated over the area covered by the region. \\
        F\_\textsc{line}\_err$^\dag$ & $\rm 10^{-17} erg\ s^{-1}\ cm^{-2}$ & Uncertainty of the F\_\textsc{line} measurements \\
        met\_scal &  & Oxygen abundance determined using the Scal prescription \citep{Pilyugin2016}$^d$ \\
        met\_scal\_err &  & Uncertainty of met\_scal \\ 
        RPAH & & \RPAH\ values (available for 19 galaxies) \\
        RPAH\_err & & \RPAH\ uncertainties \\
        RPAHst & & \RPAHst\ values \\
        RPAHst\_err & & \RPAHst\ uncertainties \\
    \hline
    \end{tabularx}
    \begin{footnotesize}
\revone{
\raggedright Notes: \\
\raggedright $^*$ \textsc{band} is one of the following JWST/MIRI bands: F770W, F770W\_ss (star-subtracted F770W), F1130W, F2100W. \\
\raggedright $^\dag$ \textsc{line} is one of the following keywords (emission lines): Ha (\Ha$\lambda6563$\AA), Hb (\Hb$\lambda4861$\AA), OIII5007 (\OIII$\lambda5007$\AA), NII6584 (\NII$\lambda6584$\AA), SII6717 (\SII$\lambda6717$\AA), SII6731 (\SII$\lambda6731$\AA), and SIII9069 (\SIII$\lambda9069$\AA). \\
\raggedright $^a$ By construction, the radii of all SNR apertures correspond to 25~pc. \\
\raggedright $^b$ Taken from Barnes et al., submitted; based on the data from \citet{Chandar2025} \\
\raggedright $^{c}$ Used to select the subsample of ``resolved'' \HII\ regions in this paper. \\
\raggedright $^{d}$ Taken from \citet{Groves2023} or \citet{Li2024} for \HII\ regions or SNRs from the 19 PHANGS-MUSE galaxies, respectively. \\
}
    \end{footnotesize}
    \end{small}
\end{table*}

\end{document}

%% file: authors.tex
%%%%%%%%%%%%%%%%%%%%%%%%%%%%%%%%%%%%%%%%%%%%%%%%%%%%%%%%%%%%%%%%%%%%%%%%%%%

\newcommand{\OSU}{\label{OSU} Department of Astronomy, The Ohio State University, 140 West 18th Avenue, Columbus, Ohio 43210, USA}

\newcommand{\Alberta}{\label{Alberta} Department of Physics, University of Alberta, Edmonton, AB T6G 2E1, Canada}

\newcommand{\ANU}{\label{ANU} Research School of Astronomy and Astrophysics, Australian National University, Canberra, ACT 2611, Australia}

\newcommand{\IPAC}{\label{IPAC} Caltech-IPAC, 1200 E. California Blvd. Pasadena, CA 91125, USA}

\newcommand{\Carnegie}{\label{Carnegi} Observatories of the Carnegie Institution for Science, 813 Santa Barbara Street, Pasadena, CA 91101, USA}

\newcommand{\CCAPP}{\label{CCAPP} Center for Cosmology and Astroparticle Physics, 191 West Woodruff Avenue, Columbus, OH 43210, USA}

\newcommand{\CfA}{\label{CfA} Harvard-Smithsonian Center for Astrophysics, 60 Garden Street, Cambridge, MA 02138, USA}

\newcommand{\CITEVA}{\label{CITEVA} Centro de Astronomía (CITEVA), Universidad de Antofagasta, Avenida Angamos 601, Antofagasta, Chile}

\newcommand{\CNRS}{\label{CNRS} CNRS, IRAP, 9 Av. du Colonel Roche, BP 44346, F-31028 Toulouse cedex 4, France}

\newcommand{\ESO}{\label{ESO} European Southern Observatory, Karl-Schwarzschild Stra{\ss}e 2, D-85748 Garching bei M\"{u}nchen, Germany}

\newcommand{\HD}{\label{HD} Astronomisches Rechen-Institut, Zentrum f\"{u}r Astronomie der Universit\"{a}t Heidelberg, M\"{o}nchhofstra\ss e 12-14, D-69120 Heidelberg, Germany}

\newcommand{\ICRAR}{\label{ICRAR} International Centre for Radio Astronomy Research, University of Western Australia, 35 Stirling Highway, Crawley, WA 6009, Australia}

\newcommand{\IRAM}{\label{IRAM} Institut de Radioastronomie Millim\'{e}trique (IRAM), 300 Rue de la Piscine, F-38406 Saint Martin d'H\`{e}res, France}

\newcommand{\ITA}{\label{ITA} Universit\"{a}t Heidelberg, Zentrum f\"{u}r Astronomie, Institut f\"{u}r Theoretische Astrophysik, Albert-Ueberle-Str 2, D-69120 Heidelberg, Germany}

\newcommand{\IWR}{\label{IWR} Universit\"{a}t Heidelberg, Interdisziplin\"{a}res Zentrum f\"{u}r Wissenschaftliches Rechnen, Im Neuenheimer Feld 205, D-69120 Heidelberg, Germany}

\newcommand{\JHU}{\label{JHU} Department of Physics and Astronomy, The Johns Hopkins University, Baltimore, MD 21218, USA}

\newcommand{\Leiden}{\label{Leiden} Leiden Observatory, Leiden University, P.O. Box 9513, 2300 RA Leiden, The Netherlands}

\newcommand{\Maryland}{\label{Maryland} Department of Astronomy, University of Maryland, College Park, MD 20742, USA}

\newcommand{\MPE}{\label{MPE} Max-Planck-Institut f\"{u}r extraterrestrische Physik, Giessenbachstra{\ss}e 1, D-85748 Garching, Germany}

\newcommand{\MPIA}{\label{MPIA} Max-Planck-Institut f\"{u}r Astronomie, K\"{o}nigstuhl 17, D-69117, Heidelberg, Germany}

\newcommand{\Nagoya}{\label{Nagoya} Department of Physics, Nagoya University, Furo-cho, Chikusa-ku, Nagoya, Aichi 464-8602, Japan}

\newcommand{\NRAO}{\label{NRAO} National Radio Astronomy Observatory, 520 Edgemont Road, Charlottesville, VA 22903-2475, USA}

\newcommand{\OAN}{\label{OAN} Observatorio Astron\'{o}mico Nacional (IGN), C/Alfonso XII, 3, E-28014 Madrid, Spain}

\newcommand{\ObsParis}{\label{ObsParis} Sorbonne Universit\'{e}, Observatoire de Paris, Universit\'{e} PSL, CNRS, LERMA, F-75014, Paris, France}

\newcommand{\Princeton}{\label{Princeton} Department of Astrophysical Sciences, Princeton University, Princeton, NJ 08544 USA}

\newcommand{\UToledo}{\label{UToledo} University of Toledo, 2801 W. Bancroft St., Mail Stop 111, Toledo, OH, 43606}

\newcommand{\Toulouse}{\label{Toulouse} Universit\'{e} de Toulouse, UPS-OMP, IRAP, F-31028 Toulouse cedex 4, France}

\newcommand{\UBonn}{\label{UBonn} Argelander-Institut f\"ur Astronomie, Universit\"at Bonn, Auf dem H\"ugel 71, 53121 Bonn, Germany}

\newcommand{\UChile}{\label{UChile} Departamento de Astronom\'{i}a, Universidad de Chile, Camino del Observatorio 1515, Las Condes, Santiago, Chile}

\newcommand{\UConn}{\label{UConn} Department of Physics, University of Connecticut, Storrs, CT, 06269, USA}

\newcommand{\UCSD}{\label{UCSD} Department of Astronomy \& Astrophysics,  University of California, San Diego, 9500 Gilman Drive, La Jolla, CA 92093, USA}

\newcommand{\UGent}{\label{UGent} Sterrenkundig Observatorium, Universiteit Gent, Krijgslaan 281 S9, B-9000 Gent, Belgium}

\newcommand{\ULyon}{\label{ULyon} Univ Lyon, Univ Lyon 1, ENS de Lyon, CNRS, Centre de Recherche Astrophysique de Lyon UMR5574,\\ F-69230 Saint-Genis-Laval, France}

\newcommand{\UMass}{\label{UMass} University of Massachusetts—Amherst, 710 N. Pleasant Street, Amherst, MA 01003, USA}

\newcommand{\UWyoming}{\label{UWyoming} Department of Physics and Astronomy, University of Wyoming, Laramie, WY 82071, USA}

\newcommand{\LAM}{\label{LAM} Aix Marseille Univ, CNRS, CNES, LAM (Laboratoire d’Astrophysique de Marseille), Marseille, France}

\newcommand{\UHawaii}{\label{UHawaii} Institute for Astronomy, University of Hawaii, 2680 Woodlawn Drive, Honolulu, HI 96822, USA}

\newcommand{\UCM}{\label{UCM} Departamento de F\'{\i}sica de la Tierra y Astrof\'{\i}sica, Universidad Complutense de Madrid, E-28040, Spain}

\newcommand{\IPARC}{\label{IPARC} Instituto de F\'{\i}sica de Part\'{\i}culas y del Cosmos IPARCOS, Facultad de Ciencias F\'{\i}sicas, Universidad Complutense de Madrid, E-28040, Spain}

\newcommand{\STScI}{\label{STScI} Space Telescope Science Institute, 3700 San Martin Drive, Baltimore, MD 21218, USA}

\newcommand{\McMaster}{\label{McMaster} Department of Physics and Astronomy, McMaster University, 1280 Main Street West, Hamilton, ON L8S 4M1, Canada}

\newcommand{\INAF}{\label{INAF} INAF -- Osservatorio Astrofisico di Arcetri, Largo E. Fermi 5, I-50157, Firenze, Italy}

\newcommand{\Sydney}{\label{Sydney} Sydney Institute for Astronomy, School of Physics A28, The University of Sydney, NSW 2006, Australia}

\newcommand{\UA}{\label{UA} Centro de Astronomía (CITEVA), Universidad de Antofagasta, Avenida Angamos 601, Antofagasta, Chile}

\newcommand{\CITA}{\label{CITA} Canadian Institute for Theoretical Astrophysics (CITA), University of Toronto, 60 St George St, Toronto, ON M5S 3H8, Canada}

\newcommand{\ASIAA}{\label{ASIAA} Institute of Astronomy and Astrophysics, Academia Sinica, No. 1, Sec. 4, Roosevelt Road, Taipei 106319, Taiwan}

\newcommand{\TKU}{\label{TKU} Department of Physics, Tamkang University, No.151, Yingzhuan Rd., Tamsui Dist., New Taipei City 251301, Taiwan}

\newcommand{\PSMA}{\label{PSMA} Penn State Mont Alto, 1 Campus Drive, Mont Alto, PA  17237, USA}

\newcommand{\ILL}{\label{ILL} ILL}

\newcommand{\Whitman}{\label{Whitman} Whitman College, 345 Boyer Avenue, Walla Walla, WA 99362, USA}

\newcommand{\Ox}{\label{Ox} Sub-department of Astrophysics, Department of Physics, University of Oxford, Keble Road, Oxford OX1 3RH, UK}

\newcommand{\stromlo}{\label{stromlo} Research School of Astronomy and Astrophysics, Australian National University, Mt Stromlo Observatory, Weston Creek, ACT 2611, Australia}

\newcommand{\FAU}{\label{FAU} Dr. Karl-Remeis Sternwarte and Erlangen Centre for Astroparticle Physics, Friedrich-Alexander Universitat Erlangen-Nurnberg, Sternwartstr. 7, 96049 Bamberg, Germany}

\newcommand{\Nice}{\label{Nice} Universit\'e C\^ote d'Azur, Observatoire de la C\^ote d'Azur, CNRS, Laboratoire Lagrange, 06000, Nice, France}

\newcommand{\Gemini}{\label{Gemini} International Gemini Observatory, NSF NOIRLab, 950 N Cherry Ave, Tucson, AZ 85719, USA}

\newcommand{\ESOchile}{\label{ESOchile} European Southern Observatory (ESO), Alonso de Córdova 3107, Casilla 19, Santiago 19001, Chile} 

\newcommand{\UNAM}{\label{UNAM}
Instituto de Astronom\'ia, Universidad Nacional Aut\'onoma de M\'exico, Ap. 70-264, 04510 CDMX, Mexico}

\newcommand{\KIPAC}{\label{KIPAC} Kavli Institute for Particle Astrophysics \& Cosmology, Stanford University, CA 94305, USA}

\newcommand{\CDU}{\label{CDU} Center for Decoding the Universe, Stanford University, CA 94305, USA}

\newcommand{\UnknownInst}{\label{UnknownInst} What is your affiliation?}
%%%%%%%%%%%%%%%%%%%%%%%%%%%%%%%%%%%%%%%%%%%%%%%%%%%%%%%%%%%%%%%%%%%%%%%%%%%

\author{
        Oleg V. Egorov\orcid{0000-0002-4755-118X}\inst{\ref{HD}}\thanks{\email{oleg.egorov@uni-heidelberg.de}}  
        \and
    Adam~K.~Leroy\orcid{0000-0002-2545-1700}
\inst{\ref{OSU}, \ref{CCAPP}}
\and
    Karin Sandstrom\orcid{0000-0002-4378-8534}\inst{\ref{UCSD}}
    \and 
    Kathryn Kreckel\orcid{0000-0001-6551-3091}\inst{\ref{HD}}
    \and
    Dalya Baron\orcid{0000-0003-4974-3481}\inst{\ref{KIPAC}, \ref{CDU}} 
    \and
    Francesco Belfiore\orcid{0000-0002-2545-5752}\inst{\ref{INAF}} 
       \and
    Ryan Chown\orcid{0000-0001-8241-7704}\inst{\ref{OSU}}
    \and 
    Jessica Sutter\orcid{0000-0002-9183-8102}\inst{\ref{Whitman}}
    \and
   M\'ed\'eric Boquien\orcid{0000-0003-0946-6176} 
\inst{\ref{Nice}}
\and 
    Mar Canal i Saguer\inst{\ref{FAU}}
    \and
    Enrico Congiu\orcid{0000-0002-8549-4083}\inst{\ref{ESOchile}}
    \and
Daniel~A.~Dale\orcid{0000-0002-5782-9093}\inst{\ref{UWyoming}}
    \and
    Evgeniya Egorova\orcid{0000-0003-2717-8784}\inst{\ref{HD}}
    \and
    Michael Huber\inst{\ref{HD}}
    \and
    Jing Li\orcid{0000-0002-4825-9367}\inst{\ref{HD}}
   \and
Thomas~G.~Williams\orcid{0000-0002-0012-2142}
    \inst{\ref{Ox}}
    \and
    J\'er\'emy Chastenet\orcid{0000-0002-5235-5589}\inst{\ref{UGent}}
    \and 
    I-Da Chiang\orcid{0000-0003-2551-7148}\inst{\ref{ASIAA}} 
    \and
    Ivan Gerasimov\orcid{0000-0001-7113-8152}\inst{\ref{Nice}}
    \and
    Hamid Hassani\orcid{0000-0002-8806-6308}\inst{\ref{Alberta}}
\and 
   Hwihyun Kim\orcid{0000-0003-4770-688X}\inst{\ref{Gemini}} 
   \and 
   Hannah Koziol\orcid{0009-0001-5949-1524}\inst{\ref{UCSD}}
   \and
   Janice C. Lee\orcid{0000-0003-0946-6176}\inst{\ref{STScI}}
   \and
   Rebecca L. McClain\orcid{0000-0002-6187-4866}\inst{\ref{OSU}, \ref{CCAPP}}
\and
Jos\'e Eduardo M\'endez Delgado\orcid{0000-0002-6972-6411}\inst{\ref{UNAM}}
   \and
Hsi-An Pan\orcid{0000-0002-1370-6964}\inst{\ref{TKU}}
\and 
   Debosmita Pathak\orcid{0000-0003-2721-487X
}\inst{\ref{OSU}}
    \and
    Erik Rosolowsky\orcid{0000-0002-5204-2259}\inst{\ref{Alberta}}
    \and 
    Sumit K. Sarbadhicary\orcid{0000-0002-4781-7291}
\inst{\ref{OSU}, \ref{CCAPP}}
\and
    Eva Schinnerer\orcid{0000-0002-3933-7677}\inst{\ref{MPIA}}
    \and 
    David Thilker\orcid{0000-0002-8528-7340}\inst{\ref{JHU}}
    \and
    Leonardo Ubeda\orcid{0000-0001-7130-2880}\inst{\ref{STScI}}
    \and
    Tony Weinbeck\orcid{0009-0005-8923-558X}\inst{\ref{UWyoming}}
% PHANGS (preliminary order)\inst{\ref{UnknownInst}}
}

\institute{\HD \and \OSU \and \CCAPP \and \UCSD  \and \KIPAC \and \CDU \and \INAF  \and \Whitman  \and  \Nice \and \FAU  \and \ESOchile \and \UWyoming \and \Ox   \and \UGent \and \ASIAA \and \Alberta \and  \Gemini \and \STScI \and \UNAM \and \TKU \and \MPIA \and \JHU %\and 
%\UnknownInst
}

%=====
% add your name, affiliation and orcid below:
%=====
% 
%commented, not signed (TBC):
% I-Da Chiang
% Jeremy Chastenet
% Ivan
% Eduardo 